\documentclass[10pt,a4paper]{article}

\usepackage[top=2cm, bottom=3cm, left=2.5cm, right=2cm]{geometry}
\usepackage{amssymb,amsmath,amsxtra,amsthm,bm}
\usepackage{amsbsy}
\numberwithin{equation}{section}

\usepackage{algorithmicx}
\usepackage[ruled]{algorithm}
\usepackage{algpseudocode}
\usepackage{algpascal}

\usepackage{graphicx}
\usepackage{epstopdf}
\usepackage{caption}
\usepackage{float}
\usepackage{subfigure}
\graphicspath{{figs/}}

\usepackage{booktabs}
\usepackage{multirow}

\newtheorem{example}{Example}

\usepackage[latin1]{inputenc}
\usepackage{tikz}
\usetikzlibrary{shapes,arrows}

\usepackage{titling} 

\title{\textbf{A Re-weighted Joint Spatial-Radon Domain CT Image Reconstruction Model for Metal Artifact Reduction}}

\author{
Haimiao Zhang
\thanks{School of Mathematics and Statistics, Wuhan University, Wuhan 430072, China(haimiaozh@whu.edu.cn). The research of this author is supported by the Fundamental Research Funds for the Central Universities under Grant 2015201020202, and NSFC grant 91530321. }
\and
 Bin Dong
 \thanks{Corresponding author. Beijing International Center for Mathematical Research, Peking University, Beijing 100871, China (dongbin@math.pku.edu.cn). The research of this author is supported by NSFC grant 91530321. }
\and
Baodong Liu
\thanks{Corresponding author. Division of Nuclear Technology and Applications, Institute of High Energy Physics, Chinese Academy of Sciences, Beijing 100049, China, and
Beijing Engineering Research Center of Radiographic Techniques and Equipment, Beijing 100049, China, and
University of Chinese Academy of Sciences, Beijing 100049, China (liubd@ihep.ac.cn). The research of this author is supported by the National Key Scientific Instrument and Equipment Development Project 2017YFF0107200,  the Instrument Developing Project of the Chinese Academy of Sciences YZ201410 and SRF for ROCS, SEM.
}
}

\usepackage[colorlinks,
            linkcolor=red,
            anchorcolor=blue,
            citecolor=red]{hyperref}

\begin{document}
\maketitle

\begin{abstract}
High density implants such as metals often lead to serious artifacts in the reconstructed CT images which hampers the accuracy of image based diagnosis and treatment planning. In this paper, we propose a novel wavelet frame based CT image reconstruction model to reduce metal artifacts. This model is built on a joint spatial and Radon (projection) domain (JSR) image reconstruction framework with a built-in weighting and re-weighting mechanism in Radon domain to repair degraded projection data. The new weighting strategy used in the proposed model makes the regularization in Radon domain by wavelet frame transform more effective. The proposed model, which will be referred to as the re-weighted JSR model, combines the ideas of the recently proposed wavelet frame based JSR model \cite{Dong2013} and the normalized metal artifact reduction model \cite{meyer2010normalized}, and manages to achieve noticeably better CT reconstruction quality than both methods. To solve the proposed re-weighted JSR model, an efficient alternative iteration algorithm is proposed with guaranteed convergence. Numerical experiments on both simulated and real CT image data demonstrate the effectiveness of the re-weighted JSR model and its advantage over some of the state-of-the-art methods.

  \textbf{Keywords:} Computerized tomography, metal artifact reduction, tight wavelet frame, joint spatial-Radon domain reconstruction.
\end{abstract}

\section{Introduction}

X-ray based computerized tomography (CT) is one of the most important imaging modalities in diagnosis and treatment planning of various diseases such as cancer. Although magnetic resonance imaging (MRI) is a safer alternative to CT, the acquisition time of MRI is normally much longer than CT, and it is prohibited for patients with metal implants. However,  metal implants often lead to serious degradations in the reconstructed CT images. This is mainly due to the commonly adopted assumption that the CT imaging process is linear with monochromatic X-ray source. This assumption works well when the imaged region contains similar types of materials such as soft tissue and muscles which have similar attenuation coefficient \cite{hubbell1995tables}. In practice, however, the X-ray source available is multi-chromatic and the detected photons are only the mean energy after attenuation. When the X-ray beam passes through high density materials such as bones and metals, the low energy components of the X-ray are attenuated more than high energy components.  As a result, the reconstructed CT image by a simple model such as the widely used filtered backprojection will suffer serious degradations \cite{buzug2008computed}. Therefore, more careful design of the CT image reconstruction model is necessary to suppress metal artifacts. In this paper, we focus on metal artifact reduction in 2-dimensional (2D) CT image reconstruction. Generalization to 3D is straightforward in theory, though requires further numerical studies.

Based on the Lambert-Beer Law, the measured photon intensity from  multi-chromatic energy X-ray can be described by
\begin{equation}\label{def-Y-with-spectral}
\bm{\mathcal{I}}= \int_{E_{m}}^{E_{M}} \bm{S}(E)e^{-\bm{\mathcal{P}}\bm{u}_{E}}dE,
\end{equation}
where $\bm{\mathcal{P}}$ denotes the imaging process, $ \bm{S}(E)=\bm{\mathcal{I}}_{0}(E)/\int_{E_{m}}^{E_{M}}\bm{\mathcal{I}}_{0}(E)dE$ with $\bm{\mathcal{I}}_{0}(E)$ being the incident photons of X-ray at energy level $E\in[E_{m},E_{M}]$, and $\bm{u}_{E}$ is the linear attenuation coefficient at energy level $E$. In this case, the imaging process is nonlinear if the photon intensity $\bm{S}(E)$ is not constant and $\bm{u}_{E}$ is varied at each energy level $E$. In practice, the exact $\bm{S}(E)$ at each energy level $E$ is not readily available and difficult to measure, and solving a huge nonlinear system is overwhelmingly challenging. Therefore, monochromatic energy is a common assumption adopted in the literature. In the monochromatic energy X-ray imaging, model \eqref{def-Y-with-spectral} is simplified to
\begin{equation}\label{multi-energy-Xray}
\bm{\mathcal{P}}\bm{u}\approx \bm{Y},
\end{equation}
where $\bm{Y}=-\ln\bm{\mathcal{I}}$ is the measured projection data with $\bm{\mathcal{I}}$ obtained from \eqref{def-Y-with-spectral}. However, since \eqref{multi-energy-Xray} is merely an approximation of \eqref{def-Y-with-spectral}, when high density components such as metal implants are present, simple models solving the linear inverse problem \eqref{multi-energy-Xray} may lead to severe metal artifacts in the reconstructed images. The regions corresponding to metals in the Radon domain, which are referred to as the metal trace, are nonlinearly degraded. Most existing metal artifact reduction methods introduce different techniques to repair the degraded projection data $\bm{Y}$ or/and to suppress metal artifacts in spatial domain using regularization.

Classical CT image reconstruction algorithms include filtered backprojection(FBP) \cite{katsevich2002theoretically}, FDK \cite{feldkamp1984practical} and Simultaneous Algebraic Reconstruction Technique(SART) \cite{andersen1984simultaneous,wang1996iterative}.  Since the problem \eqref{multi-energy-Xray} is an ill-posed linear inverse problem, when metals are present in the reconstructed images, images reconstructed by these classical approaches may be seriously degraded. In the past three decades, different strategies for example, hardware filtering, dual energy CT (DECT) \cite{dilmanian1997single,bamberg2011metal} and statistical reconstruction methods are combined with the aforementioned classical reconstruction algorithms to reduce the metal artifacts. Nonetheless, suppressing metal artifact in CT images remains a great challenge.

In image processing, regularization based approach has been playing an important role, since most of the image processing tasks, such as super-resolution, deblurring, inpanting, etc., are ill-posed inverse problems. Total variation (TV) \cite{rudin1992nonlinear,aubert2006mathematical,ChanShen} and wavelet frame based approach \cite{Chan2004Wavelet,Chan2004Tight,cai2012image,Dong2010IASNotes} are two successful examples among others. In low-dose CT imaging, the  iterative reconstruction model with TV regularization \cite{sidky2008image,ritschl2011improved,Jia2010gpu} and wavelet frame based regularization \cite{tian2011low,liu2012adaptive,Jia2011GPU,zhang2013l0,zhou2013adaptive,Dong2013,zhan2016ct,choi2016limited} were proposed and achieved better visual quality of the reconstructed images than traditional methods. However, these methods are effective in suppressing artifacts caused by noise and lack of projection data, while they are less effective in suppressing metal artifacts.

Recently in \cite{meyer2010normalized}, the authors proposed a normalized metal artifact reduction (NMAR) model which outperforms classical reconstruction methods such as FBP. It first normalizes the measured projection data using the projection of a segmented image which is obtained from thresholding of a roughly reconstructed image. Then, interpolation is conducted on the normalized projection data to repair the degradations caused by metals. Finally, it reconstructs the CT image from the repaired projection data using FBP. However, the NMAR model only works well when noise level of the projection data is low. In \cite{mehranian2013x}, a nonconvex CT image metal artifact reduction model with sparsity based interpolation in wavelet domain was shown to outperform the NMAR model. But the convergence of this model is not guaranteed since the $\ell_0$-norm was used to promote sparsity. A TV regularization based metal artifact reduction model was proposed in \cite{zhang2016iterative} and it outperforms the NMAR method on simulated and real data set at the presence of noise. Other TV based metal artifact reduction models can be found in \cite{zhang2013hybrid,zhang2011new} and wavelet frame based model in \cite{zhao2000x}. Note that, the TV based model works well for piecewise constant images, whereas textures in the image may be smeared out. The prior image constrained iterative compressed sensing(PICCS) reconstruction technique  is able to reduce metal artifacts \cite{bannas2016prior} by estimating the removed metal degraded projection data with prior image constrains, which also can be used to reduce the limited angle artifacts \cite{chen2015synchronized}.

Another strategy to reduce the metal artifact is based on the analysis of the geometric property of the metal trace in Radon domain, which does not require any interpolation. Metal artifacts in the reconstructed CT image is characterized mathematically by the wavefront set through microlocal analysis of the projection data \cite{park2014characterization}.  Based on such characterization, a beam hardening corrector is proposed in \cite{park2016metal} to reduce the metal artifacts. However, this model requires the knowledge of the X-ray energy spectrum which is generally hard to acquire in practice. One may refer to  \cite{gjesteby2016metal} for a systematic review of the metal artifact reduction techniques in CT image reconstruction.

In this work, we propose a new model for metal artifact reduction in CT image reconstruction. This model, which will be referred to as the re-weighted JSR model, has a weighting and re-weighting mechanism naturally embedded in a framework of joint Radon domain inpainting and spatial domain regularization. The re-weighted JSR model is composed of different terms specialized in different tasks, and yet highly collaborative with each other. In one of the terms, weights are applied to the projection data, which are calculated from the projection of a segmented CT image. After weighting the projection data, the inpainting in Radon domain becomes easier since the weighted projection data has sparser representation in the wavelet frame domain than the original projection data. The effect of weighting is cancelled out in another term of the re-weighted JSR model to ensure a correct reconstruction in spatial domain. Such weighting strategy also ensures that the corrected projection data and the unknown CT image obey the linear model \eqref{multi-energy-Xray} better than the original projection data. Regularization by wavelet frame transforms is applied in both spatial and Radon domain to suppress noise and preserve features, and to facilitate a stable image reconstruction.

The rest of the paper is organized as follows. In Section \ref{section-initial-model}, we briefly review the concept of wavelet frames and wavelet frame transforms, followed by a detailed description of how metal trace and weights that will be needed in the re-weighted JSR model are estimated. Section \ref{section-JSR-model} introduces the re-weighted JSR model and an efficient optimization algorithm solving the proposed model. Numerical experiments are presented in Section \ref{section-numerical} for phantoms and in Section \ref{section-numerical-real} for a real scan data. Finally, we summarize our contributions and findings in Section \ref{sec-conclusion}.

\section{Preliminaries}\label{section-initial-model}

\subsection{Tight wavelet frames}

In this subsection, we briefly recall some of the basics of tight wavelet frames that will be used in later sections. Tight wavelet frame systems are redundant systems. Their redundancy not only grants robustness to the system, but also grants vast flexibility in designing frame systems satisfying properties that are desirable in various applications (e.g. short support, symmetry and high order of vanishing moments). Together with their multiscale structure, tight wavelet frame systems can robustly decompose piecewise smooth functions (such as images) into smooth and sparse components, which is the key to their success in image restoration. The interested readers should consult \cite{ron1997affine,ron1997affine2,daubechies1992ten,Daubechies2003Framelets} for theories of wavelet frames, \cite{shen2010wavelet,dong2015image} for a short survey on the theory and applications of wavelet frames, and \cite{Dong2010IASNotes} for a more detailed survey.

Given a set of compactly supported functions $\Psi=\{ \psi_{\ell}\in L_{2}(\mathbb{R}^{d}): 1\le \ell\le r\}$, with $d\in\mathbb{N}$, the quasi-affine system $X(\Psi)$ generated by $\Psi$ is defined by the collection of dilations and shifts of $\Psi$ as
\begin{equation}\label{frame-MRA-system}
X(\Psi)=\{\psi_{\ell,j,\bm{k}}\in L_{2}(\mathbb{R}^{d}):1\le \ell\le r, j\in \mathbb{Z}, \bm{k}\in \mathbb{Z}^{d}\}
\end{equation}
where $\psi_{\ell,j,\bm{k}}$ is defined by
\begin{eqnarray}
\psi_{\ell,j,\bm{k}}=\left\{\begin{array}{ll}
2^{\frac{jd}{2}}\psi_{\ell}(2^{j}\cdot-\bm{k}), & j\ge 0,\\
2^{2jd}\psi_{\ell}(2^{j}\cdot-2^{j}\bm{k}), & j<0.
\end{array} \right.
\end{eqnarray}
Then, $X(\Psi)$ is called a tight wavelet frame of $L_{2}(\mathbb{R}^{d})$ if
$$
\|f\|_{L_{2}(\mathbb{R}^{d})}^{2}= \sum_{\psi\in X(\Psi)}|\langle f,\psi\rangle|^{2}.
$$
In the tight wavelet frame system $X(\Psi)$, each generator function $\psi_{\ell}$ is called a framelet.

The constructions of compactly supported and desirably (anti)symmetric framelets $\Psi$ are usually based on the multiresolution analysis (MRA) generated by some refinable function $\phi$ with refinement mask $\bm{h}_0$ satisfying
\begin{equation*}\label{phimaskphi}
\phi=2^d\sum_{\bm{k}\in\mathbb{Z}^d}\bm{h}_0[\bm{k}]\phi(2\cdot-\bm{k}).
\end{equation*}
The idea of an MRA-based construction of framelets $\Psi=\{\psi_1,\ldots, \psi_r\}$ is to find masks $\bm{h}_\ell$, which are finite sequences (or filters), such that
\begin{equation}\label{psimaskphi}
\psi_{\ell} =
2^{d}\sum_{\bm{k}\in\mathbb{Z}^d}{\bm{h}_{\ell}[\bm{k}]\phi(2\cdot-\bm{k})},\quad \ell=1,2,\ldots,r.
\end{equation}
The sequences $\bm{h}_1,\ldots,\bm{h}_r$ are called wavelet frame masks, or the high pass filters associated to the tight wavelet frame system, and $\bm{h}_0$ is also known as the low pass filter.

The unitary extension principle (UEP) \cite{ron1997affine} provides a rather general characterization of MRA-based tight wavelet frames. Roughly speaking, as long as $\{\bm{h}_1,\ldots,\bm{h}_r\}$ are finitely supported and their Fourier series $\widehat{\bm{h}}_\ell$ satisfy
\begin{equation}\label{UEPCondition}
\sum_{\ell=0}^{r}|\widehat{\bm{h}}_{\ell}(\bm{\xi})|^2=1
\quad\text{and}\quad
\sum_{\ell=0}^{r}\widehat{\bm{h}}_{\ell}(\bm{\xi})\overline{\widehat{\bm{h}}_{\ell}(\bm{\xi}+\bm{\nu})}=0,
\end{equation}
for all $\bm{\nu}\in\{0,\pi\}^d\setminus\{\bm{0}\}$ and $\bm{\xi}\in[-\pi,\pi]^d$, the quasi-affine system $X(\Psi)$ with $\Psi=\{\psi_1,\ldots, \psi_r\}$ defined by \eqref{psimaskphi} forms a tight frame of $L_2(\mathbb{R}^d)$. Note that the quasi-affine system is shift-invariant and it corresponds to the undecimated wavelet (frame) transform \cite{ron1997affine,coifman1995translation}, which performs better than the traditional wavelet system in image restoration. Furthermore, undecimated wavelet frame transform has a more natural link to differential operators in the framework of variational and PDEs models \cite{cai2012image,cai2016image,dong2017image,dong2017general}.

\begin{example}\label{linear-B-spline}
Let $\bm{h}_{0}=\frac{1}{4}[1,2,1]$ be the refinement mask of the piecewise linear B-spline $B_{2}(x)=\max\{1-|x|,0\}$. Define the high pass filters $\bm{h}_{1}=\frac{\sqrt{2}}{4}[1,0,-1]$ and $\bm{h}_{2}=\frac{1}{4}[-1,2,-1]$. Then $\bm{h}_{0}, \bm{h}_{1}, \bm{h}_{2}$ satisfy the UEP \eqref{UEPCondition}. Hence, the system $X(\Psi)$ with $\Psi=\{\psi_{1},\psi_{2}\}$ defined by \eqref{frame-MRA-system} is a tight frame of $L_{2}(\mathbb{R})$.
\end{example}

\begin{example}\label{cubic-B-spline}
Let $\bm{h}_{0}=\frac{1}{16}[1,4,6,4,1]$ be the refinement mask of the piecewise cubic B-spline $B_{4}$. Define the high pass filters $h_{1}$, $h_{2}$, $h_{3}$, $h_{4}$ as follows:
\begin{eqnarray}
\bm{h}_{1}=\frac{1}{16}[1,-4,6,-4,1], \quad \bm{h}_{2}=\frac{1}{8}[-1,2,0,-2,1],\\
\bm{h}_{3}=\frac{\sqrt{6}}{16}[1,0,-2,0,1], \quad \bm{h}_{4}=\frac{1}{8}[-1,-2,0,2,1].
\end{eqnarray}
Then $\bm{h}_{0}, \bm{h}_{1}, \bm{h}_{2}, \bm{h}_{3}, \bm{h}_{4}$ satisfy the UEP \eqref{UEPCondition} and the system $X(\Psi)$ with $\Psi=\{\psi_{1},\psi_{2}, \psi_{3}, \psi_{4}\}$ defined by \eqref{frame-MRA-system} is a tight frame of $L_{2}(\mathbb{R})$.
\end{example}

In the discrete setting, the $L$-level wavelet frame transform/decomposition with filter banks $\{\bm{h}_{0},\bm{h}_{1},...,\bm{h}_{r}\}$ is defined by
$$
\bm{W}\bm{u}=\{\bm{W}_{\ell,l}\bm{u}:\; 1\le \ell \le r, 0\le l \le L-1 \}\cup\{\bm{W}_{0,L-1}\bm{u}\},\quad \forall \bm{u} \in \ell_2(\mathbb{Z}^{2}).
$$
The wavelet frame coefficients of image $\bm{u}$ are computed by $\bm{W}_{\ell,l}\bm{u}=\bm{h}_{\ell,l}[-\cdot]\circledast \bm{u},$ where $\circledast$ denotes the convolution operator with a certain boundary condition, e.g., periodic boundary condition, and $\bm{h}_{\ell,l}$ is defined by
\begin{eqnarray}
\bm{h}_{\ell,l}=\widetilde{\bm{h}}_{\ell,l}\circledast \widetilde{\bm{h}}_{l-1,0} \circledast...\circledast \widetilde{\bm{h}}_{0,0} \; \mbox{with} \quad
\widetilde{\bm{h}}_{\ell,l}[\bm{k}]=\left\{\begin{array}{ll}
\bm{h}_{\ell}[2^{-l}\bm{k}], &  \bm{k}\in 2^{l}\mathbb{Z}^{2},\\
0, & \bm{k}\notin 2^{l}\mathbb{Z}^{2}.
\end{array} \right.
\end{eqnarray}
Denote the inverse wavelet frame transform (or wavelet frame reconstruction operator) as $\bm{W}^{\top}$, which is the adjoint operator of $\bm{W}$, then the UEP leads to a perfect reconstruction formula as
$$
\bm{u}=\bm{W}^{\top}\bm{W}\bm{u}, \quad \forall \bm{u}.
$$

For the 2D image processing, fast decomposition/reconstruction algorithms are available for the quasi-affine system $X(\Psi)$, which is constructed by the tensor product from the univariate wavelet frame.
We briefly recall how a set of 2D filters can be constructed from a given set of 1D filters by tensor product. Given a set of univariate masks $\{\bm{h}_\ell: \ell=0,1,\ldots,r\}$, define the 2D masks
$\bm{h}_{i,j}[k_1,k_2]$ as
\begin{equation*}
\bm{h}_{i,j}[k_1,k_2]:=\bm{h}_i[k_1]\bm{h}_j[k_2],\quad 0\le i,j\le r;\
(k_1,k_2)\in\mathbb{Z}^2.
\end{equation*}
It is known that if the set of 1D filters satisfies the UEP conditions, so does the corresponding set of 2D filters constructed by tensor product \cite{Dong2010IASNotes}.

Among many different choices of framelets, the ones constructed from B-splines are popular in image processing. This is mainly due to their short supports and symmetry which are desirable in many applications. A tight frame system constructed from a low order B-spline has less filters and each filter has shorter support than the tight frame systems constructed from high order B-splines. Thus, low order B-spline framelets are more computationally efficient to use than high order B-spline framelets, whereas the latter can capture richer image features than the former. Furthermore, since high order B-spline framelets have larger supports, they may introduce more numerical viscosity which often lead to smoother reconstructions in image restoration tasks such as denoising, deblurring, inpainting, etc. Therefore, the choice of B-spline framelets really depends on the task and the computation load one can afford. In all the models we use in the following sections, we choose the Haar framelets for spatial domain regularization and the piecewise cubic B-pline framelets for projection domain regularization. The reason for such choice is simply because most CT images can be well approximated by piecewise constant functions and their corresponding projection images have higher regularity. Throughout the rest of the paper, we fix the level of decomposition to $L=3$. We finally note that, choices of $\bm{W}$ (e.g. choice of framelets and levels of decomposition) indeed affect the reconstruction. For example, using a data-driven tight frame can generate better reconstructions than B-spline framelets in sparse view \cite{zhan2016ct} and limited view \cite{choi2016limited} CT reconstruction. However, we forgo a more detailed discussion on the choices of $\bm{W}$ in order not to dilute the main focus of this paper.

\subsection{Initialization}

The proposed re-weighted JSR model requires a pre-estimation of the metal trace and weights in projection domain. They can be obtained fairly easily from a roughly reconstructed CT image using a simple reconstruction model. In this paper, we use the tight wavelet frame based analysis model \cite{cai2009split}. This subsection describes the details on how metal trace and weights are computed using the NURBS-based cardiac-torso(NCAT) phantom \cite{segars2008realistic}. Two metal components are implanted in the NCAT phantom as shown in Figure \ref{clear-NCAT-with-marked-metal} and the simulated projection data is obtained from a multi-chromatic X-ray source. Details on the settings of the imaging system are postponed to Section \ref{section-simulate-phantom-design}.

\subsubsection{Metal trace estimation}
Given the measured projection data $\bm{Y}$ from the multi-chromatic X-ray source imaging system \eqref{multi-energy-Xray} and the projection operator $\bm{\mathcal{P}}$ , the unknown CT image can be approximately reconstructed by the following  tight wavelet frame based analysis model
\begin{equation}\label{analysis-model}
  \min_{\bm{u} } \frac{1}{2}||\bm{\mathcal{P}u-Y}||^2+ ||\bm{\lambda}\cdot \bm{W}\bm{u}||_{1,2},
\end{equation}
where $\bm{W}$ is the tight wavelet frame transform as reviewed in the previous subsection, and $\|\cdot\|$ is the $\ell_{2}$ norm.
The second term of \eqref{analysis-model} is defined by
$$||\bm{\lambda} \cdot \bm{Wu}||_{1,2}=\left\Vert\sum_{l=0}^{L-1}\left(\sum_{\ell=1}^r\lambda_{\ell,l}|(\bm{Wu})_{\ell,l}|^{2}\right)^{\frac{1}{2}}\right\Vert_{1}.$$ The definition of the $\ell_{1,2}$ norm was first introduced by \cite{cai2012image}, whose corresponding proximal operator is the isotropic soft shrinkage \eqref{define-soft-threshold-2}.

\begin{figure}[h!]
\center
    \subfigure[ ]{\includegraphics[scale=0.38,trim=0cm 0cm 0cm 0cm]{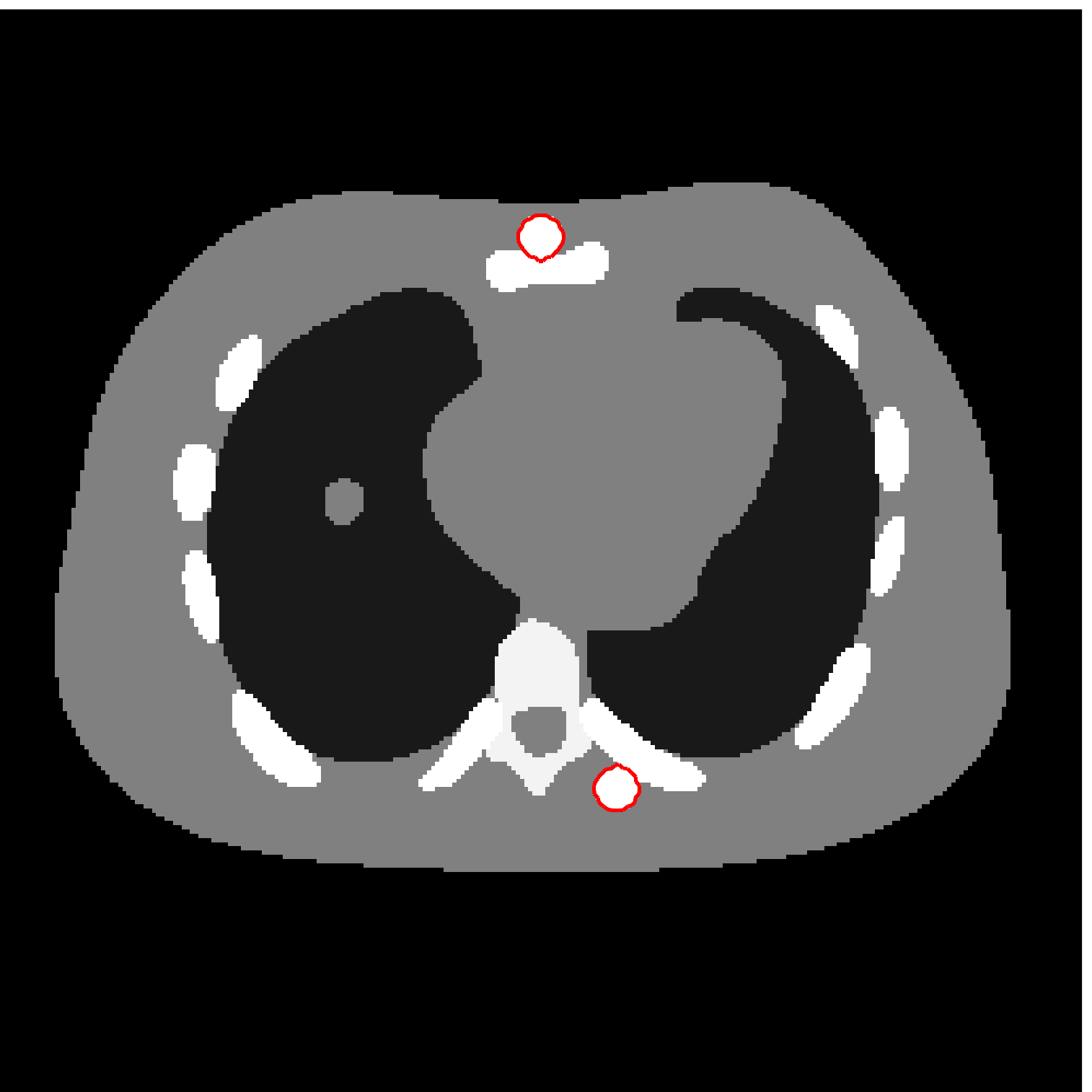}\label{clear-NCAT-with-marked-metal}}
    \subfigure[ ]{\includegraphics[scale=0.38,trim=0cm 0cm 0cm 0cm]{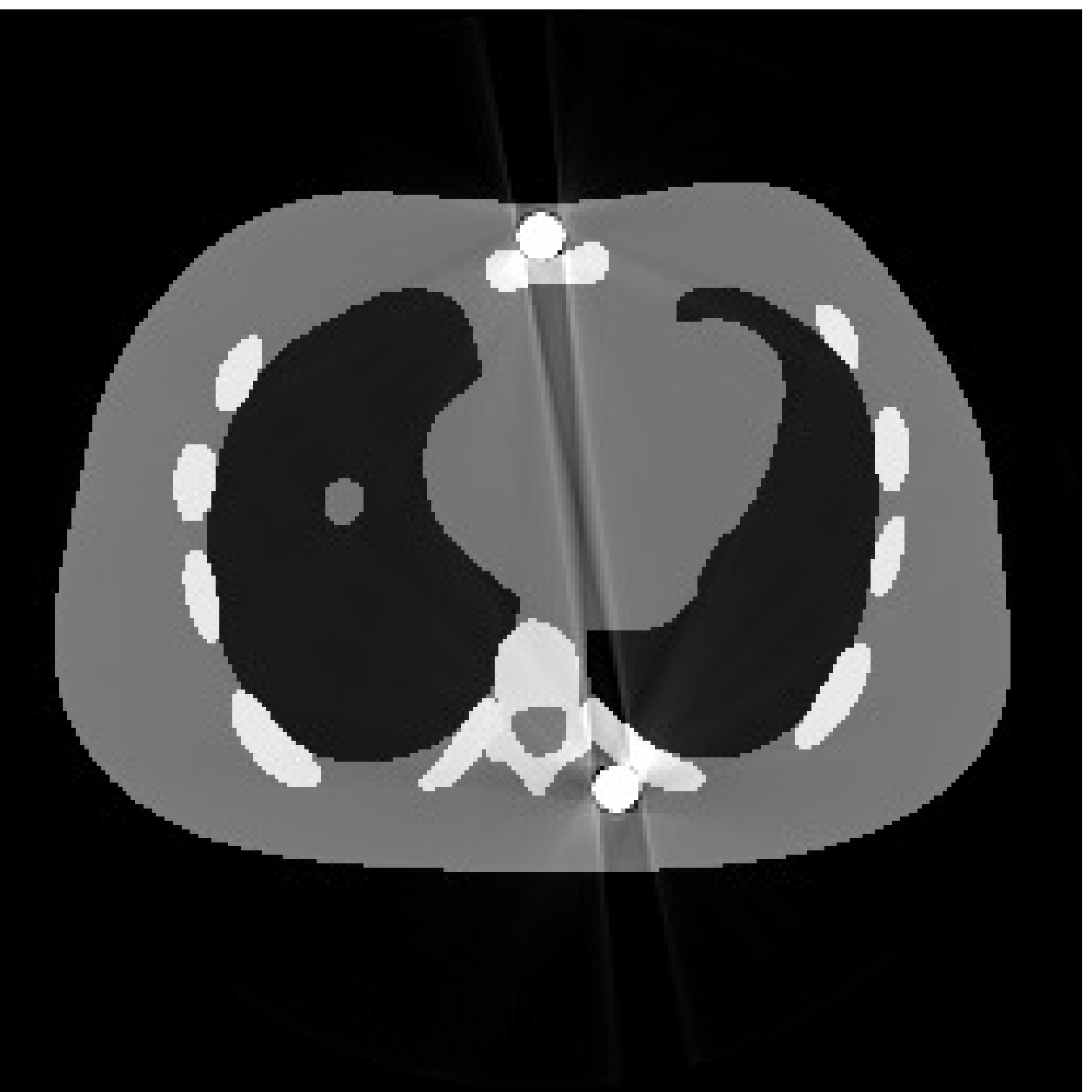}\label{no-noise-NCAT}}
    \subfigure[ ]{\includegraphics[scale=0.38,trim=0cm 0cm 0cm 0cm]{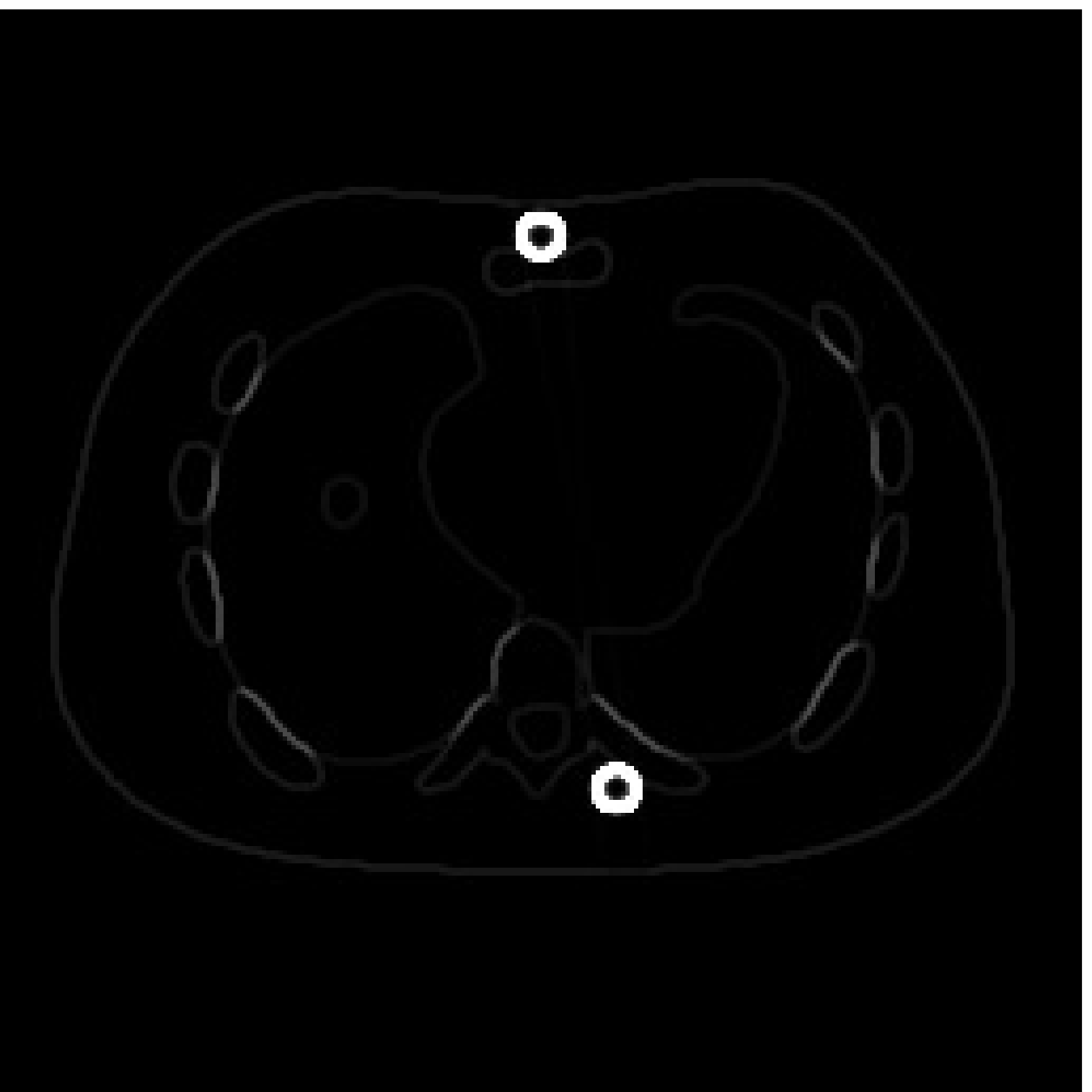}\label{Coeff-mag-NCAT}}
    \subfigure[ ]{\includegraphics[scale=0.38,trim=0cm 0cm 0cm 0cm]{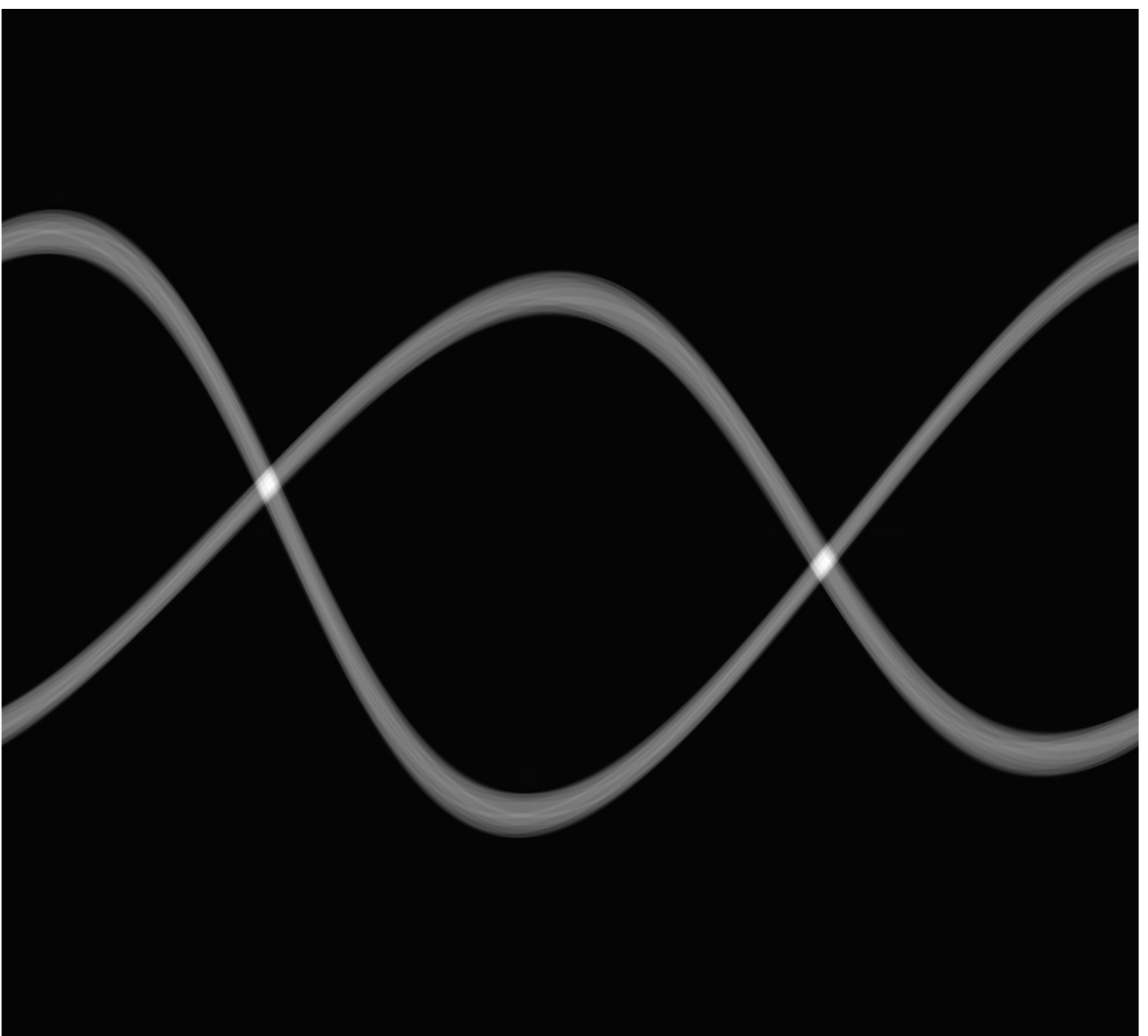}\label{metalTrace_NCAT}}
    \caption{
    \subref{clear-NCAT-with-marked-metal} The reference NCAT phantom with red curve marked metal components.
    \subref{no-noise-NCAT} The reconstructed image $\bm{u}_a$ from the analysis model \eqref{analysis-model}.
    \subref{Coeff-mag-NCAT} Summation of high frequency wavelet coefficients of reconstructed image.
    \subref{metalTrace_NCAT} Projection data of the estimated metal position indicator function in Radon domain.
    }\label{metal-trace-NCAT}
\end{figure}

The optimization problem \eqref{analysis-model} can be solved by the split Bregman algorithm \cite{goldstein2009split,cai2009split} efficiently, which is also equivalent to the alternating direction method of multipliers (ADMM) \cite{Esser2009Applications,Gabay1976A,Glowinski1975Sur}. The reconstructed phantom image by model \eqref{analysis-model}, denoted by $\bm{u}_{a}$, is shown in Figure \ref{no-noise-NCAT}. Metal location in image domain can be robustly estimated by the summation of the high frequency wavelet frame coefficients (Figure \ref{Coeff-mag-NCAT}) followed by a simple thresholding. Then, the index of the metal trace in Radon domain, denoted by $\bm{\Gamma}$, can be identified by the projection of the indicator function associated to the metal location (Figure \ref{metalTrace_NCAT}).

Note that one may estimate the metal location by simply thresholding the initially reconstructed image. However, the metal artifacts may have a significant influence on the estimation if the threshold is not properly chosen. Thanks to the multiscale structure of the wavelet frame transform, we are able to robustly detect features from poorly reconstructed images based on the summation of high frequency tight framelet coefficients. This has already been observed in the past \cite{dong2010frame,cai2016image}. In Table \ref{diff-seg-rec-NCAT}, we show that the quality of the reconstructed image using the proposed re-weighted JSR model is not very sensitive to the choice of the threshold (denoted by $\tau$) on the summation of high frequency tight framelet coefficients. Furthermore, the proposed approach is better than directly thresholding on the initially reconstructed image.

\begin{table}[h]
  \caption{ NCAT phantom reconstructed by the proposed re-weighted JSR model using different thresholding parameter $\tau$ for metal trace estimations.}  \label{diff-seg-rec-NCAT}
\centering
  \begin{tabular}{|c|c|c|c|}
  \hline
  \multirow{6}{*}{Thresholding wavelet} &
\multicolumn{1}{|c|}{Threshold} &
\multicolumn{1}{c|}{SSIM } &
\multicolumn{1}{c|}{RelErr } \\
  \cline{2-4}
  \multirow{6}{*}{frame coefficients}
  &$\tau=0.1$   &   0.9826         &   0.0835      \\
  \cline{2-4}
  \multirow{6}{*}{(proposed)}
  &$\tau=0.2$   &   0.9835         &   0.0842     \\
  \cline{2-4}
  &$\tau=0.3$  &   0.9835         &   0.0845      \\
  \cline{2-4}
  &$\tau=0.4$    &   0.9839         &   0.0860    \\
  \cline{2-4}
  &$\tau=0.5$  &   0.9830         &   0.0877     \\
  \cline{2-4}
  &$\tau=0.6$   &   0.9821         &   0.0887        \\
\hline
  \cline{2-4}
  \multirow{5}{*}{Thresholding initially}
  &$\tau=1.1$  &   0.9758         &   0.0873     \\
  \cline{2-4}
  \multirow{5}{*}{reconstructed image}
  &$\tau=1.2$  &   0.9816         &   0.0883      \\
  \cline{2-4}
  &$\tau=1.3$  &   0.9824         &   0.0892    \\
  \cline{2-4}
  &$\tau=1.4$  &   0.9823         &   0.0894     \\
  \cline{2-4}
  &$\tau=1.5$  &   0.9824         &   0.0890        \\
  \cline{2-4}
  &$\tau=1.6$  &   0.9825         &   0.0893      \\
\hline
\end{tabular}
\end{table}

\subsubsection{Weights estimation }

To suppress the effect of data inconsistency in the Radon domain near the metal trace $\bm{\Gamma}$, we will weight the projection data in the proposed re-weighted JSR model. This subsection describes how the weights are estimated.

\begin{figure}[h]
\center
 \subfigure[ ]{\includegraphics[scale=0.35,trim=0cm 0cm 0cm 0cm]{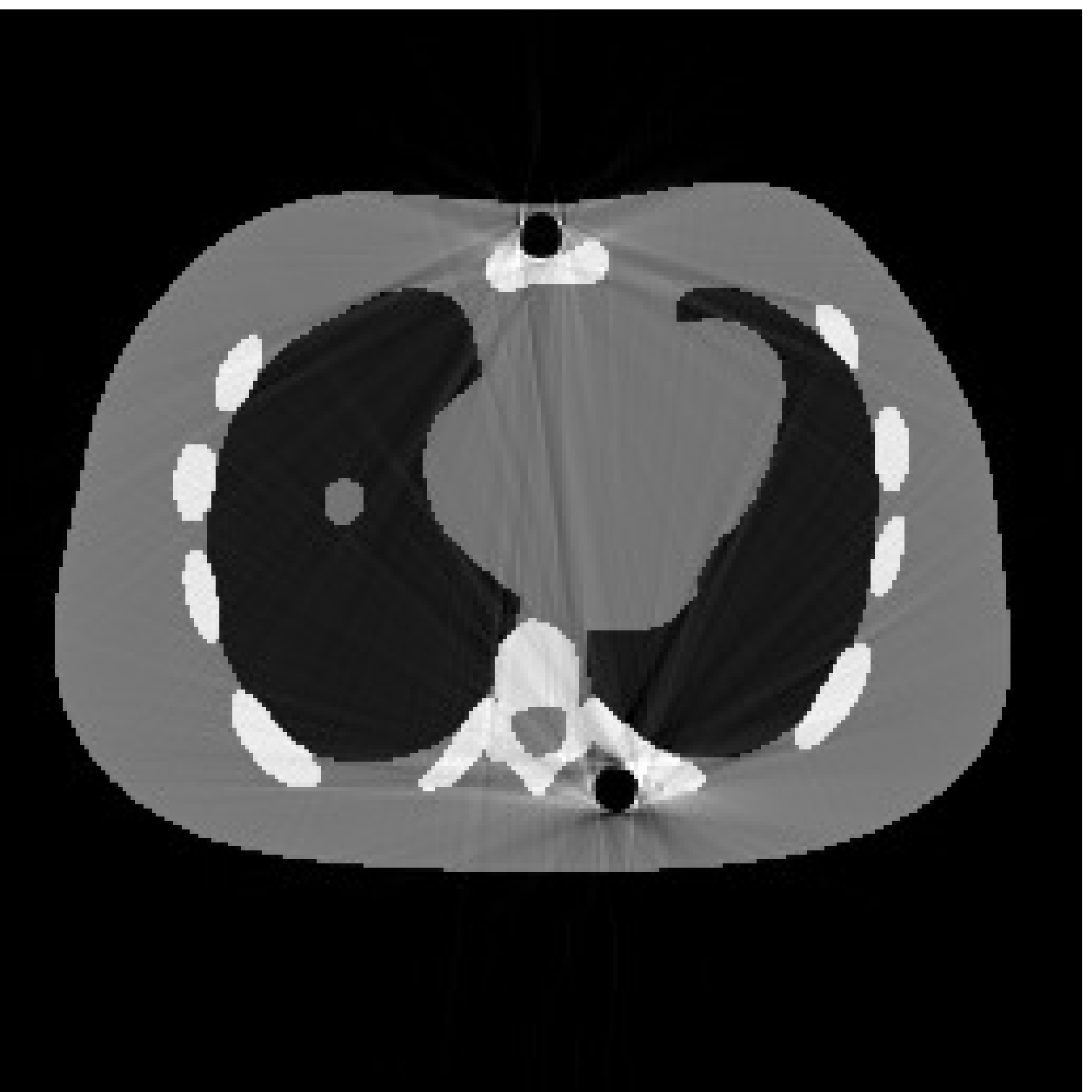}\label{NCAT-inpaint-no-noise}}
 \subfigure[ ]{\includegraphics[scale=0.35,trim=0cm 0cm 0cm 0cm]{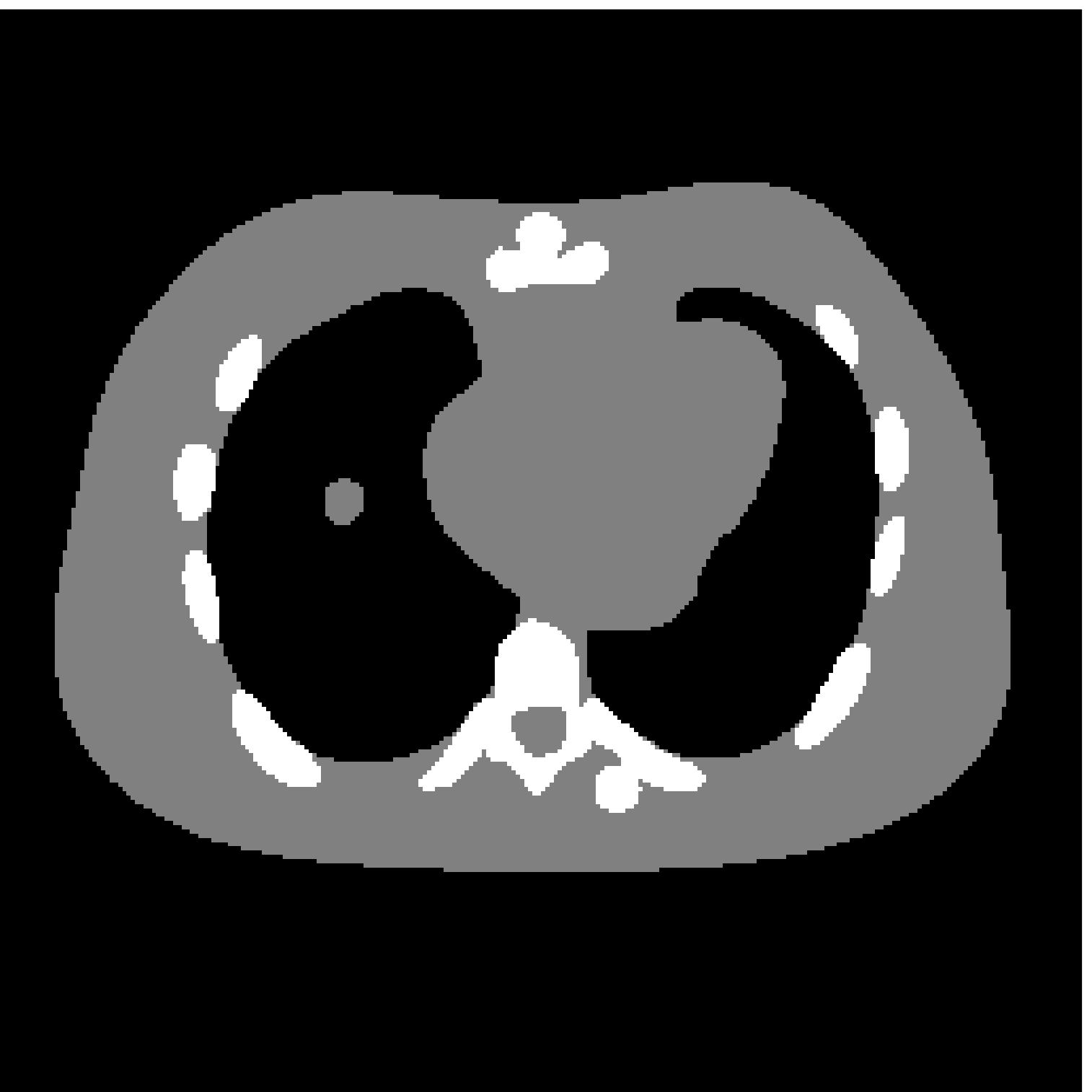}\label{seg-NCAT}}
    \caption{
    \subref{NCAT-inpaint-no-noise} Image $\bm{u}_r$ reconstructed by model \eqref{inpaint-model}.
    \subref{seg-NCAT} Segmented image from $\bm{u}_c$.
    }\label{NCAT-FBP-inpaint-no-noise}
\end{figure}

Given the metal trace $\bm{\Gamma}$ obtained from the previous subsection, we first solve the following tight wavelet frame based analysis model
\begin{equation}\label{inpaint-model}
  \min_{\bm{u}} \frac{1}{2} ||R_{\bm{\Gamma}^c}(\bm{\mathcal{P}u}-\bm{Y})||^2+ ||\bm{\lambda}\cdot \bm{Wu}||_{1,2},
\end{equation}
where $ R_{\bm{\Gamma}^c}$ is the restriction operator with respect to the domain $\bm{\Gamma}^c$. The restriction operator $R_{\bm{\Lambda}}$ is defined as: $R_{\bm{\Lambda}}\bm{v}[\bm{k}]=\bm{v}[\bm{k}]$ for $\bm{k}\in\bm{\Lambda}$; and $R_{\bm{\Lambda}}\bm{v}[\bm{k}]=0$ for $\bm{k}\in\bm{\Lambda}^c$. The optimization problem \eqref{inpaint-model} can be solved similar to \eqref{analysis-model} by the split Bregman algorithm. We denote the solution of \eqref{inpaint-model} as $\bm{u}_r$ (Figure \ref{NCAT-inpaint-no-noise}).

The weights will be computed by projecting a segmented image that approximates tissue classification of the unknown CT image. One may obtain such approximated tissue classification by segmenting image $\bm{u}_a$ obtained from \eqref{analysis-model} or $\bm{u}_r$ obtained from \eqref{inpaint-model}. However, the approximation from either image will be rather inaccurate since $\bm{u}_a$ has severe artifacts in between metal locations (Figure \ref{no-noise-NCAT}), while the metal components are missing from $\bm{u}_r$ (Figure \ref{NCAT-inpaint-no-noise}) though there are less artifacts in between metal locations. Therefore, we propose to segment a combined image defined by
\begin{equation}\label{ua-up}
\bm{u}_{c}=(1-\sigma)\bm{u}_{a}+\sigma\bm{u}_{r}, \quad 0\le \sigma\le 1,
\end{equation}
with $\sigma$ a tuning parameter. In this paper, the segmentation of $\bm{u}_c$ is obtained by the algorithm proposed by \cite{li2011level,li2010distance}.
The segmented image, denoted as $\bm{u}_{s}$ (Figure \ref{seg-NCAT}), contains three components: the air, the low density components such as soft tissues, and the high density components such as bones and metals. The intensity values of the segmented image $\bm{u}_{s}$ from prior image $\bm{u}_{c}$ are assumed to be constant for each segmented component. We propose to use the mean values of $\bm{u}_c$ in the segmented regions as the constants. After obtaining $\bm{u}_s$, the weight that will be used in our re-weighted JSR model is defined by $\bm{Y}_s:=\bm{\mathcal{P}}\bm{u}_{s}$. Finally, we note that the reconstructed image using the re-weighted JSR model is relatively insensitive to the choice of the parameter $\sigma$ in \eqref{ua-up}, which is demonstrated in Table \ref{tab-diff-seg-rec-NCAT} using NCAT phantom.

\begin{table}[h!]
  \caption{NCAT phantom reconstructed by the re-weighted JSR model from different segmentations.}  \label{tab-diff-seg-rec-NCAT}
\centering
  \begin{tabular}{|c|c|c|c|}
  \hline
\multicolumn{1}{|c|}{$\sigma$ } &
\multicolumn{1}{c|}{SSIM } &
\multicolumn{1}{c|}{RelErr } \\
  \cline{1-3}
 $\sigma =0$     &   0.9611         &   0.1127    \\
  \cline{1-3}
  $\sigma =0.2$  &   0.9678         &   0.1020     \\
  \cline{1-3}
$\sigma=0.4 $    &   0.9764         &   0.0918  \\
  \cline{1-3}
  $\sigma=0.6$   &   0.9764         &   0.0918     \\
  \cline{1-3}
 $\sigma=0.8$    &   0.9810         &   0.0938       \\
    \cline{1-3}
 $\sigma=1$      &   0.9794         &   0.1003     \\
\hline
\end{tabular}
\end{table}

\section{A re-weighted joint spatial-Radon domain reconstruction for metal artifact reduction}\label{section-JSR-model}

In this section, we propose a new re-weighted joint spatial-Radon domain reconstruction (re-weighted JSR) model to reconstruct CT images from the multi-chromatic X-ray imaging system with reduced metal artifacts. Efficient algorithm is adopted to solve the proposed model. Intuition of the weighting strategy is discussed at the end of this section.

\subsection{The re-weighted JSR model}

To reduce metal artifacts and reconstruct high quality CT images, we propose the following re-weighted JSR model
 \begin{equation}\label{re-weight-JSR-model}
 \min_{\bm{u}, \bm{f}}\frac{1}{2}||\bm{\mathcal{P}u}-\bm{Y}_{s}\bm{f}||^{2}+||\bm{\lambda}_{1}\cdot \bm{W}_{1}\bm{u}||_{1,2}+\frac{\alpha}{2}||R_{\bm{\Gamma}^c}(\bm{f}-\frac{\bm{Y}}{\bm{Y}_{s}})||^{2} +||\bm{\lambda}_{2}\cdot \bm{W}_{2}\bm{f}||_{1,2}.
 \end{equation}
Here, $\bm{Y}$ is the measured projection data from the multi-chromatic X-ray source imaging system which is contaminated by Poisson noise, $\bm{W}_{1}$ is the Haar framelet transform (multilevel with $L=3$), $\bm{W}_{2}$ is the piecewise cubic B-spline framelet transform (multilevel with $L=3$), $R_{\bm{\Gamma}^{c}}$ is the restriction operator that extracts the projection data in the complement of metal trace, and $\bm{Y}_{s}=\bm{\mathcal{P}}\bm{u}_{s}$. The division $\frac{\bm{Y}}{\bm{Y}_{s}}$ is defined point-wisely.

The novelty of the proposed re-weighted JSR model \eqref{re-weight-JSR-model} is how the weights $\bm{Y}_s$ are introduced to the joint spatial-Radon reconstruction framework. In \eqref{re-weight-JSR-model}, the projection data $\bm{Y}$ is weighted by $\bm{Y}_{s}$ so that the inpainting mechanism induced by the last two terms of \eqref{re-weight-JSR-model} is more effective. The weighting is cancelled out during the reconstruction of $\bm{u}$ from the first two terms of \eqref{re-weight-JSR-model} to ensure a correct reconstruction. An empirical explanation of such weighting strategy will be presented in Section \ref{sect-model-analysis}.

A special case of the re-weighed JSR model \eqref{re-weight-JSR-model} is
\begin{equation}\label{unweighted-JSR-model}
\min_{\bm{u} , \bm{f}} \frac{1}{2}||\bm{\mathcal{P}}\bm{u}-\bm{f}||^{2}+||\bm{\lambda}_{1}\cdot \bm{W}_{1}\bm{u}||_{1,2}+\frac{\alpha}{2}||R_{\bm{\Gamma}^c}(\bm{f}-\bm{Y})||^{2} +||\bm{\lambda}_{2}\cdot \bm{W}_{2}\bm{f}||_{1,2},
\end{equation}
where the weight $\bm{Y}_{s}$ is removed. The model \eqref{unweighted-JSR-model} was proposed in \cite{Dong2013} for sparse angle CT image reconstruction. Although in principle, model \eqref{unweighted-JSR-model} can also be applied to metal artifact reduction, it introduces new artifacts in the reconstructed images. For convenience, we refer to the model \eqref{unweighted-JSR-model} as the unweighted JSR model. In Section \ref{section-numerical}, we shall compare the unweighted JSR model with the re-weighted JSR model to demonstrate the effectiveness of the proposed weighting strategy.

To solve the re-weighted JSR model, we first rewrite it into a more succinct form. For convenience of notation, we temporarily let $\alpha=1$ and $\bm{\lambda}=\bm{\lambda}_{1}=\bm{\lambda}_{2}$. Denote
$ \bm{A}=\left(
\begin{matrix}
\bm{\mathcal{P}}&-\bm{Y}_{s}\\
\bm{0}&R_{\bm{\Gamma}^{c}}
\end{matrix}
\right),
$
$\bm{X}=\left(\begin{matrix}
\bm{u}\\
\bm{f}
\end{matrix}
\right),$
$\bm{B}=\left(\begin{matrix}
\bm{0}\\
R_{\bm{\Gamma}^{c}}(\frac{\bm{Y}}{\bm{Y}_{s}})
\end{matrix}
\right),$
$\bm{W}=\left(\begin{matrix}
\bm{W}_{1}&  0\\
0 & \bm{W}_{2}
\end{matrix}
\right).$
Then, model \eqref{re-weight-JSR-model} can be rewritten as
\begin{equation}\label{ReWJSR-simple-model}
\min_{\bm{X} } \frac{1}{2}\|\bm{A}\bm{X}-\bm{B}\|^{2}+\|\bm{\lambda}\cdot \bm{W}\bm{X}\|_{1,2},
\end{equation}
where the $\ell_{1,2}$ norm is defined block-wisely. Model \eqref{ReWJSR-simple-model} is a standard analysis based model and can be efficiently solved by the split Bregman algorithm \cite{goldstein2009split,cai2009split} with guaranteed convergence. The split Bregman algorithm solving \eqref{ReWJSR-simple-model} takes the following form
\begin{eqnarray}
\bm{X}^{k+1}&=&\arg\min_{\bm{X} } \frac{1}{2}\|\bm{A}\bm{X}-\bm{B}\|^{2}+\frac{\mu}{2}\|\bm{W}\bm{X}-\bm{Z}^{k}+\bm{C}^{k}\|^{2} \notag\\
\bm{Z}^{k+1}&=&\arg\min_{\bm{Z}} \|\bm{\lambda}\cdot \bm{Z}\|_{1,2}+\frac{\mu}{2}\|\bm{W}\bm{X}^{k+1}-\bm{Z}+\bm{C}^{k}\|^{2}  \notag\\
\bm{C}^{k+1}&=& \bm{C}^{k}+(\bm{A}\bm{X}^{k+1}-\bm{Z}^{k+1}).
\end{eqnarray}
Each subproblem has a closed form solution.

Details of the algorithm is presented in Algorithm \ref{JSR-alg-Jacobi}. {Since the $\bm{u},\bm{f}$ sub-variables are coupled together in the $\bm{X}$-subproblem, they can be updated alternatively.} The linear system in the step solving for $\bm{u}^{k+1}$ is solved using the conjugate gradient (CG) algorithm. The thresholding operator $\bm{\mathcal{T}}_{\bm{\lambda_j}/\mu_j}$, $j=1,2$, is the isotropic soft shrinkage operator \cite{cai2012image}. Now, we recall the definition of the isotropic soft shrinkage operator. Given the wavelet frame coefficients $\bm{v}:=\{\bm{v}_{\ell,l}: 1\le\ell\le r, 0\le l\le L-1\}\cup\{\bm{v}_{0,L-1}\}=:\bm{W}\bm{u}$ and thresholds $\bm{\lambda}:=\{\lambda_{\ell,l}: 1\le\ell\le r, 0\le l\le L-1\}\cup\{\lambda_{0,L-1}\}$ with $\lambda_{\ell,l}\ge0$, the shrinkage operator $\bm{\mathcal{T}}_{\bm{\lambda}}(\bm{v})$ used in Algorithm \ref{JSR-alg-Jacobi} is the isotropic shrinkage operator \cite{cai2012image} defined by
\begin{eqnarray}\label{define-soft-threshold-2}
(\bm{\mathcal{T}}_{\bm{\lambda}}(\bm{\nu}))_{\ell,l}=\frac{\bm{\nu}_{\ell,l}}{V_l}\max\{V_l-\lambda_{\ell,l},0\}
\end{eqnarray}
where $V_l=\left( \sum_{1\le\ell\le r}|\bm{\nu}_{\ell,l}|^{2}\right)^{1/2}$. As convention, we choose $\lambda_{0,L-1}=0$.

\begin{algorithm}
\caption{ Re-weighted JSR algorithm \label{JSR-alg-Jacobi}}
\begin{algorithmic}
\State{
\textbf{Input}: Set $\bm{u}^{0}=\bm{0}$, $\bm{f}^{0}=\bm{0}$, $\alpha>0,$ $\mu_{1}>0$, $\mu_{2}>0$, $\bm{\lambda}_{1}>0$. $\bm{\lambda}_{2}>0 $. $\bm{Y}_{s}=\bm{\mathcal{P}}\bm{u}_{s}$, where $\bm{u}_{s}$ is the segmented image.
}
\While{Stopping criteria is not met }
\State{Update $\bm{X}^{k+1}$
     \begin{eqnarray}
   \bm{u}^{k+1} &=& ( \bm{\mathcal{P}}^{\top}\bm{\mathcal{P}}+\mu_{1} \bm{I})^{-1} \left[ \bm{\mathcal{P}}^{\top}(\bm{Y}_{s}\bm{f}^{k})+\mu_{1}\bm{W}_{1}^{\top}(\bm{d}_{1}^{k}-\bm{b}_{1}^{k}) \right] \notag \\
   \bm{f}^{k+1} &=& \left[\alpha R_{\bm{\Gamma}^c}+ \bm{Y}_{s}^{2}+\mu_{2}\bm{I}\right]^{-1} \left[\alpha R_{\bm{\Gamma}^c}\frac{\bm{Y}}{\bm{Y}_{s}}+ \bm{Y}_{s} \bm{\mathcal{P}u}^{k}+\mu_{2}\bm{W}_{2}^{\top}(\bm{d}_{2}^{k}- \bm{b}_{2}^{k})\right], \notag
    \end{eqnarray}
    }
\State{Update $\bm{Z}^{k+1}$
 \begin{eqnarray}
     \bm{d}_{1}^{k+1} &=& \bm{\mathcal{T}}_{\bm{\lambda_{1}}/\mu_{1}}(\bm{W}_{1}\bm{u}^{k+1}+\bm{b}_{1}^{k}),\notag \\
    \bm{d}_{2}^{k+1} &=& \bm{\mathcal{T}}_{\bm{\lambda_{2}}/\mu_{2}}(\bm{W}_{2}\bm{f}^{k+1}+\bm{b}_{2}^{k}), \notag
    \end{eqnarray}
    }
\State{Update $\bm{C}^{k+1}$
 \begin{eqnarray}
    \bm{b}_{1}^{k+1} &=& \bm{b}_{1}^{k}+(\bm{W}_{1}\bm{u}^{k+1}-\bm{d}_{1}^{k+1}), \notag\\
    \bm{b}_{2}^{k+1} &=& \bm{b}_{2}^{k}+(\bm{W}_{2}\bm{f}^{k+1}-\bm{d}_{2}^{k+1}). \notag
    \end{eqnarray}
    }
\State{$k=k+1;$ }
\EndWhile
\Return $\{\bm{u}^{\ast},\bm{f}^{\ast}\}$.
\end{algorithmic}
\end{algorithm}

For a better presentation of the proposed CT image reconstruction method with reduced metal artifacts, we summarize the entire procedure in the flow chart shown in Figure \ref{flow-chart}.

\tikzstyle{block1} = [rectangle, draw, fill=blue!20,
    text width=8em, text centered, rounded corners, minimum height=2.5em]
\tikzstyle{block} = [rectangle, draw, fill=blue!20,
    text width=10.5em, text centered, rounded corners, minimum height=2.5em]
\tikzstyle{line} = [draw, -latex']
\tikzstyle{cloud} = [draw, ellipse,fill=red!20, node distance=3cm,
    minimum height=2em]

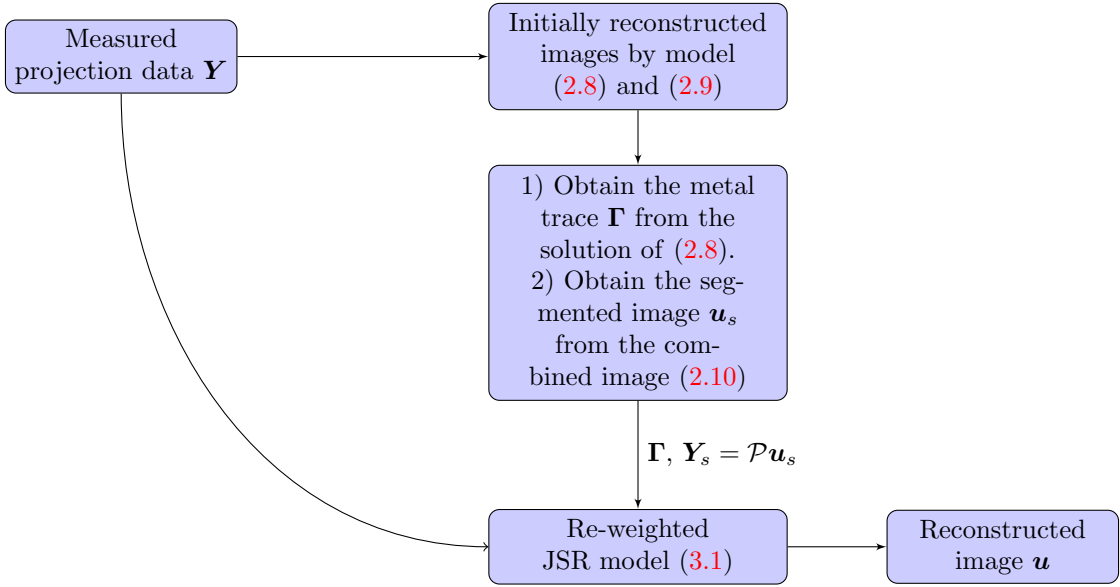
\begin{figure}[h!]
\centering
\begin{tikzpicture}[node distance = 2.5cm, auto]
    \node [block1] (init) {Measured projection data $\bm{Y}$};
    \node [block, right of=init, node distance=6.8cm] (identify) {Initially reconstructed images by model \eqref{analysis-model} and \eqref{inpaint-model}};
    \node [block, below of=identify, node distance=3cm] (evaluate) {1) Obtain the metal trace $\bm{\Gamma}$ from the solution of \eqref{analysis-model}. \\ 2) Obtain the segmented image $\bm{u}_{s}$ from the combined image \eqref{ua-up}};
    \node [block, below of=evaluate, node distance=3.5cm] (decide) { Re-weighted JSR model \eqref{re-weight-JSR-model} };
    \node [block1, right of=decide, node distance=4.8cm] (stop) {Reconstructed image $\bm{u}$};
    \path [line] (init) --  (identify);
    \path [line] (identify) -- (evaluate);
    \path [line] (evaluate) -- node {$\bm{\Gamma}$, $\bm{Y}_{s}=\mathcal{P}\bm{u}_{s}$ } (decide) ;
     \draw[->] (init) to[out=-90,in=180]  (decide);
    \path [line] (decide) -- (stop);
\end{tikzpicture}
\caption{Flow chart of the proposed CT image reconstruction with reduced metal artifacts.}\label{flow-chart}
\end{figure}

\subsection{Why does the weighting strategy work ?}\label{sect-model-analysis}
In this subsection, we give some empirical observations on the weighting strategy used in the proposed model \eqref{re-weight-JSR-model}. {For simplicity, all the numerical results in this subsection are obtained using the NCAT phantom.} The re-weighted JSR model can be viewed as a combination of the following two models:
the spatial domain reconstruction model
\begin{equation}\label{spatial-domain-model}
 \min_{\bm{u} }\frac{1}{2}||\bm{\mathcal{P}}\bm{u}-\bm{Y}_{s}\bm{f}||^{2}+||\bm{\lambda}_{1}\cdot \bm{W}_{1}\bm{u}||_{1,2},
\end{equation}
and the Radon domain inpainting model
\begin{equation}\label{unconst-inpaint-model}
 \min_{\bm{f}}\frac{1}{2}||R_{\bm{\Gamma}^c}(\bm{f}-\frac{\bm{Y}}{\bm{Y}_{s}})||^{2} +||\bm{\lambda}_{2}\cdot \bm{W}_{2}\bm{f}||_{1,2}.
\end{equation}
Obviously, \eqref{spatial-domain-model} is an analysis model with tight wavelet  frame regularization. The Radon domain inpainting model \eqref{unconst-inpaint-model} is another analysis based model  aims at repairing the degraded projection data around the metal trace.

\begin{figure}[t!]
\center
    \subfigure{\includegraphics[scale=0.4,trim=0cm 0cm 0cm 0cm]{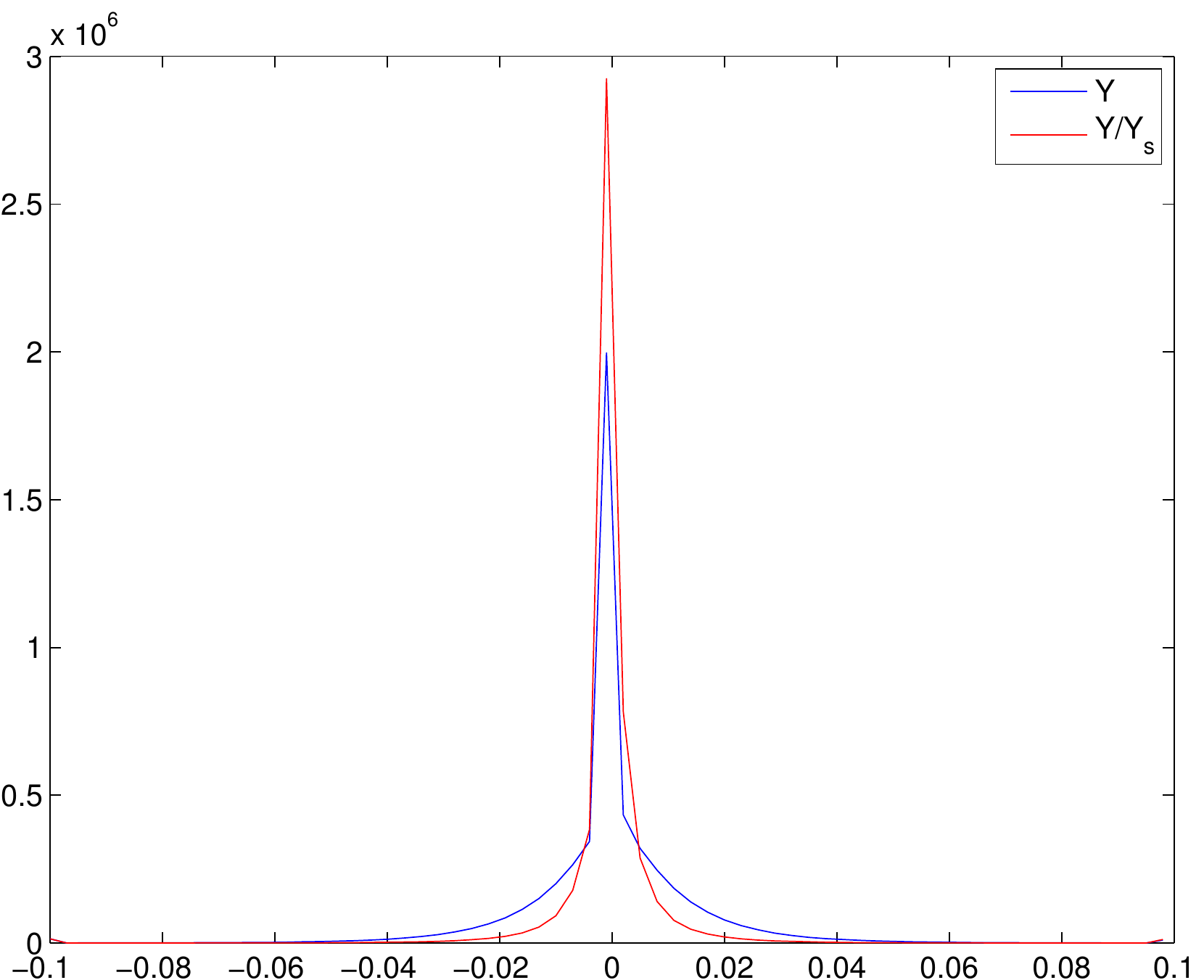}\label{NCAT-hist}}
    \caption{
     The distribution of the wavelet frame coefficients of the measured projection data $\bm{Y}$ and the weighted data $\frac{\bm{Y}}{\bm{Y}_{s}}$.
    }\label{NCAT-hist-marked-columns}
\end{figure}

We observe that the weighted projection data $\frac{\bm{Y}}{\bm{Y}_{s}} $ admits a sparse representation than the measured projection data $\bm{Y}$ in wavelet frame domain. The distributions of wavelet frame coefficients (using piecewise linear tight wavelet frame system) are plotted in Figure \ref{NCAT-hist-marked-columns}. To have a fair comparison, $\bm{Y}$ and $\frac{\bm{Y}}{\bm{Y}_{s}}$ are normalized to the same scale, and the coefficients near the metal trace are removed. From the distributions we can see that the weighted projection data $\frac{\bm{Y}}{\bm{Y}_{s}}$ has a sparser representation than the original projection data $\bm{Y}$. Therefore, the sparsity prior $||\bm{\lambda}_{2}\cdot \bm{W}_{2}\bm{f}||_{1,2}$ in the inpainting model \eqref{unconst-inpaint-model} is more effective than directly using $\bm{Y}$ as input.

By combining \eqref{unconst-inpaint-model} with \eqref{spatial-domain-model}, the proposed re-weighted JSR model is able to repair the degraded projection so that the repaired data $\bm{Y}_s \bm{f}_{\text{ReWJSR}}$ is closer to $\bm{\mathcal{P}}\bm{u}_{\text{true}}$, where $\bm{f}_{\text{ReWJSR}}$ is the Radon domain solution from \eqref{re-weight-JSR-model} and $\bm{u}_{\text{true}}$ is the unknown ground-truth CT image.  {Note that it is hard to obtain the ground truth image for polychromatic energy CT with varying attenuation coefficients with respect to the energy level, we choose the NCAT phantom (Figure \ref{clear-NCAT-with-marked-metal}) as the approximated ground truth CT image $\bm{u}_{\text{true}}$.}  {Empirically, the repaired projection data $\bm{Y}_s \bm{f}_{\text{ReWJSR}}$ fits the linear inverse problem \eqref{multi-energy-Xray} better than the repaired projection data $\bm{f}_{\text{NMAR}}$ and $\bm{f}_{\text{JSR}}$ computed from the NMAR and the unweighted JSR model \eqref{unweighted-JSR-model} respectively. This is the key to the success of the re-weighted JSR model since the linear model \eqref{multi-energy-Xray} is what we commonly assume for CT imaging. Such linear assumption is not correct (though reasonable) for a multi-chromatic imaging system. }

\begin{figure}[h!]
\center
     \subfigure[$450$th column ]{\includegraphics[scale=0.45]{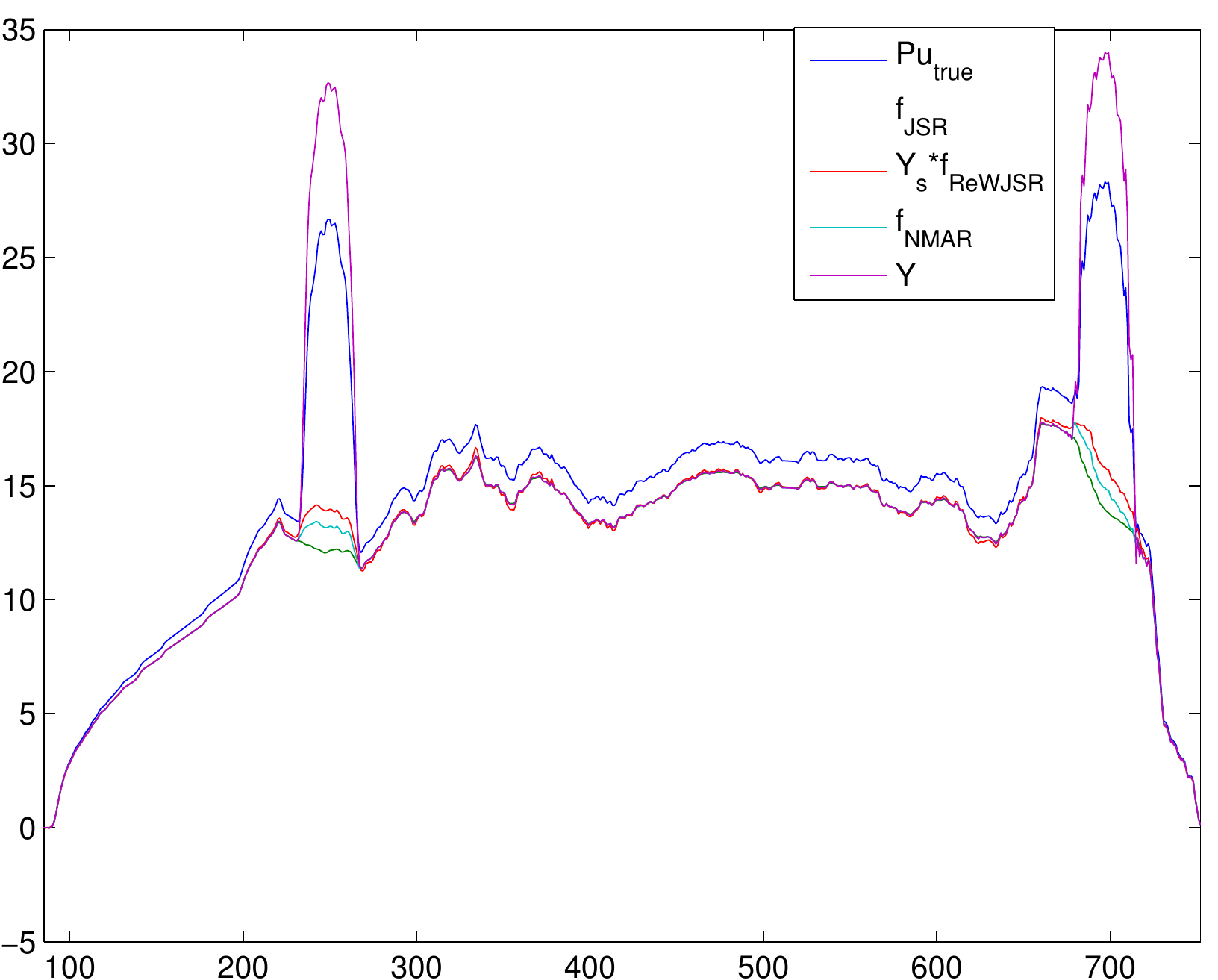}\label{NCAT-450-column}}
     \subfigure[$550$th column ]{\includegraphics[scale=0.45]{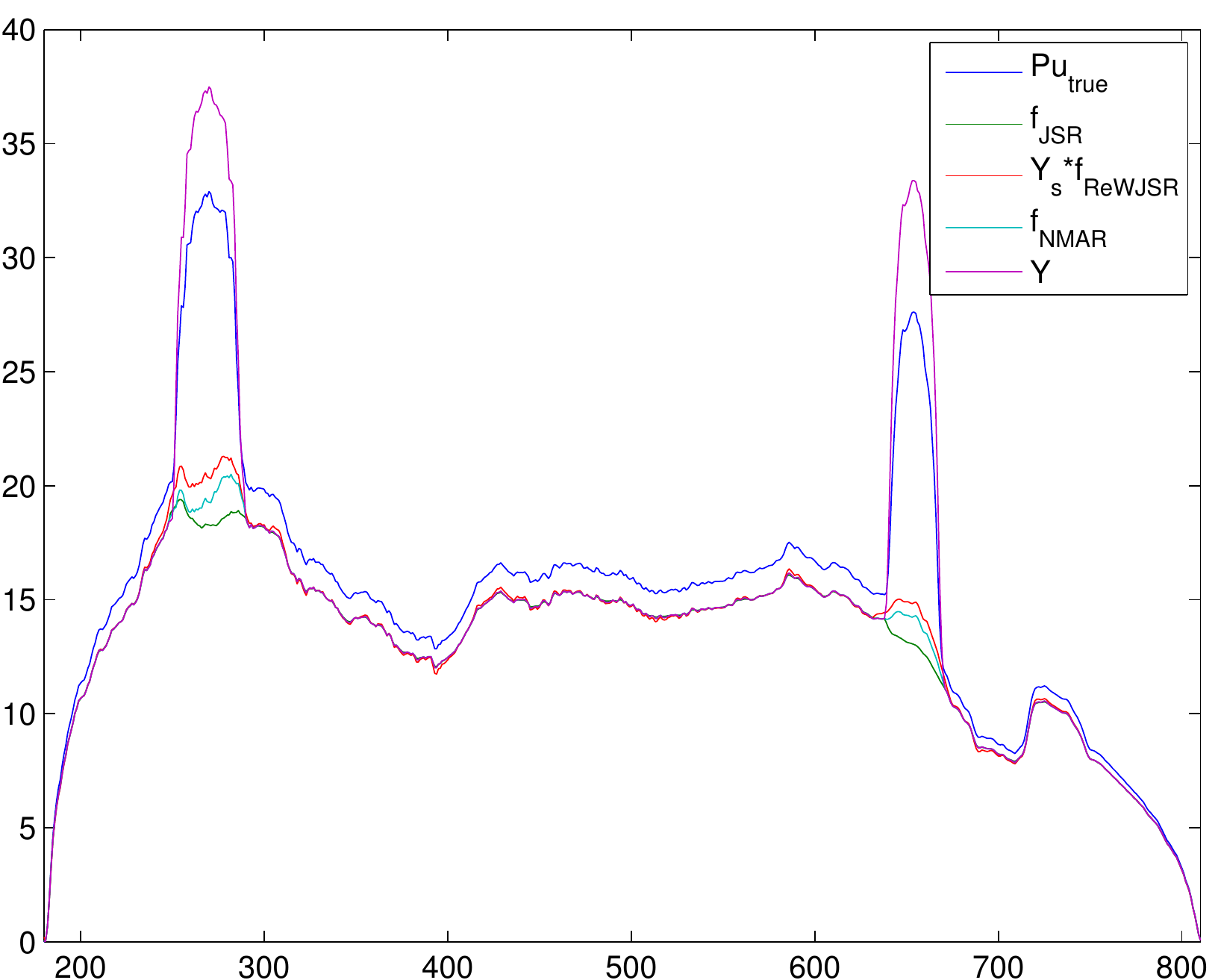}\label{NCAT-550-column}}
     \subfigure[Zoom-in of $450$th column ]{\includegraphics[scale=0.45]{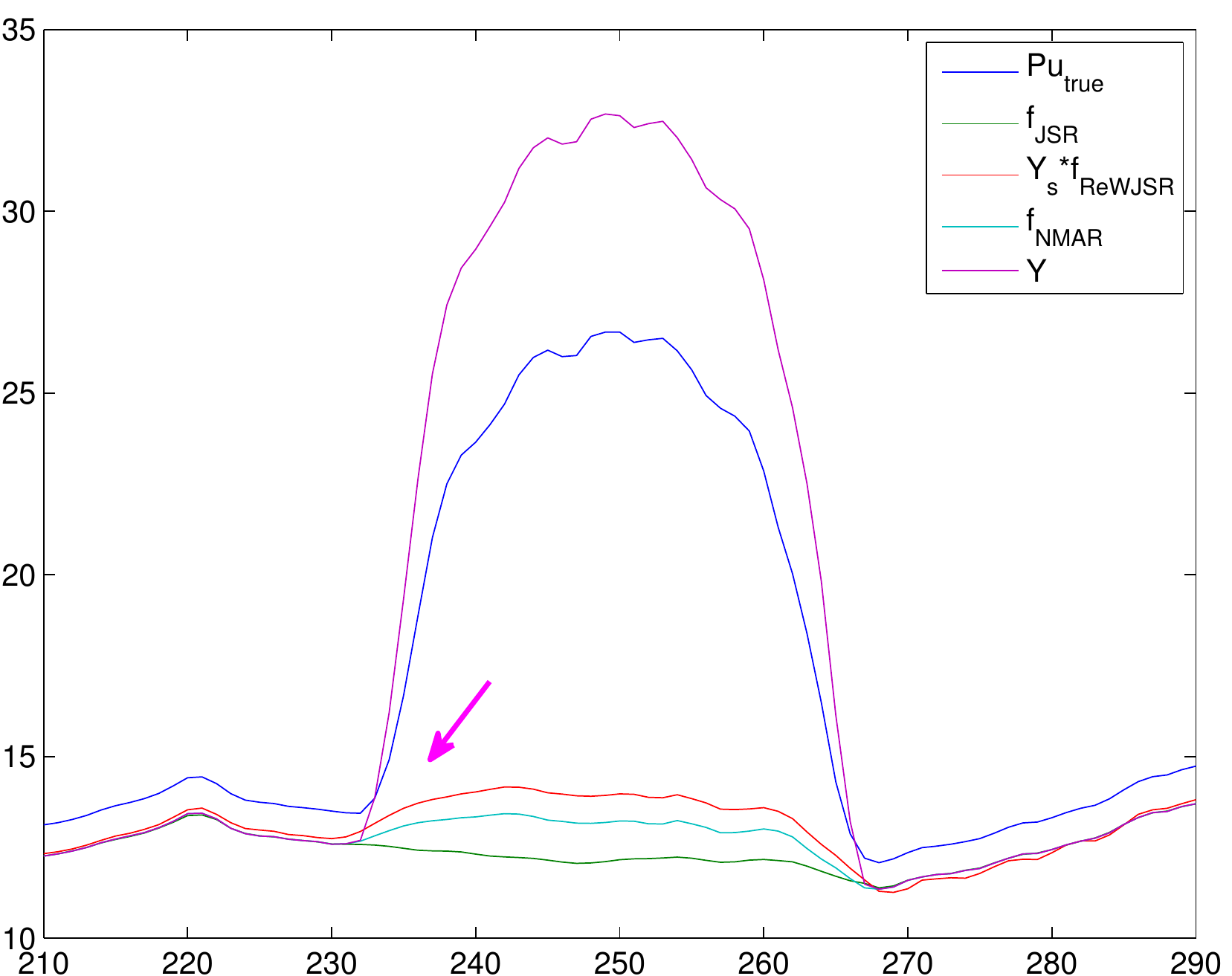}\label{NCAT-450-column-zoom-in1}}
     \subfigure[Zoom-in of $450$th column ]{\includegraphics[scale=0.45]{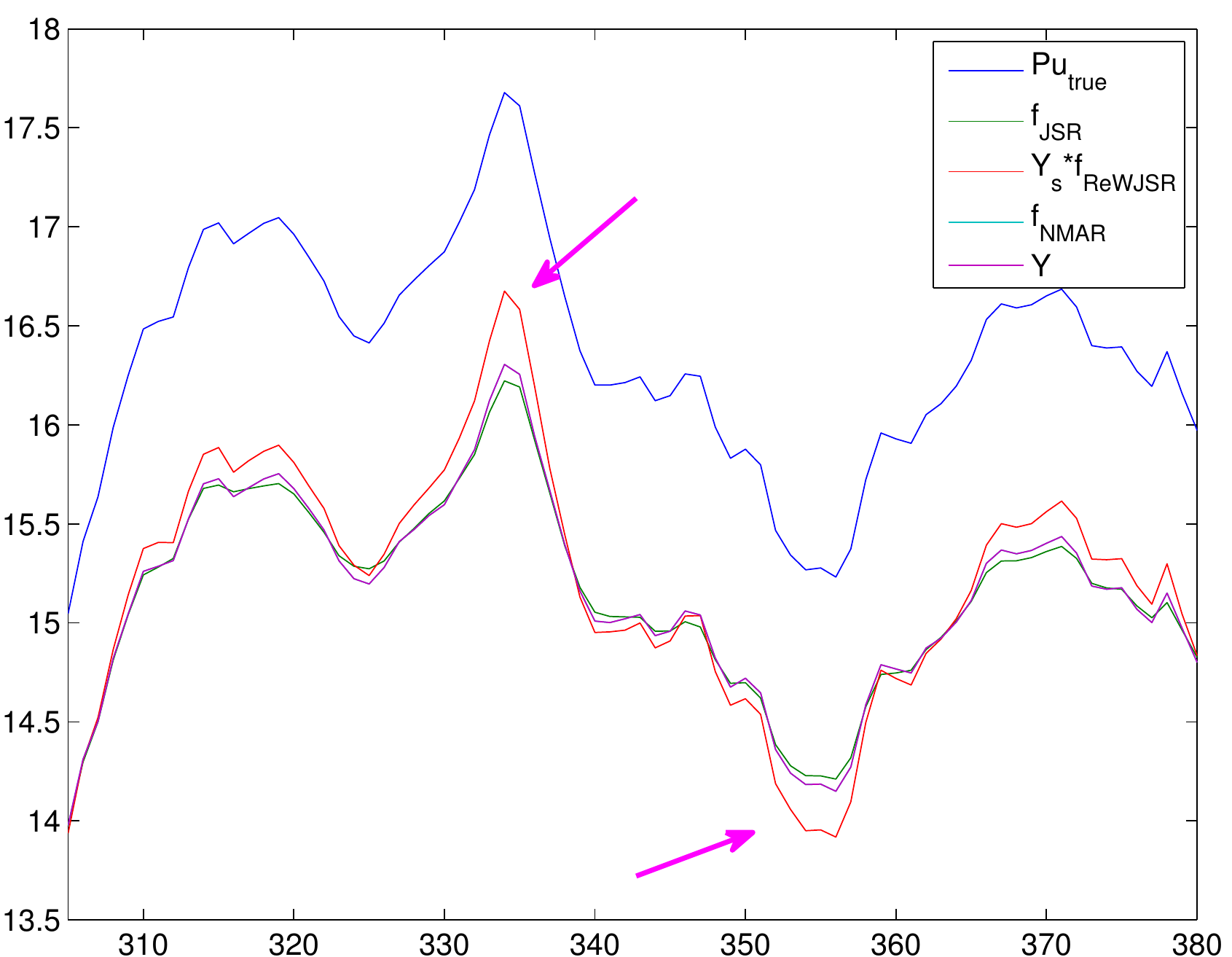}\label{NCAT-450-column-zoom-in2}}
    \caption{
    \subref{NCAT-450-column} The $450$th column of the NCAT phantom projection data.
    \subref{NCAT-550-column} The $550$th column of the NCAT phantom  projection data.
    \subref{NCAT-450-column-zoom-in1} Zoom-in view of the curve in \subref{NCAT-450-column} .
    \subref{NCAT-450-column-zoom-in2} Zoom-in view of the curve in \subref{NCAT-450-column}.
    $\bm{\mathcal{P}u}_{\text{true}}$ is the projection of the reference image $\bm{u}_{\text{true}}$. $\bm{f}_{\text{JSR}}$ is the inpainted projection data from unweighted JSR model \eqref{unweighted-JSR-model}, and  $\bm{f}_{\text{ReWJSR}}$ is the inpainted projection data from the re-weighted JSR model \eqref{re-weight-JSR-model}.
    }\label{NCAT-weighted-columns}
\end{figure}

 {
To support such claim, we present comparisons of $\bm{\mathcal{P}}\bm{u}_{\text{true}}$ with $\bm{Y}$, $\bm{f}_{\text{NMAR}}$, $\bm{f}_{\text{JSR}}$ and $\bm{Y}_s \bm{f}_{\text{ReWJSR}}$ in Figure \ref{NCAT-weighted-columns} using the NCAT phantom. We observe that $\bm{Y}_{s}\bm{f}_{\text{ReWJSR}}$ is a better approximation to the projection of the reference image $\bm{\mathcal{P}u}_{\text{true}}$ than the repaired projection data from the JSR and NMAR model (see Figure \ref{NCAT-weighted-columns}(c),(d)). The NMAR model also generates a better repaired projection $\bm{f}_{\text{NMAR}}$ than the unweighted JSR model due to its re-weighting strategy. However, the unweighted JSR model is still able to reduce the majority of the metal artifacts in the reconstructed image due to its sparsity based joint regularization. These observations, together with the reconstruction results in Section 4, show that the re-weighted JSR model combines the merits of the NMAR model's weighting strategy and the sparsity based joint regularization of the unweighted JSR model.}

To quantitatively measure the difference between $\bm{\mathcal{P}u}_{\text{true}}$ and the repaired projection data from different models and the measured projection data $\bm{Y}$, we calculate the $\ell_2$-norms $\|\bm{Y}_{s}\bm{f}_{\text{ReWJSR}}-\bm{\mathcal{P}u}_{\text{true}}\|$, $\|\bm{f}_{\text{JSR}}-\bm{\mathcal{P}u}_{\text{true}}\|$, $\|\bm{f}_{\text{NMAR}}-\bm{\mathcal{P}u}_{\text{true}}\|$ and $\|\bm{Y}-\bm{\mathcal{P}u}_{\text{true}}\|$. Since the region of the metal trace $\bm{\Gamma}$ has major contribution to these quantities, we also compute the $\ell_2$-norms excluding the regions of the metal trace. Results are shown in Table \ref{L2-norm-compare-Radon-domain}. {Obviously, the repaired projection data from the re-weighted JSR model is closer to $\bm{\mathcal{P}u}_{\text{true}}$ than that from the NMAR and JSR model. However, although $\bm{Y}_{s}\bm{f}_{\text{ReWJSR}}$ is closer to $\bm{\mathcal{P}u}_{\text{true}}$ in regions outside of $\bm{\Gamma}$, $\bm{Y}$ is overall closer to $\bm{\mathcal{P}u}_{\text{true}}$ than $\bm{Y}_{s}\bm{f}_{\text{ReWJSR}}$ due to the inaccurate recovery of the projection data inside the metal trace $\bm{\Gamma}$ by the re-weighted JSR model (see Figure \ref{NCAT-weighted-columns}(c)). This is probably why the re-weighted JSR model still cannot fully remove metal artifacts, though it improves over the unweighted JSR and the NMAR model.}

\begin{table}[h]
\caption{The $\ell_{2}$-norm of the difference between the projection data of the reference image, i.e. $\bm{\mathcal{P}u}_{\text{true}}$, and the corrected projection data from different models.}  \label{L2-norm-compare-Radon-domain}
\centering
  \begin{tabular}{|c|c|c|c|c|c|c|}
\hline
\multirow{5}{*}{NCAT } &
\multicolumn{1}{c|}{Models} &
\multicolumn{1}{c|}{Including $\bm{\Gamma}$} &
\multicolumn{1}{c|}{Excluding $\bm{\Gamma}$ }     \\
\cline{2-4}
  &  Measured Data     &     1547.6       &   713.8                       \\
\cline{2-4}
  &  NMAR                 & 2802.1       &   713.4                   \\
\cline{2-4}
  &  JSR     &   3048.2        &   709.7                      \\
\cline{2-4}
  & Re-weighted JSR     &   2777.5       &   688.3                      \\
\hline
\end{tabular}
\end{table}

\section{Numerical experiments}\label{section-numerical}

In this section, we validate the effectiveness of the re-weighted JSR model using both simulated phantoms and real data. All numerical experiments are implemented in MATLAB running on a platform with 16 GB RAM and Intel(R) Core(TM) i7-6700T CPU at 2.8-GHz with 4 cores.

The regularization parameter $\bm{\lambda}$ appeared in any of the models presented in the previous sections takes the form $\bm{\lambda}=\{(\frac{1}{2})^{l}\lambda\ |\ l =0,1,...,L-1\}$ where $\lambda>0$ is a tuning parameter.
The stopping criterion for Algorithm \ref{JSR-alg-Jacobi} is
$$
\frac{\|\bm{u}^{k+1}-\bm{u}^{k}\| }{\|\bm{u}^{k+1}\|} \le tol=2\times 10^{-3}.
$$
The unweighted JSR model and the TV-FADM model are solved using the algorithms proposed in \cite{Dong2013} and \cite{zhang2016iterative} respectively with the same stopping criterion as above. We set the maximum allowable number of iteration of all the iterative algorithms to 700. We use the relative error (RelErr) and structural similarity (SSIM) index \cite{wang2004image} to quantitative evaluate the reconstructed images from different methods.

\subsection{Experiments on image phantoms}\label{section-simulate-phantom}

\subsubsection{Experimental design}\label{section-simulate-phantom-design}

The fan-beam CT imaging system similar to \cite{wang2006penalized} is chosen for all the simulations in this subsection. The source to detector distance is 949.075 mm, the distance from the source to the iso-center is 541 mm and the strip width is 1.024 mm. The source trajectory covers the full circular orbit of $360^{\circ}$ with 984 angular views and the number of bins per view is 888. The tube potential is 140kvp with 2.5mm aluminum and 0.5mm copper filters.

The NCAT phantom (shown in Figure \ref{clear-NCAT-with-marked-metal}) and the cerebral phantom\footnote{\url{http://see.xidian.edu.cn/vipsl/database_CTMR.html}} (shown in Figure \ref{CThead-512-ref}) are chosen as image phantoms. For the NCAT phantom, it has $256\times 256$ pixels. Two metal components (Titanium) are implanted in the image, which is shown in Figure \ref{clear-NCAT-with-marked-metal} with red curves labeling the locations of the metals. For the cerebral phantom, it has $512\times 512$ pixels and three  metal components (Titanium) are implanted. Both of the phantoms contain three major components, i.e. soft tissue, bone and metal components, and their linear attenuation coefficients can be found in \cite{hubbell1995tables}.

\begin{figure}[t!]
\center
    \subfigure{\includegraphics[scale=0.4,trim=0cm 0cm 0cm 0cm]{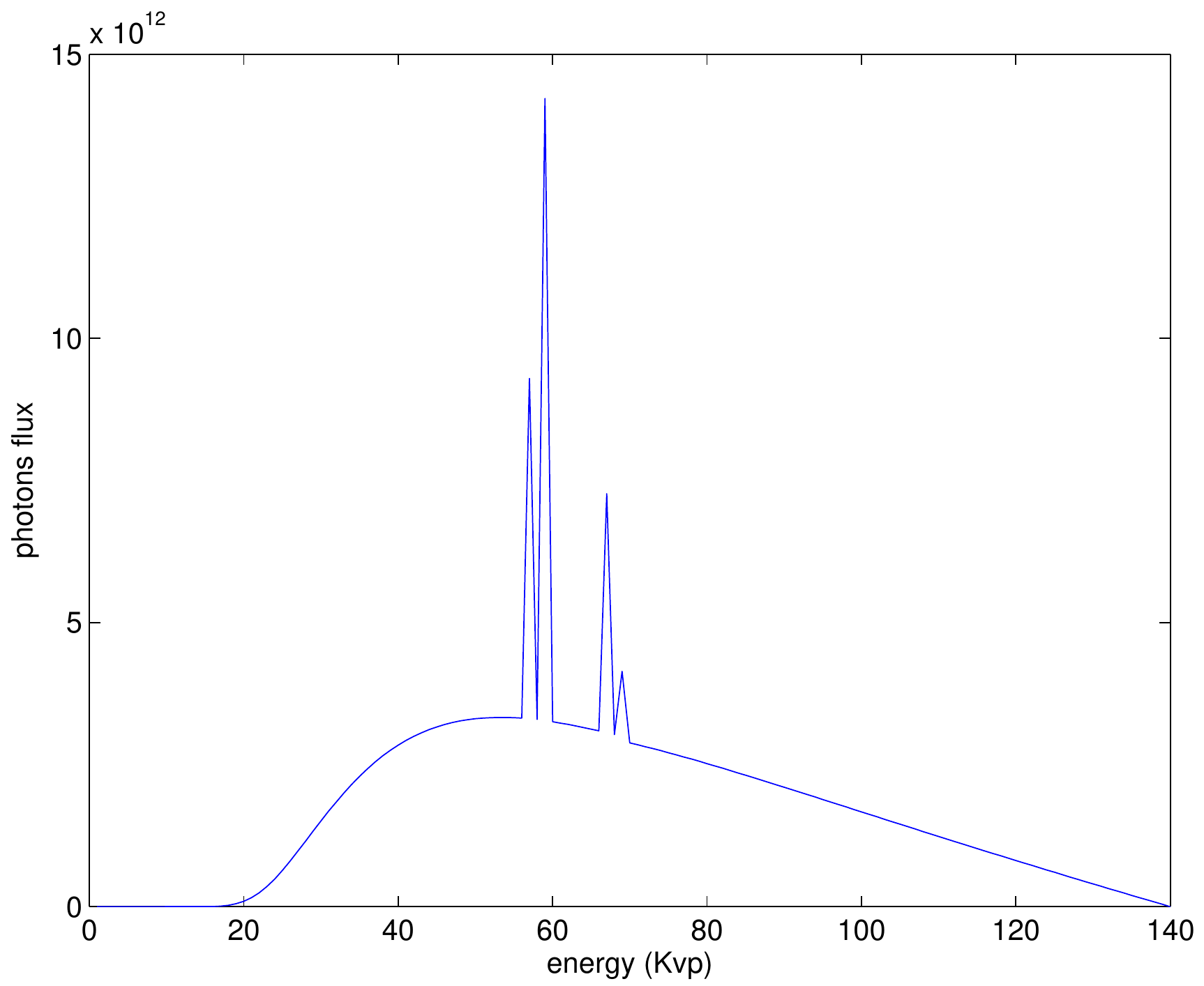}\label{energy-spectrum-140keV}}
    \caption{Energy spectrum $\bm{\mathcal{I}}_0(E)$ of X-ray source used to simulate the projection data $\bm{Y}_{0}$.
    }\label {fig-energy-spectrum-140keV}
\end{figure}

The projection data $\bm{Y}_{0}$ is obtained from a multi-chromatic X-ray imaging system definition by \eqref{def-Y-with-spectral} with the energy spectrum $\bm{\mathcal{I}}_0(E)$ shown in Figure \ref{fig-energy-spectrum-140keV}. The measured projection data $\bm{Y}$ contaminated by Poisson noise is generated in the following way
\begin{equation}\label{noise-poisson}
\bm{Y}=-\log(max\{\text{Poissrnd}(I_{0}\exp(-\bm{Y}_{0}))/I_{0},1/I_{0}\}),
\end{equation}
where $\text{Poissrnd}(\cdot)$ is used to add the Poisson noise, $I_{0}$ is the incident photons' number and the term $1/I_{0}$ is used to replace the $0$ pixel value after adding Poisson noise. In this subsection, we select $I_{0}=10^{5}$ in \eqref{noise-poisson}.

\subsubsection{Reconstruction results}
\begin{figure}[t]
\centering
    \subfigure[FBP]{\includegraphics[scale=0.4,trim=0cm 0cm 0cm 0cm]{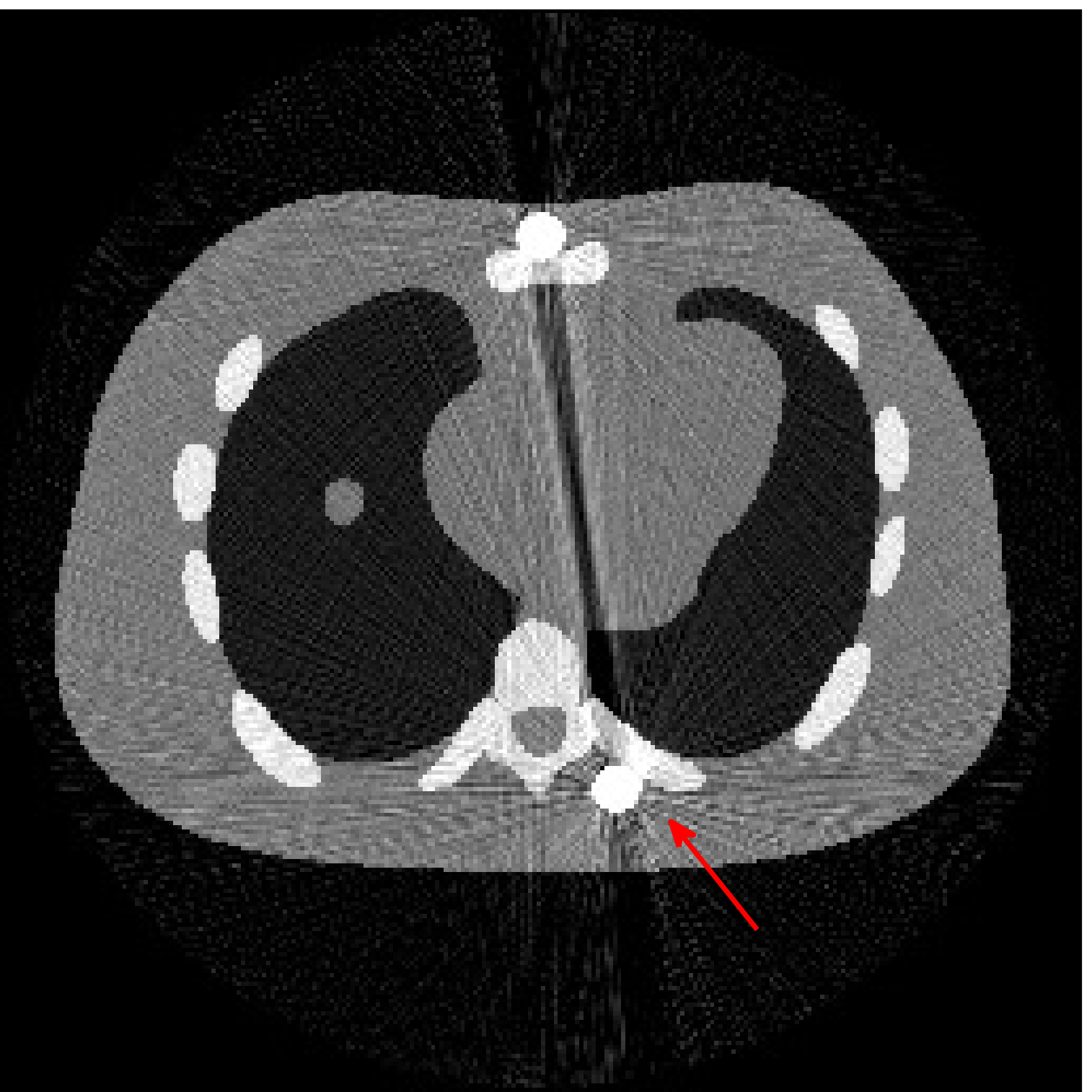}\label{NCAT-low-FBP-recU}}\\
    \subfigure[Unweighted JSR ]{\includegraphics[scale=0.4,trim=0cm 0cm 0cm 0cm]{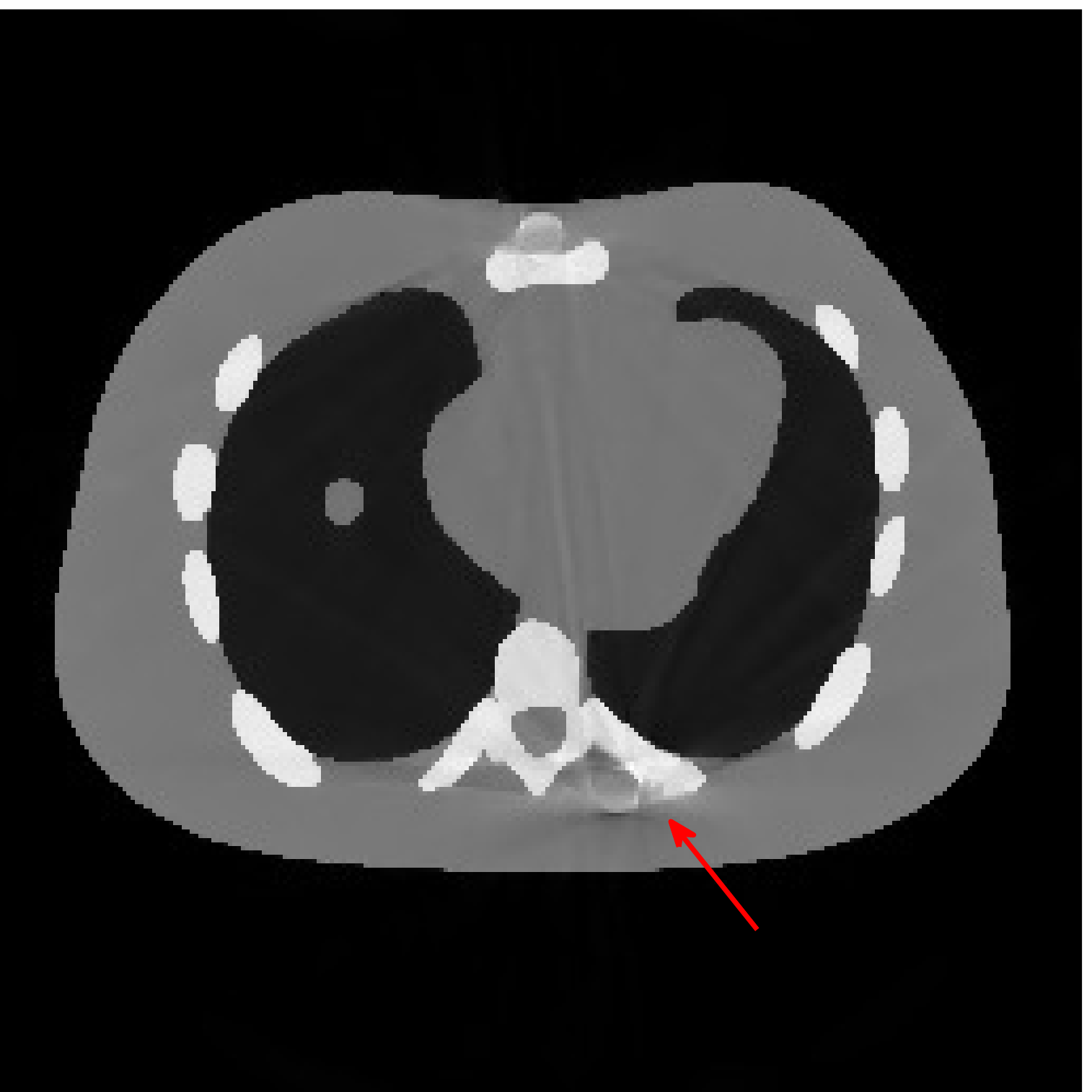}\label{NCAT-low-JSR-recU}}
    \subfigure[NMAR ]{\includegraphics[scale=0.4,trim=0cm 0cm 0cm 0cm]{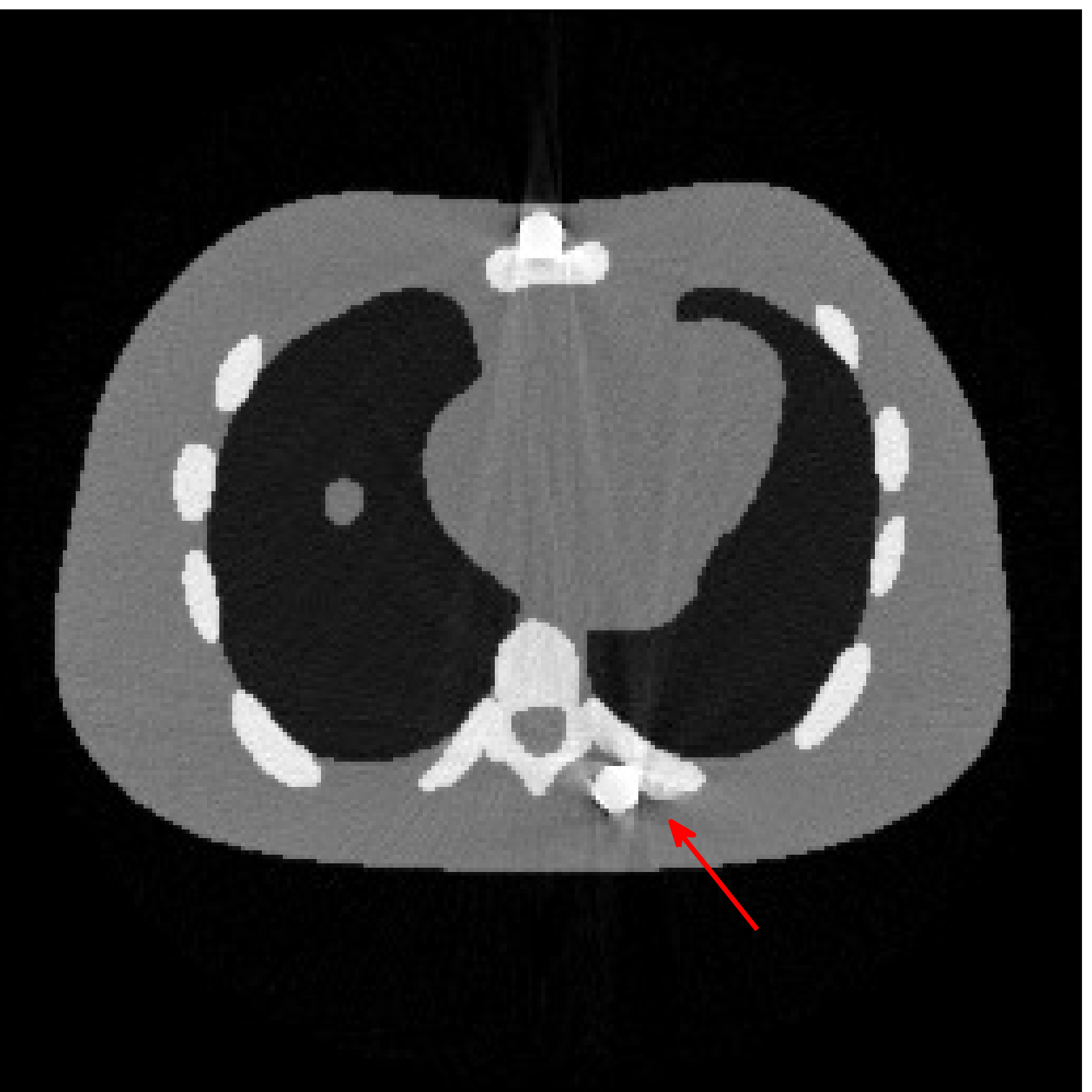}\label{NCAT-low-NMAR-recU}}
    \subfigure[TV-FADM ]{\includegraphics[scale=0.4,trim=0cm 0cm 0cm 0cm]{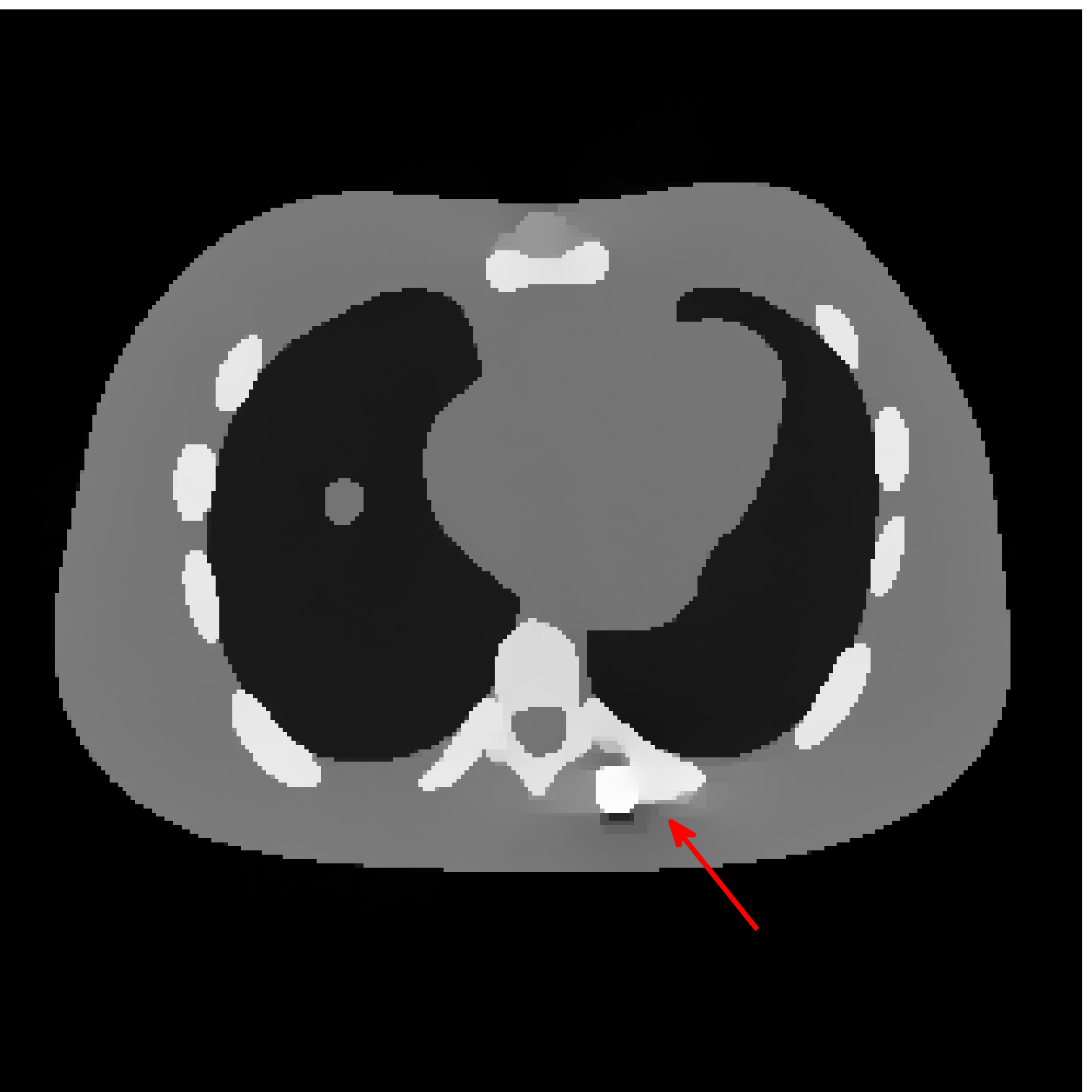}\label{NCAT-low-TVFADM-recU}}
    \subfigure[re-weighted JSR]{\includegraphics[scale=0.4,trim=0cm 0cm 0cm 0cm]{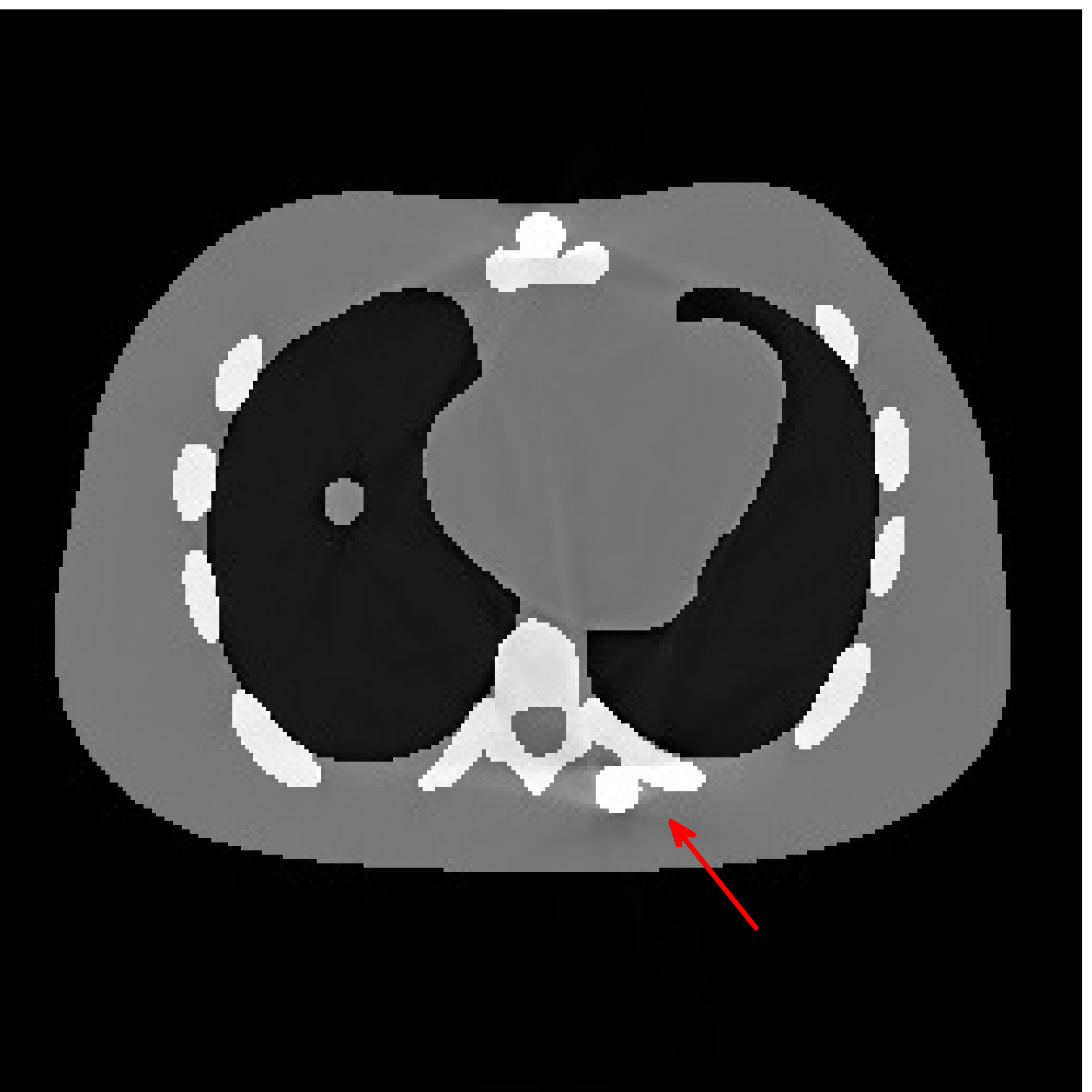}\label{NCAT-low-reWeighted-JSR-recU}}
    \caption{Comparison of reconstruction results using NCAT phantom.
    }\label{fig-NCAT-FBP-NMAR-TV-JSR}
\end{figure}

\begin{figure}[h]
\centering
    \subfigure[Original]{\includegraphics[scale=0.36,trim=0cm 0cm 0cm 0cm]{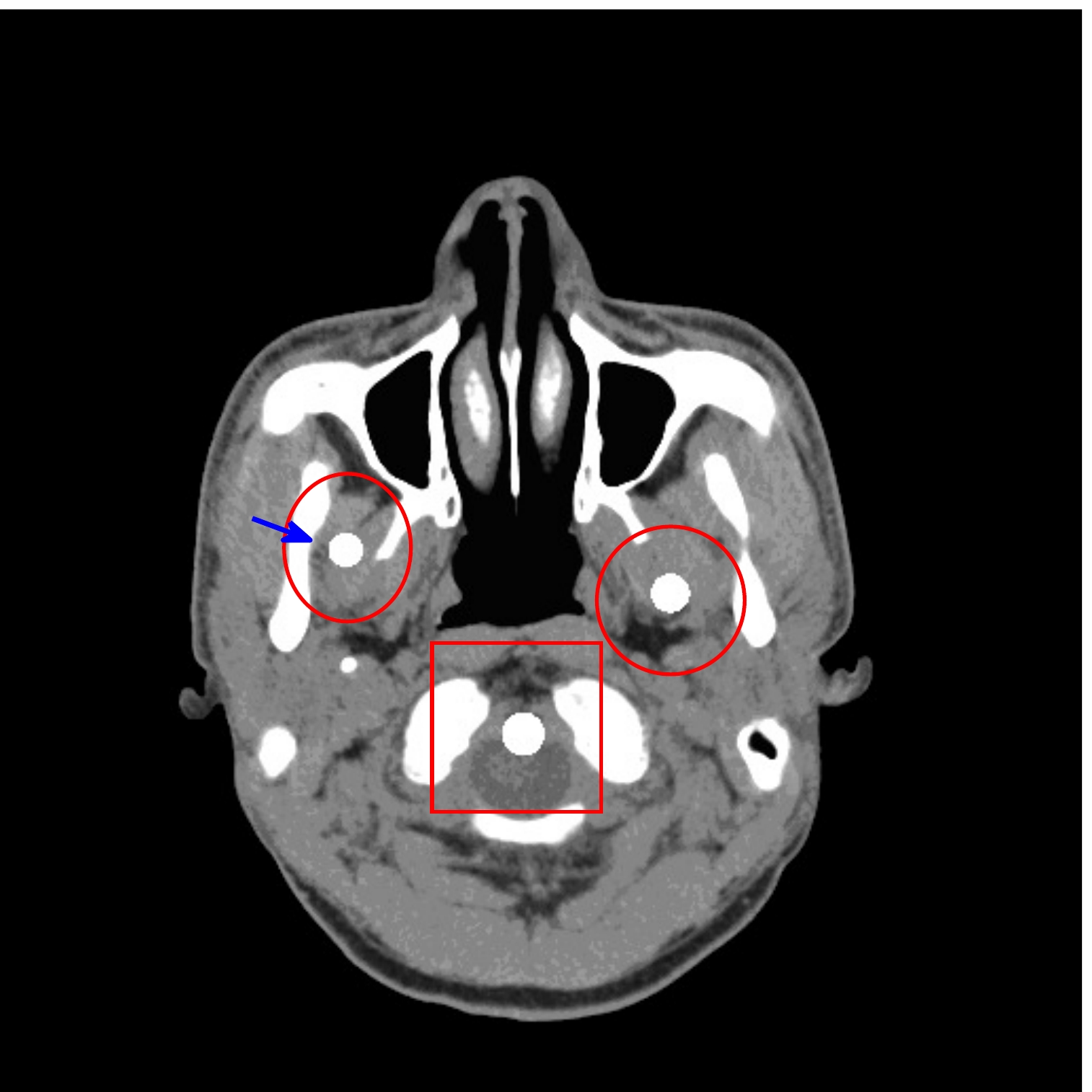}\label{CThead-512-ref}}
    \subfigure[FBP ]{\includegraphics[scale=0.36,trim=0cm 0cm 0cm 0cm]{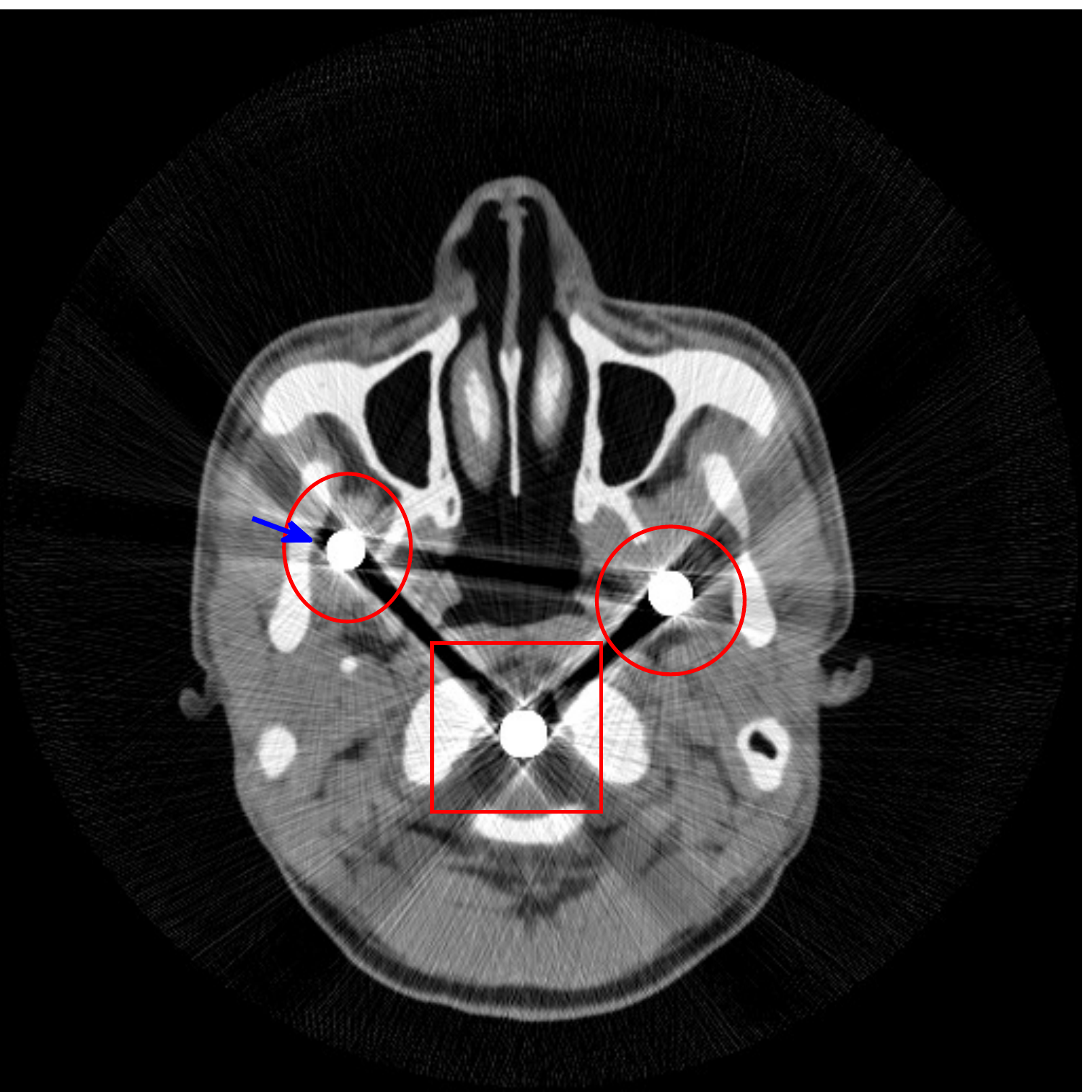}\label{Head-512-FBP-low-mark}}
    \subfigure[Unweighted JSR ]{\includegraphics[scale=0.36,trim=0cm 0cm 0cm 0cm]{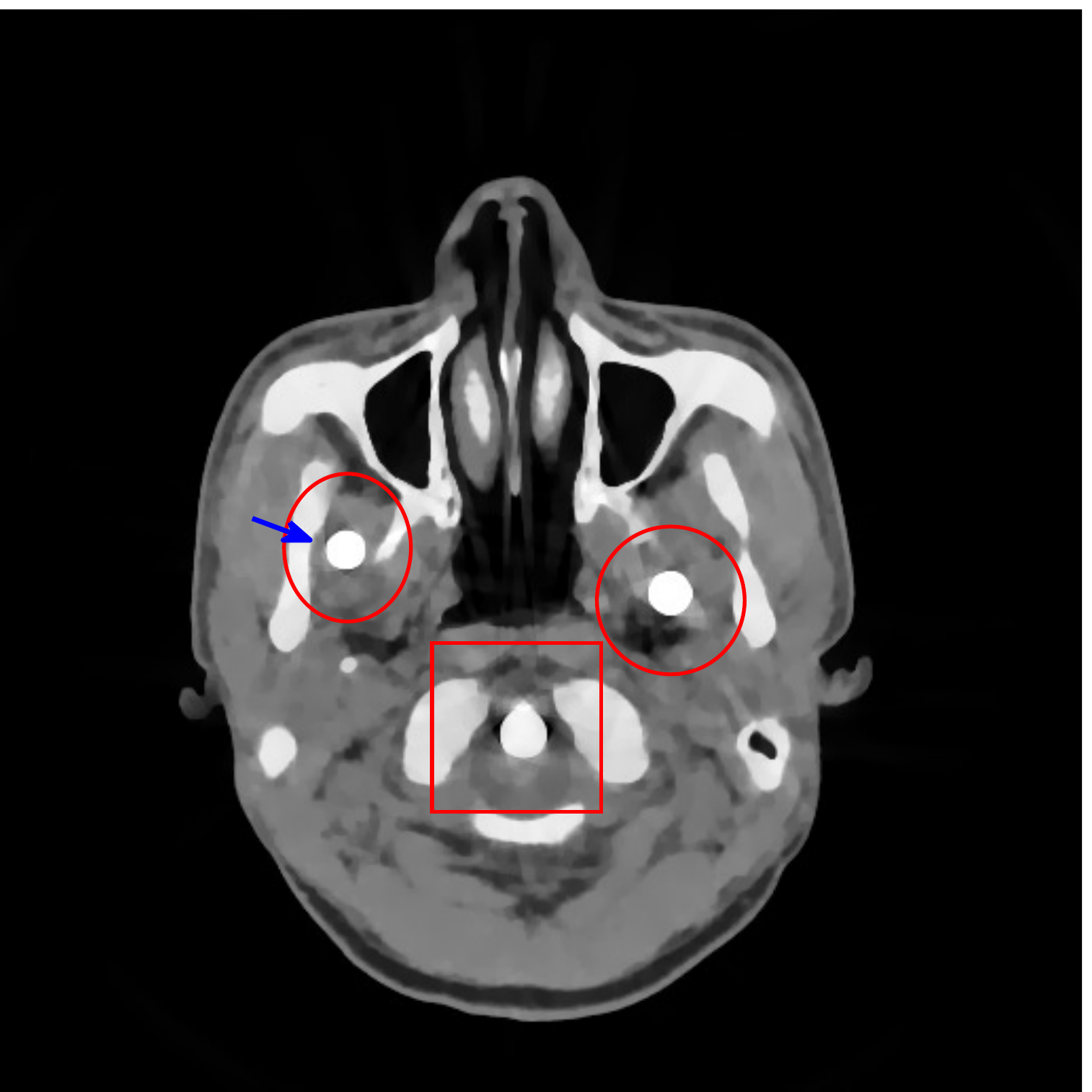}\label{Head-512-JSR-low-mark}}
    \subfigure[NMAR  ]{\includegraphics[scale=0.36,trim=0cm 0cm 0cm 0cm]{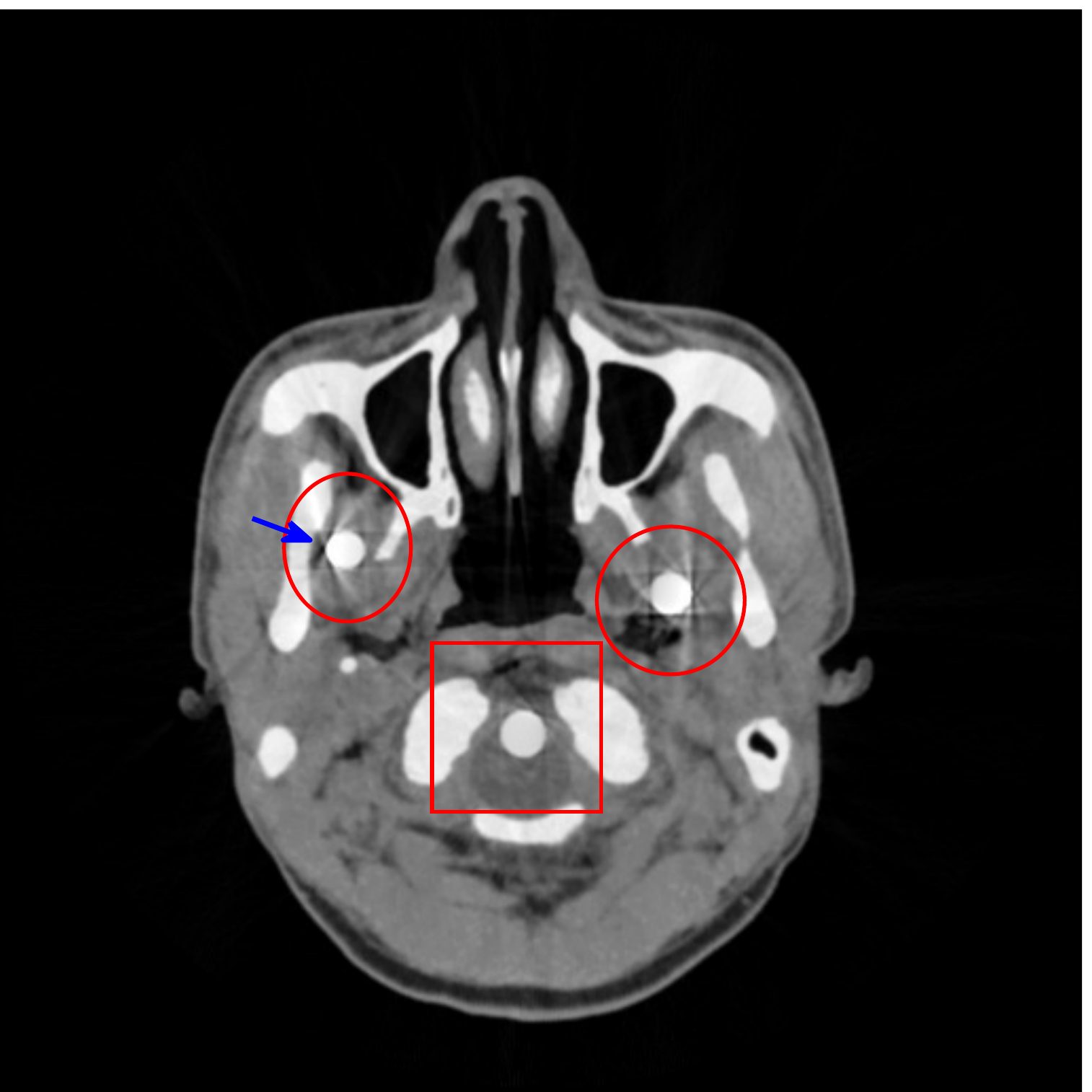}\label{Head-512-NMAR-low-mark}}
    \subfigure[TV-FADM  ]{\includegraphics[scale=0.36,trim=0cm 0cm 0cm 0cm]{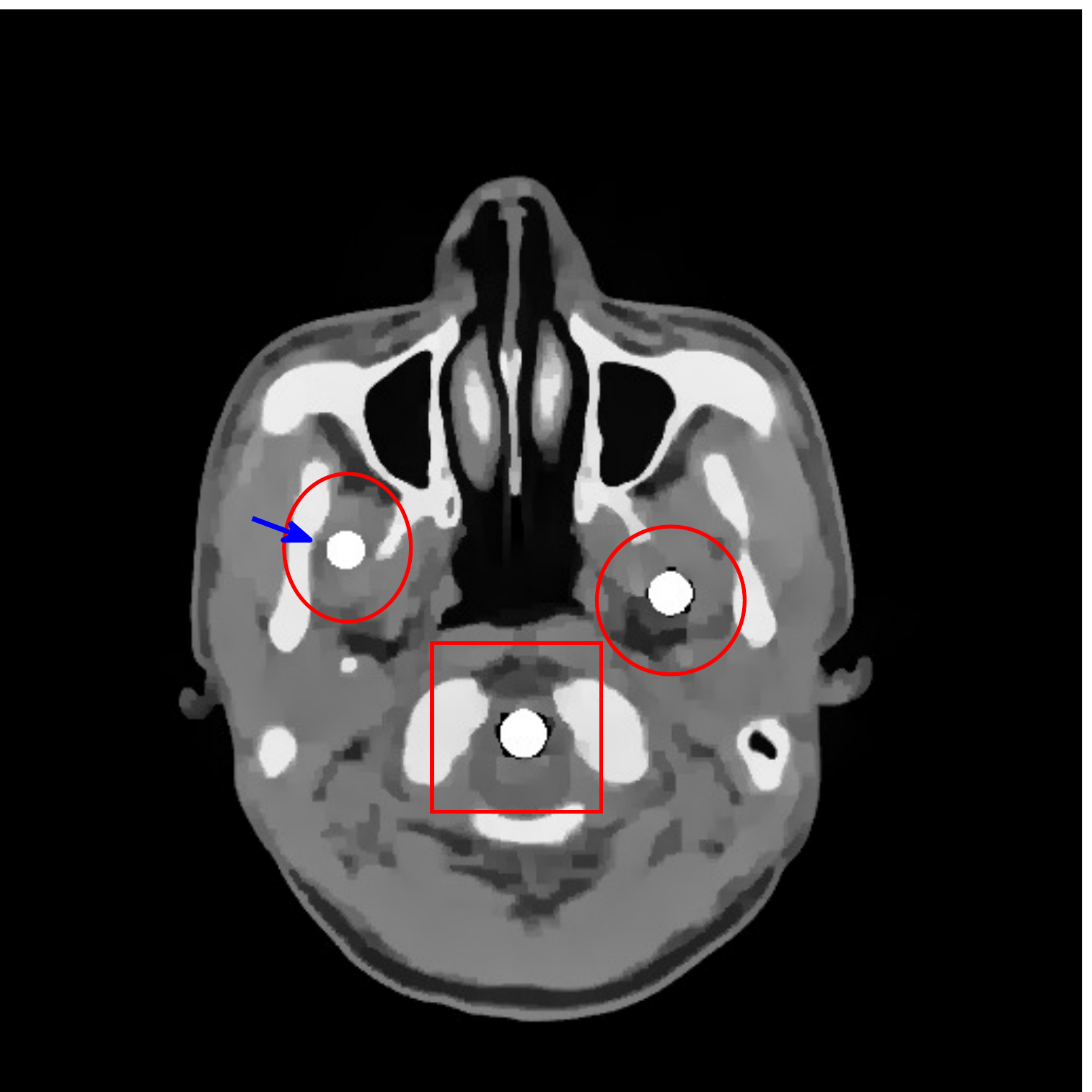}\label{Head-512-TVFADM-low-mark}}
    \subfigure[re-weighted JSR ]{\includegraphics[scale=0.36,trim=0cm 0cm 0cm 0cm]{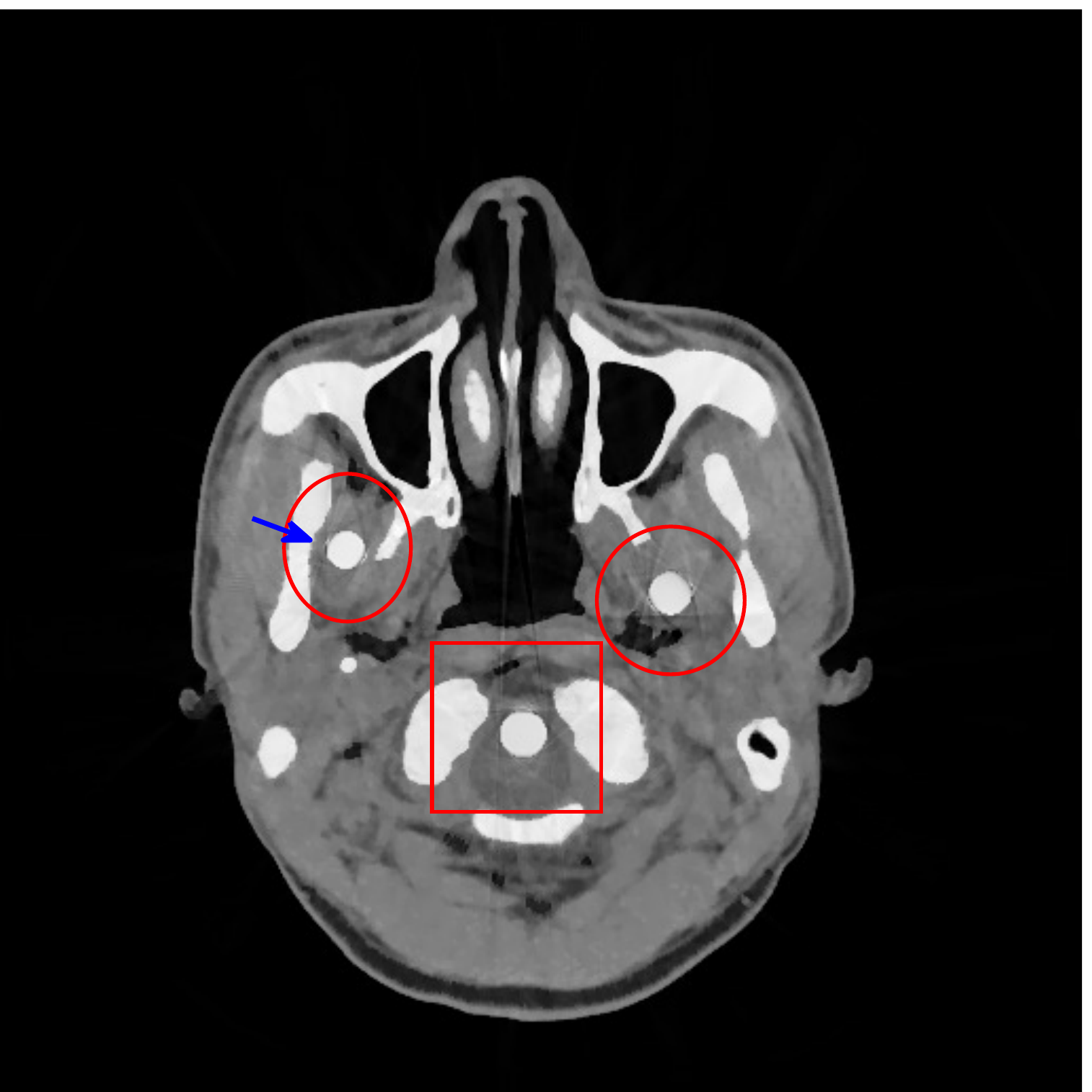}\label{Head-512-JSRDR-low-mark}}
    \caption{Comparison of reconstruction results using cerebral phantom.
    }\label{fig-CThead512-FBP-analysis-inpaint-seg}
\end{figure}

Table \ref{para-4-NCAT} is the parameter setting of re-weighted JSR model \eqref{re-weight-JSR-model} and unweighed JSR model \eqref{unweighted-JSR-model} for both phantoms.  The analysis model \eqref{analysis-model}, the inpainting model \eqref{inpaint-model}, the unweighted JSR model \eqref{unweighted-JSR-model} and the re-weighted JSR model \eqref{re-weight-JSR-model} are all initialized with $\bm{u}_{0}=0$. The number of CG iteration in all the algorithms involved is set to $5$.
We compare the re-weighted JSR model \eqref{re-weight-JSR-model} with the classical FBP algorithm \cite{katsevich2002theoretically}, the unweighted JSR model \eqref{unweighted-JSR-model}, NMAR \cite{meyer2010normalized} and  TV-FADM \cite{zhang2016iterative}. The codes of TV-FADM are provided by the authors of \cite{zhang2016iterative}. Comparisons of the aforementioned methods using the NCAT and cerebral phantoms are shown in Figure \ref{fig-NCAT-FBP-NMAR-TV-JSR} and Figure \ref{fig-CThead512-FBP-analysis-inpaint-seg}, respectively.

\begin{table}[t]
  \caption{Parameter setting in Model \eqref{re-weight-JSR-model} for NCAT and cerebral phantoms.}  \label{para-4-NCAT}
\centering
(a) NCAT\\
  \begin{tabular}{|c|c|c|c|c|c|c|}
\hline
\multirow{2}{*}{re-weighted JSR } &
\multicolumn{1}{c|}{paremeters} &
\multicolumn{1}{c|}{$\alpha$ } &
\multicolumn{1}{c|}{$\lambda_{1}$ }  &
\multicolumn{1}{c|}{$\lambda_{2}$ }  &
\multicolumn{1}{c|}{$\mu_{1}$ }  &
\multicolumn{1}{c|}{$\mu_{2}$ }  \\
\cline{2-7}
  &  value      &   1       &   3.6        & 0.15            &  0.3       &   0.3             \\
\hline
\multirow{2}{*}{unweighted JSR } &
\multicolumn{1}{c|}{paremeters} &
\multicolumn{1}{c|}{$\alpha$ } &
\multicolumn{1}{c|}{$\lambda_{1}$ }  &
\multicolumn{1}{c|}{$\lambda_{2}$ }  &
\multicolumn{1}{c|}{$\mu_{1}$ }  &
\multicolumn{1}{c|}{$\mu_{2}$ }  \\
\cline{2-7}
  &  value      &   1        &   1        & 0.6            &  0.1       &   9           \\
\hline
\end{tabular}

\vspace{10pt}
 (b) Cerebral\\
\begin{tabular}{|c|c|c|c|c|c|c|}
\hline
\multirow{2}{*}{re-weighted JSR } &
\multicolumn{1}{c|}{paremeters} &
\multicolumn{1}{c|}{$\alpha$ } &
\multicolumn{1}{c|}{$\lambda_{1}$ }  &
\multicolumn{1}{c|}{$\lambda_{2}$ }  &
\multicolumn{1}{c|}{$\mu_{1}$ }  &
\multicolumn{1}{c|}{$\mu_{2}$ }  \\
\cline{2-7}
  &  value      &   18        &   1.6        & 0.15            &  0.6       &  1.2            \\
\hline
\multirow{2}{*}{unweighted JSR } &
\multicolumn{1}{c|}{paremeters} &
\multicolumn{1}{c|}{$\alpha$ } &
\multicolumn{1}{c|}{$\lambda_{1}$ }  &
\multicolumn{1}{c|}{$\lambda_{2}$ }  &
\multicolumn{1}{c|}{$\mu_{1}$ }  &
\multicolumn{1}{c|}{$\mu_{2}$ }  \\
\cline{2-7}
  & value      &   1        &   0.8        & 0.45            &  0.1       &  2            \\
\hline
\end{tabular}
\end{table}

Figure \ref{NCAT-low-FBP-recU} shows that the reconstructed NCAT phantom from FBP has severe metal artifacts and is noisy. The reconstructed image from the unweighted JSR model \eqref{unweighted-JSR-model} shown in Figure \ref{NCAT-low-JSR-recU} has a better visual effect with noticeably less noise and metal artifacts. Sharp edges are also well preserved except for the blurry effects in the region surrounding the metals. The reconstructed image from NMAR shown in Figure \ref{NCAT-low-NMAR-recU} also has most of the metal artifacts suppressed and the regions surrounding the metals are much less blurry than the unweighted JSR. However, the unweighted JSR does a better job than NMAR in suppressing noise and preserving sharp image features away from the metals. TV-FADM is able to reconstruct images with minimum metal artifacts and noise, as shown in Figure \ref{NCAT-low-TVFADM-recU}. However, the metal components are fused with nearby structures which is highlighted by the red arrow. The reconstructed image from proposed re-weighted JSR model has the best overall quality with rather minor metal artifacts.

Figure \ref{fig-CThead512-FBP-analysis-inpaint-seg} shows the reconstructed cerebral phantom from different methods. We highlight some regions with more distinct differences with red contours. Since the cerebral phantom contains more textures, it is more challenging than the NCAT phantom. The pros and cons of these methods are mostly the same as the previous example. However, we note that the reconstructed image from TV-FADM shown in Figure \ref{Head-512-TVFADM-low-mark} has severe artifact, which is due to the well-known staircase artifact of TV regularization. We found that TV-FADM is relatively sensitive to the choice of its parameters. It is not easy to balance between sharpness of image features and metal artifacts reduction. The soft tissue around metal components is also not well preserved by the NMAR method as indicated by the blue arrow in Figure \ref{Head-512-NMAR-low-mark}. Furthermore, the circled areas in Figure \ref{Head-512-NMAR-low-mark} show that there are still some artifacts around the metal. Same as the NCAT phantom, the proposed re-weighted JSR model has the best overall performance. {Notice that the intensity of metals in Figure \ref{Head-512-NMAR-low-mark} and \ref{Head-512-JSRDR-low-mark} seems lower than the rest of the reconstructed images. This is because we set the intensity of the metal components in the segmentation $\bm{u}_{s}$ with the same mean value as that of bones. Increasing the value of metal components of $\bm{u}_s$ can increase the intensity of metals in the reconstructed images, whereas it also introduces more artifacts around the metals.}

\begin{table}[t]
\renewcommand{\arraystretch}{1.3}
\caption{Relative error, SSIM index and the CPU time (sec.) of the NCAT and cerebral phantoms reconstructed by different algorithms, i.e. FBP, unweighted JSR \eqref{unweighted-JSR-model}, NMAR \cite{meyer2010normalized}, TVFADM \cite{zhang2016iterative} and the proposed re-weighted JSR \eqref{re-weight-JSR-model}. \label{tab-NCAT-Head-RelErr-SSIM-CPU}}
\centering
(a) NCAT\\
\begin{tabular}{|c|c|c|c|}
\hline
\multicolumn{1}{|c|}{Model } &
\multicolumn{1}{c|}{Relative error } &
\multicolumn{1}{c|}{SSIM  }  &
\multicolumn{1}{c|}{Time} \\ \hline
 FBP               & 0.1927          & 0.4903    &  -    \\\hline
 Unweighted JSR    & 0.0947          & 0.9645    &  356   \\\hline
  NMAR             & 0.1109          & 0.8431    &  -     \\\hline
    TV-FADM        & 0.0911          & 0.9870    &  69    \\\hline
re-weighted JSR    & 0.0860          & 0.9839    &  239    \\\hline
\end{tabular}

\vspace{10pt}
(b) Cerebral \\
\begin{tabular}{|c|c|c|c|}
\hline
\multicolumn{1}{|c|}{Model } &
\multicolumn{1}{c|}{Relative error } &
\multicolumn{1}{c|}{SSIM  }  &
\multicolumn{1}{c|}{Time}  \\ \hline
 FBP               & 0.2816          & 0.4847    &  -     \\ \hline
 Unweighted JSR    & 0.1410         & 0.9006    &  812   \\ \hline
  NMAR             & 0.1330          & 0.8532    &  -     \\ \hline
   TV-FADM         &  0.3336         &  0.9090   &  516    \\ \hline
 re-weighted JSR   &  0.1129         &  0.9155   &  309   \\ \hline
\end{tabular}
\end{table}

\begin{figure}[h!]
\center
	\subfigure[ ]{\includegraphics[scale=0.3,trim=0cm 0cm 0cm 0cm]{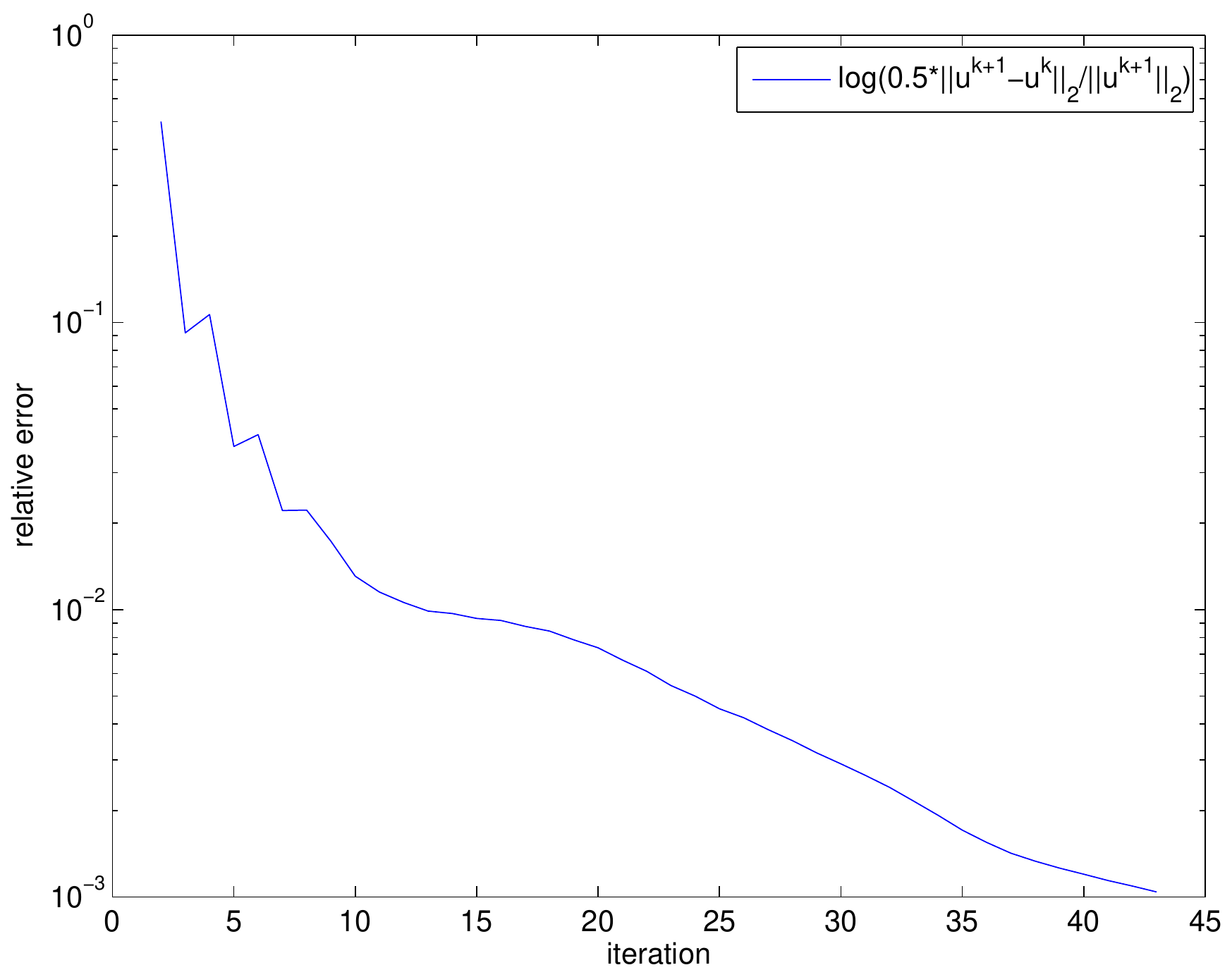}\label{error-curve-log-uk}}
	\subfigure[ ]{\includegraphics[scale=0.3,trim=0cm 0cm 0cm 0cm]{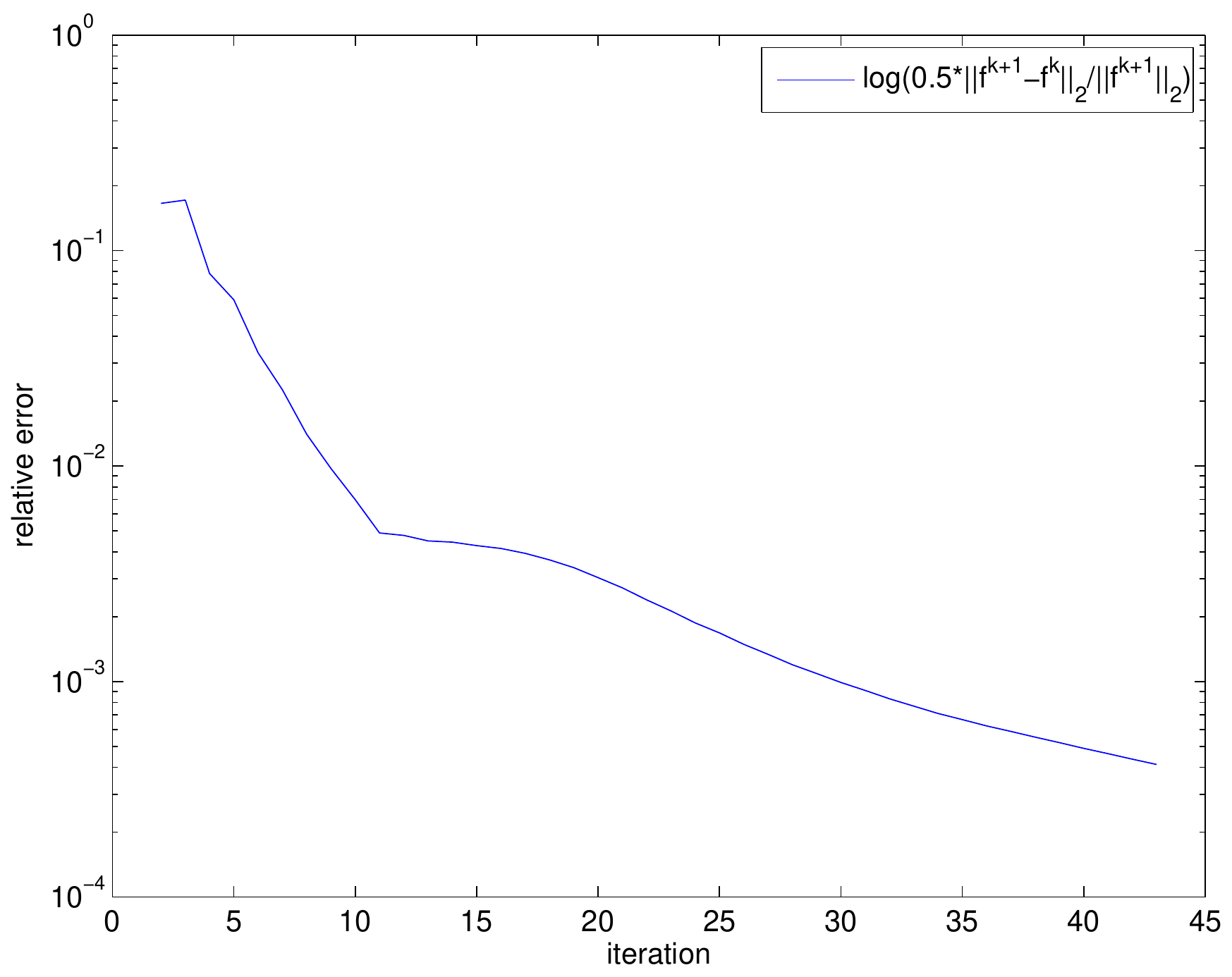}\label{error-curve-log-fk}}
    \subfigure[ ]{\includegraphics[scale=0.3,trim=0cm 0cm 0cm 0cm]{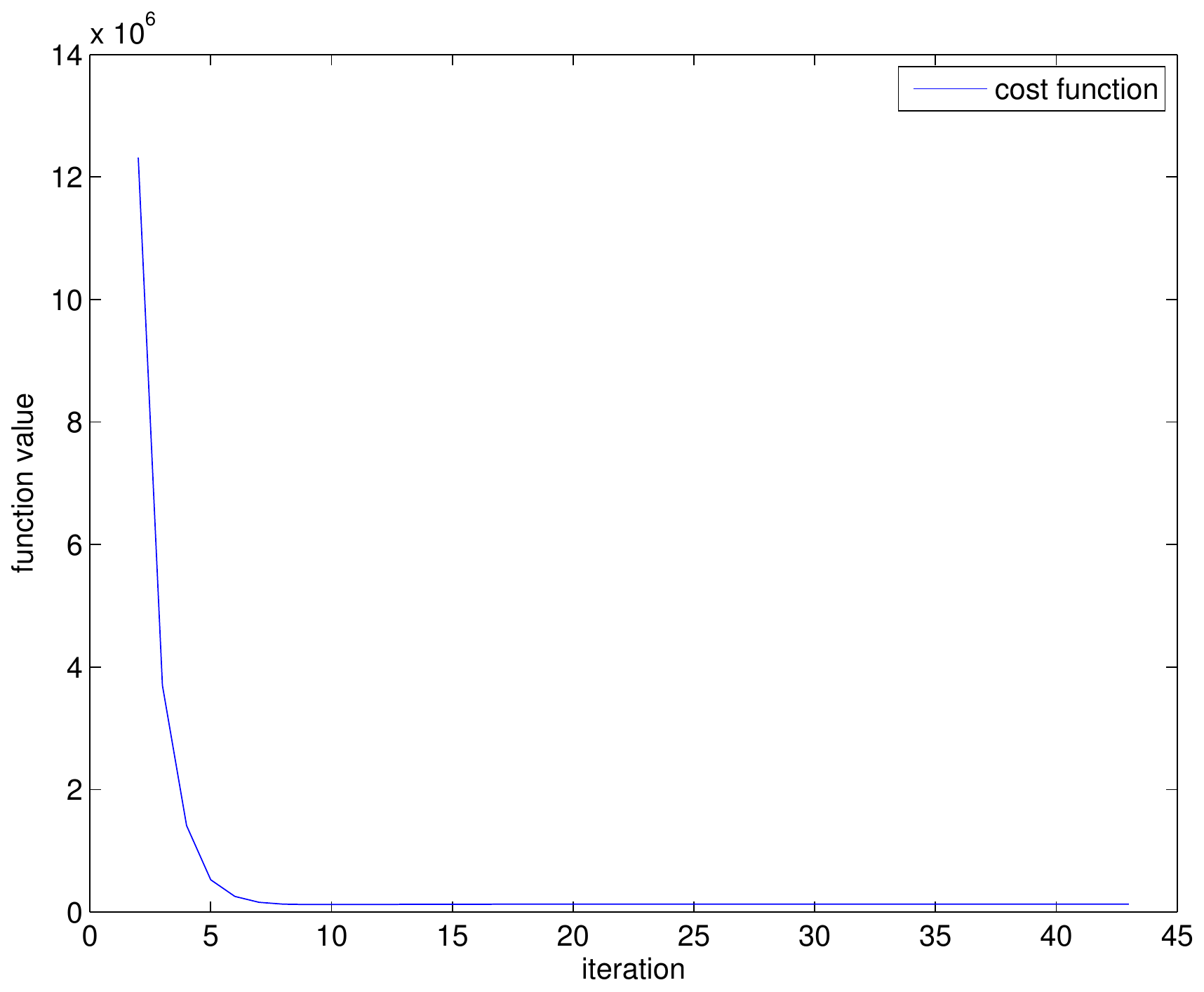}\label{error-curve-costfunc}}
    \caption{
    Convergence of Algorithm 1 on the NCAT phantom.
    }\label{fig-error-curve}
\end{figure}

Quantitative assessments of the reconstruction quality of these methods are given in Table \ref{tab-NCAT-Head-RelErr-SSIM-CPU}, where the relative errors, SSIM values and computation time are presented. Although Algorithm \ref{JSR-alg-Jacobi} that solves the re-weighted JSR model is relatively time consuming, we are able to gain on quality of the reconstructed images. { Finally, to numerically demonstrate the convergence of the Algorithm 1, we present the decay of $\log\left(0.5\frac{\|\bm{u}^{k+1}-\bm{u}^k\|}{\|\bm{u}^{k+1}\|}\right)$, $\log\left(0.5\frac{\|\bm{f}^{k+1}-\bm{f}^k\|}{\|\bm{f}^{k+1}\|}\right)$ and the cost function of the re-weighted JSR model in Figure \ref{error-curve-log-uk}, \ref{error-curve-log-fk} and \ref{error-curve-costfunc} respectively.}

\subsection{Numerical experiments: real data}\label{section-numerical-real}

\begin{figure}[h!]
\center
  \subfigure[Chicken leg with nails placed on each side.]{\includegraphics[scale=0.12,trim=0cm 0cm 0cm 0cm]{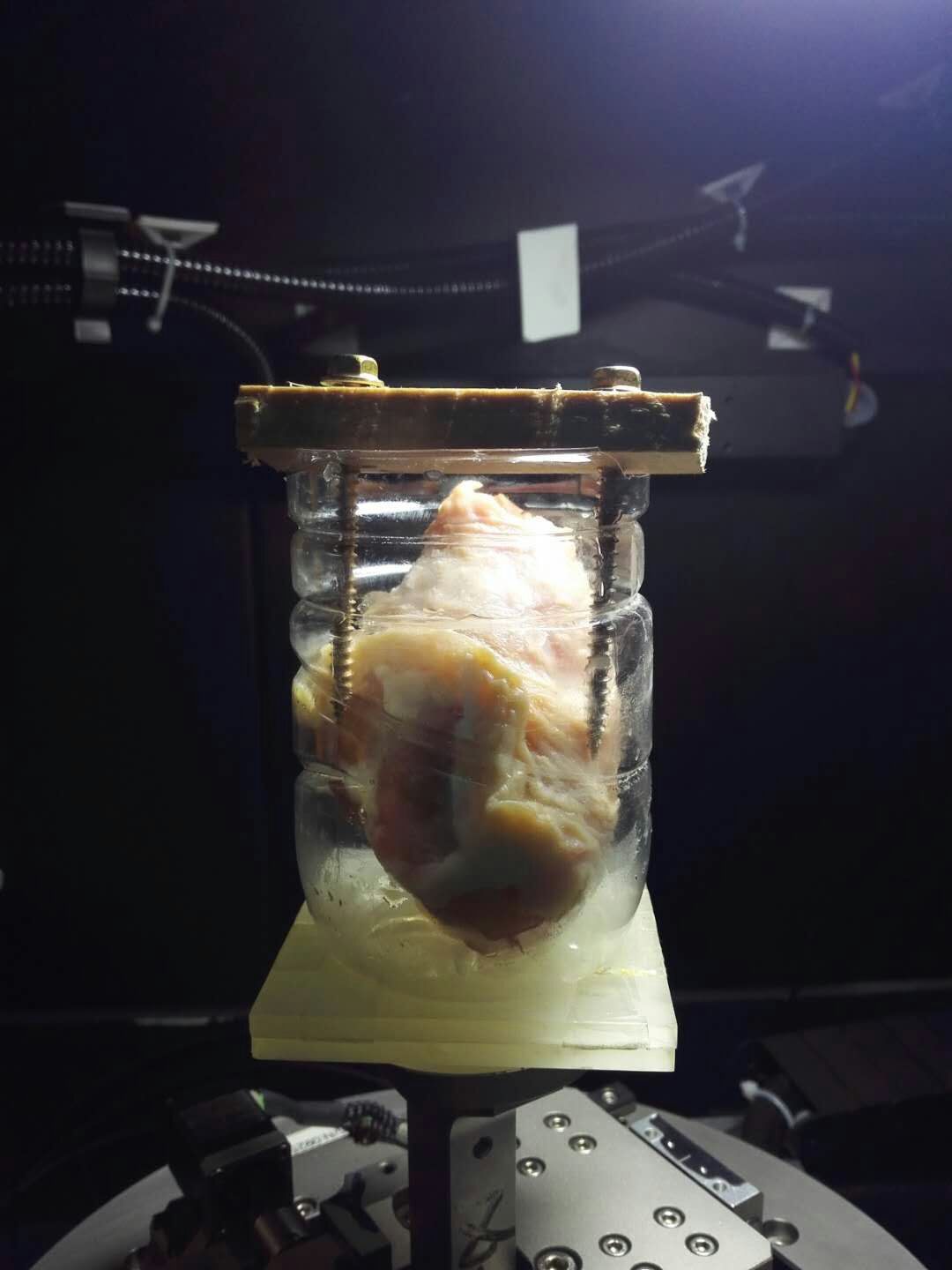}\label{scan-1}}
  \subfigure[Scanning of the reference image (without metal)]{\includegraphics[scale=0.12,trim=0cm 0cm 0cm 0cm]{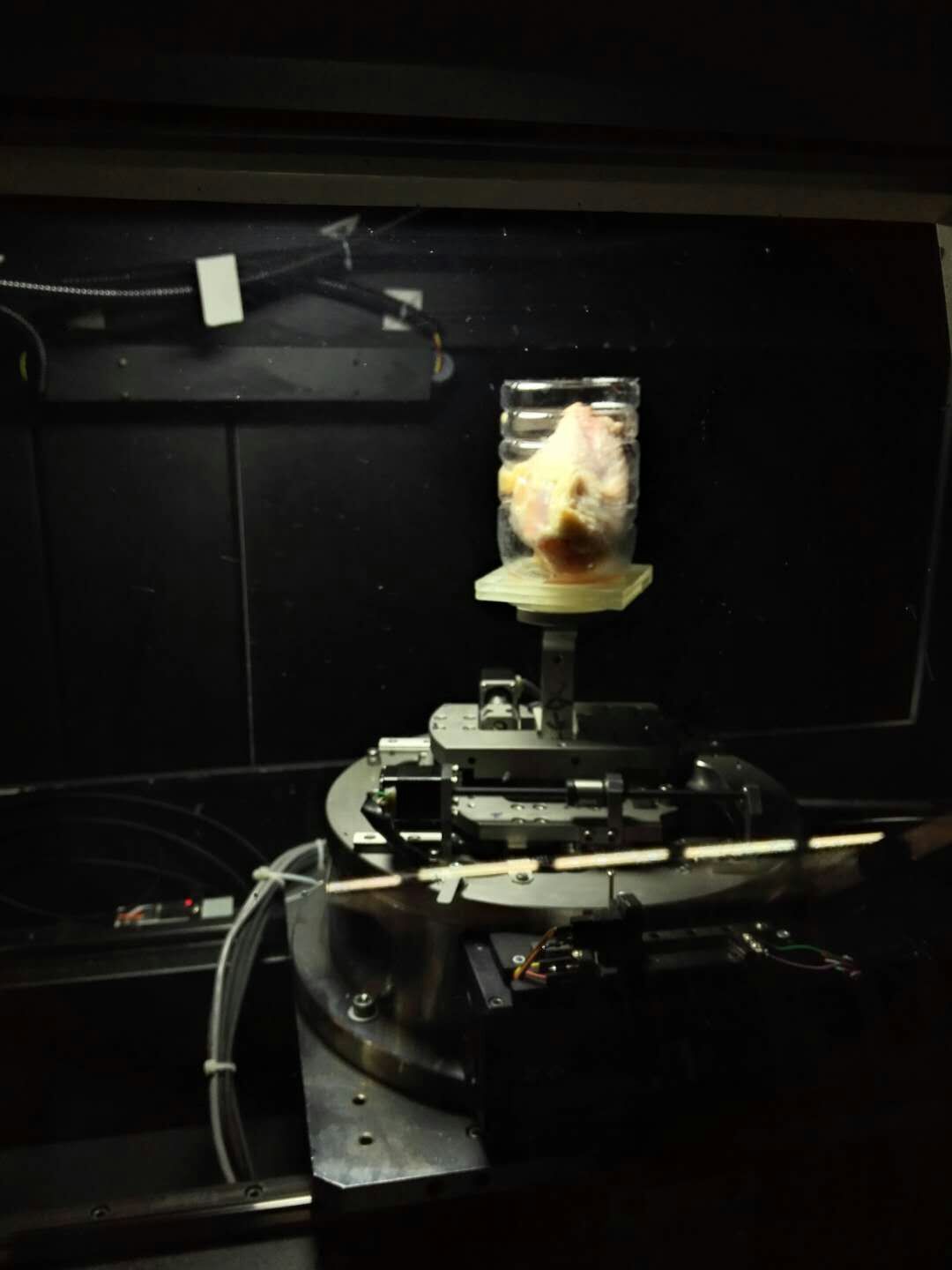}\label{scan-2}}
  \subfigure[Scanning again with metals]{\includegraphics[scale=0.197,trim=0cm 0cm 0cm 0cm]{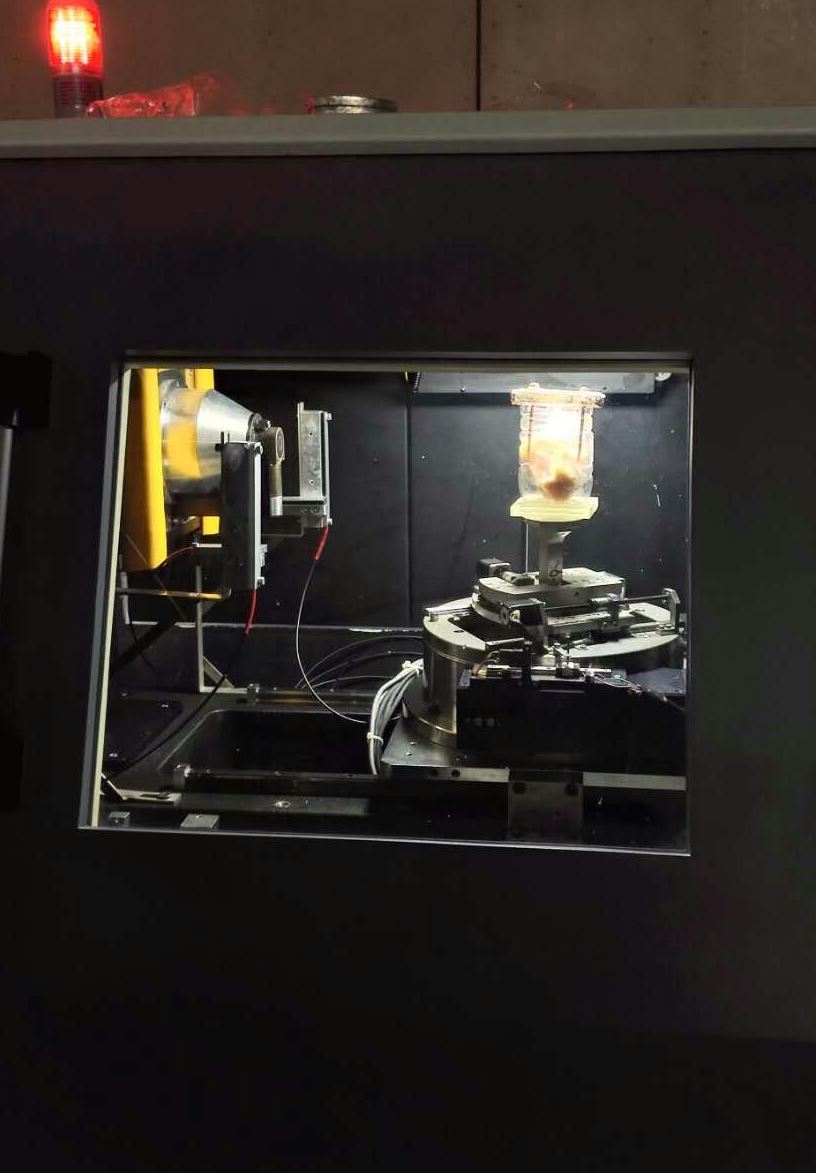}\label{scan-4}}
  \caption{  Real data scanning.
    }\label{fig-scan-protocol}
\end{figure}

We perform a CT scan of a chicken leg placed in a disposable cup (Figure \ref{scan-1}). We first scan the chicken leg without metals (Figure \ref{scan-2}) to create a reference image using FBP algorithm. Then, we place two steel thread nails on each side of the chicken leg and scan the subject again using the same scanning protocol (Figure \ref{scan-4}). The projection data is acquired from a MicroCT scanner equipped at the Division of Nuclear Technology and Applications, Institute of High Energy Physics, Chinese Academy of Sciences. The X-ray source is with 90 kV and 70 mA energy and the flat plane detector contains $1024\times 1024$ pixels. The scanning trajectory is a full circle with equally spaced views at $1^\circ$ per view. The physical size of each detector unit is $0.127mm  \times 0.127mm $. The distance from the X-ray source to the detector is $789mm$. In order to conduct a 2D experiment, we choose the 512th row of the detector array.

\begin{figure}[h]
\centering
    \subfigure[Reference]{\includegraphics[scale=0.4,trim=0cm 0cm 0cm 0cm]{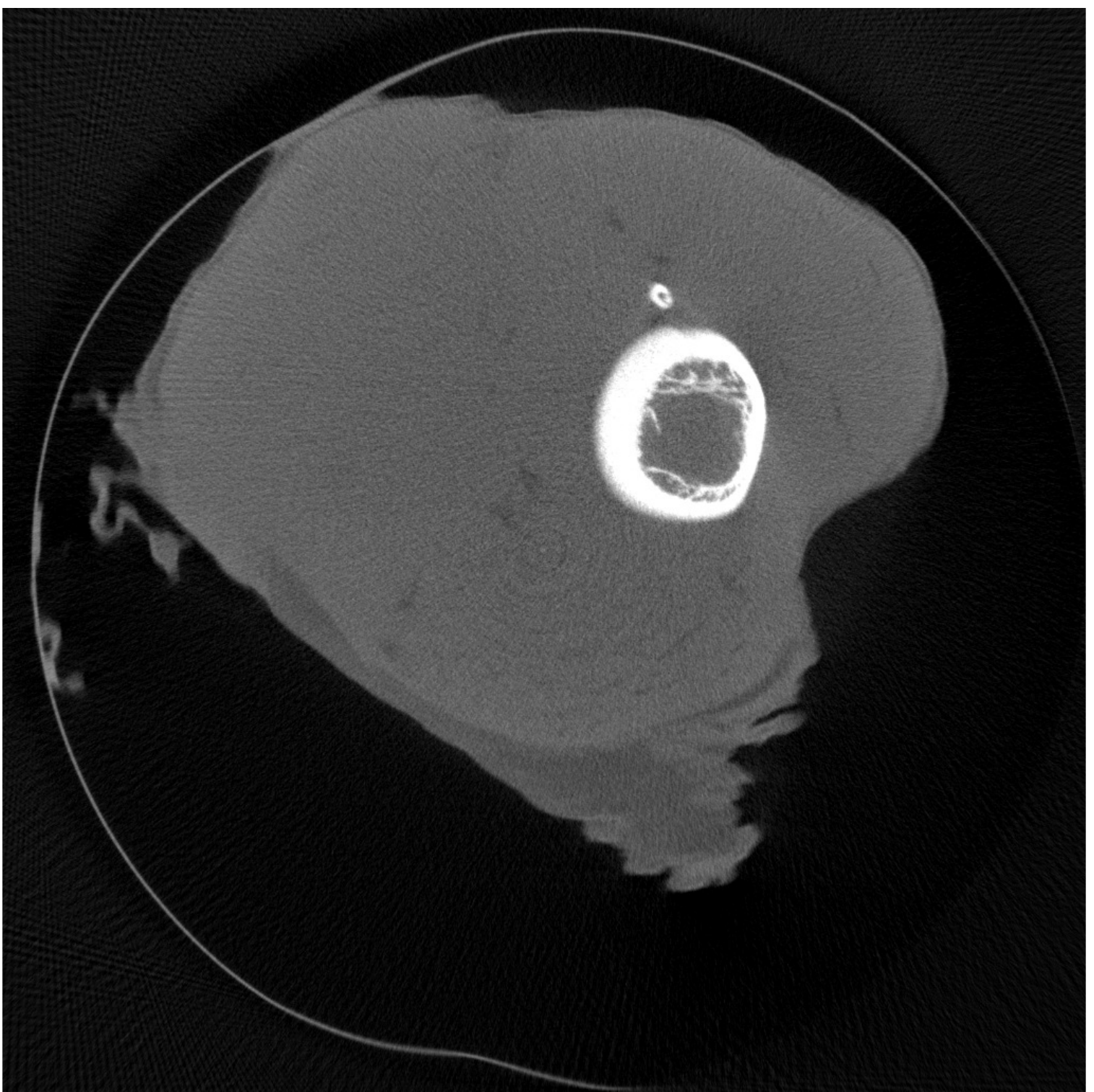}\label{FBP-CW-reference}}\\
    \subfigure[FBP]{\includegraphics[scale=0.4,trim=0cm 0cm 0cm 0cm]{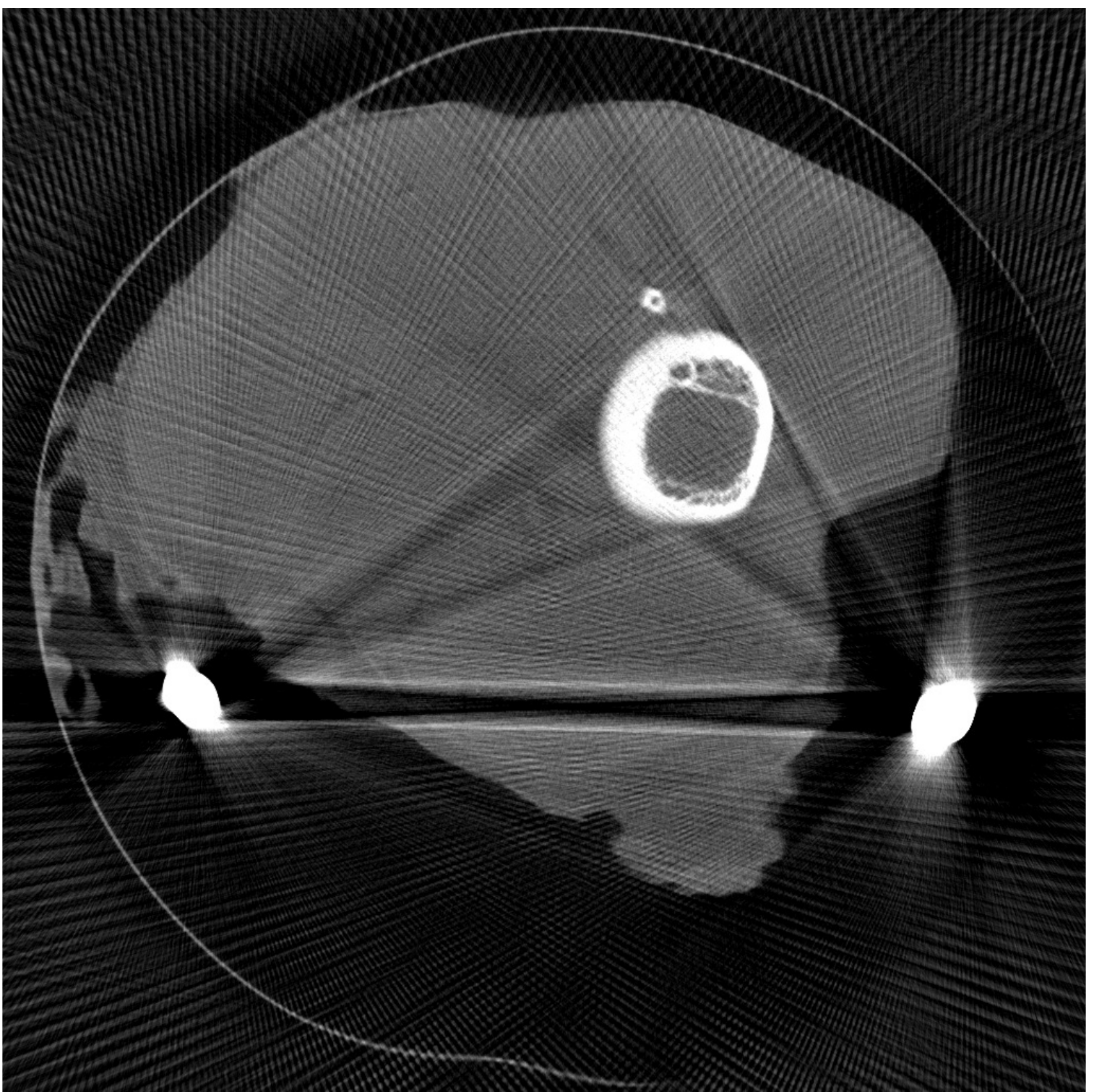}\label{FBP-CW}}
    \subfigure[Model \eqref{analysis-model}]{\includegraphics[scale=0.4,trim=0cm 0cm 0cm 0cm]{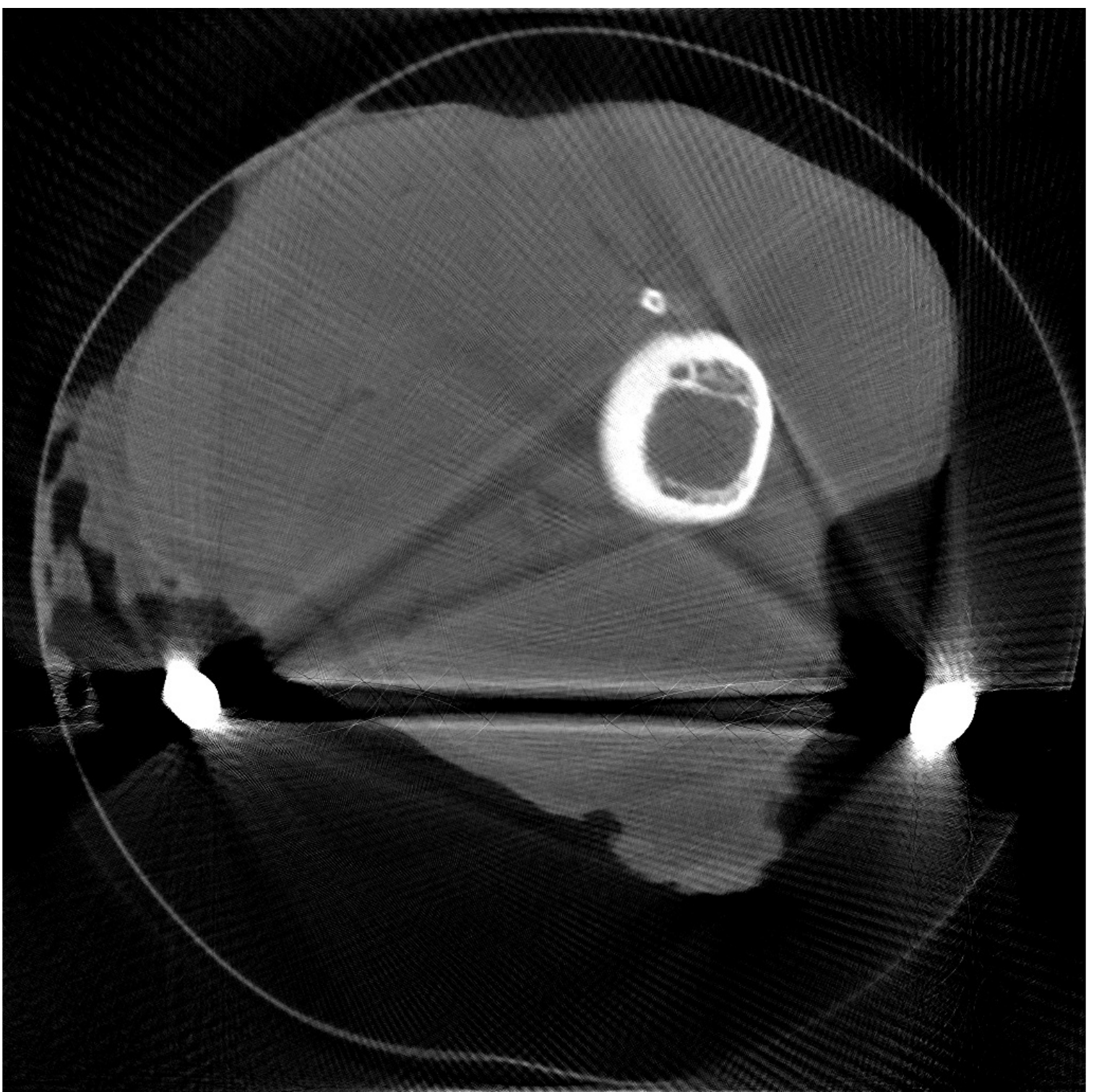}\label{analysis-CW}}
    \subfigure[Model \eqref{inpaint-model}]{\includegraphics[scale=0.4,trim=0cm 0cm 0cm 0cm]{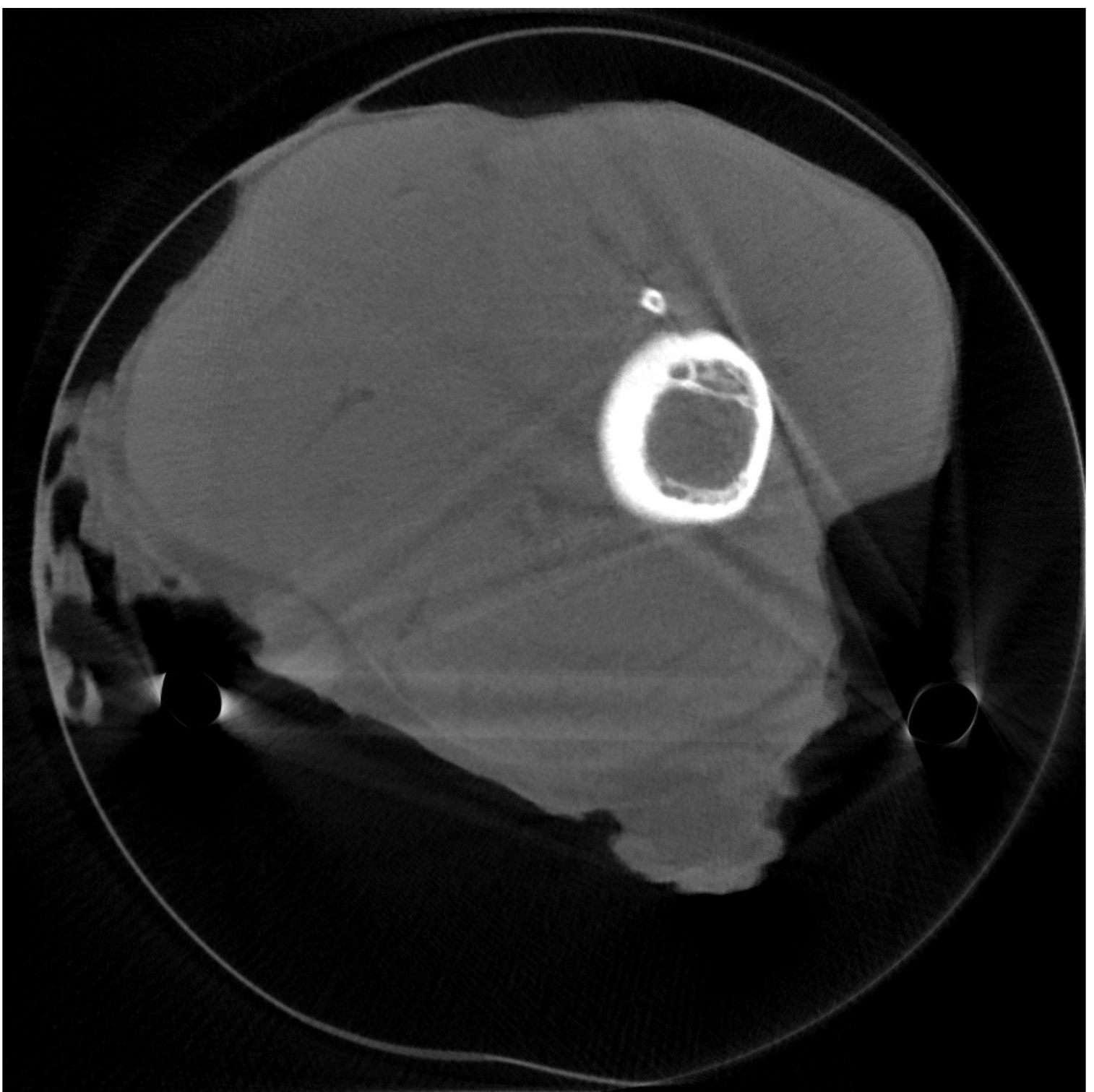}\label{inpaint-CW}}
    \subfigure[Segmented image]{\includegraphics[scale=0.4,trim=0cm 0cm 0cm 0cm]{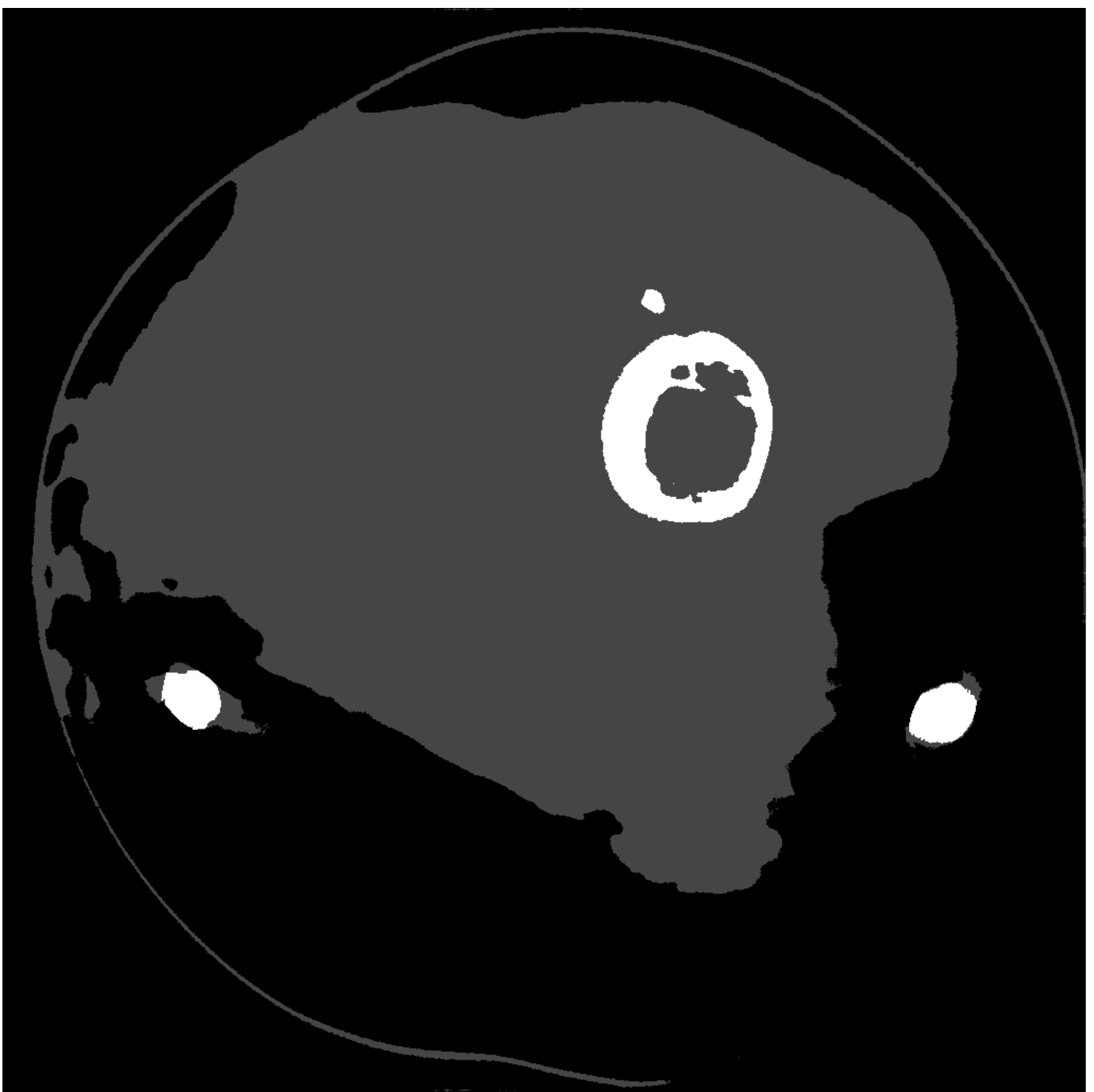}\label{seg-CW}}
    \caption{
    \subref{FBP-CW-reference} Reference image without meal implants.
    \subref{FBP-CW} FBP reconstructed image with $1024\times 1024$ pixels.
    \subref{analysis-CW} Reconstructed image from the analysis model \eqref{analysis-model}.
    \subref{inpaint-CW} Reconstructed image from the inpainting model \eqref{inpaint-model}.
    \subref{seg-CW} Segmented image from the image \eqref{ua-up}.
    }\label {fig-CW-FBP}
\end{figure}

Figure \ref{fig-CW-FBP} shows the images reconstructed using FBP, the analysis model \eqref{analysis-model}, the inpainting model \eqref{inpaint-model} and the segmented image from the image obtained by \eqref{ua-up}. The reference image without metal implants are shown in Figure \ref{FBP-CW-reference}. All the images in this subsection are displayed within the grayscale interval $[0,0.05]$. The segmented image $\bm{u}_{s}$ shown in Figure \ref{seg-CW} is used to estimate the weights needed in NMAR and the re-weighted JSR model.

\begin{figure}[p]
  \subfigure[unweighted JSR]{
  \begin{minipage}{0.64\textwidth}\label{CW-JSR}
  \includegraphics[width=0.9\textwidth]{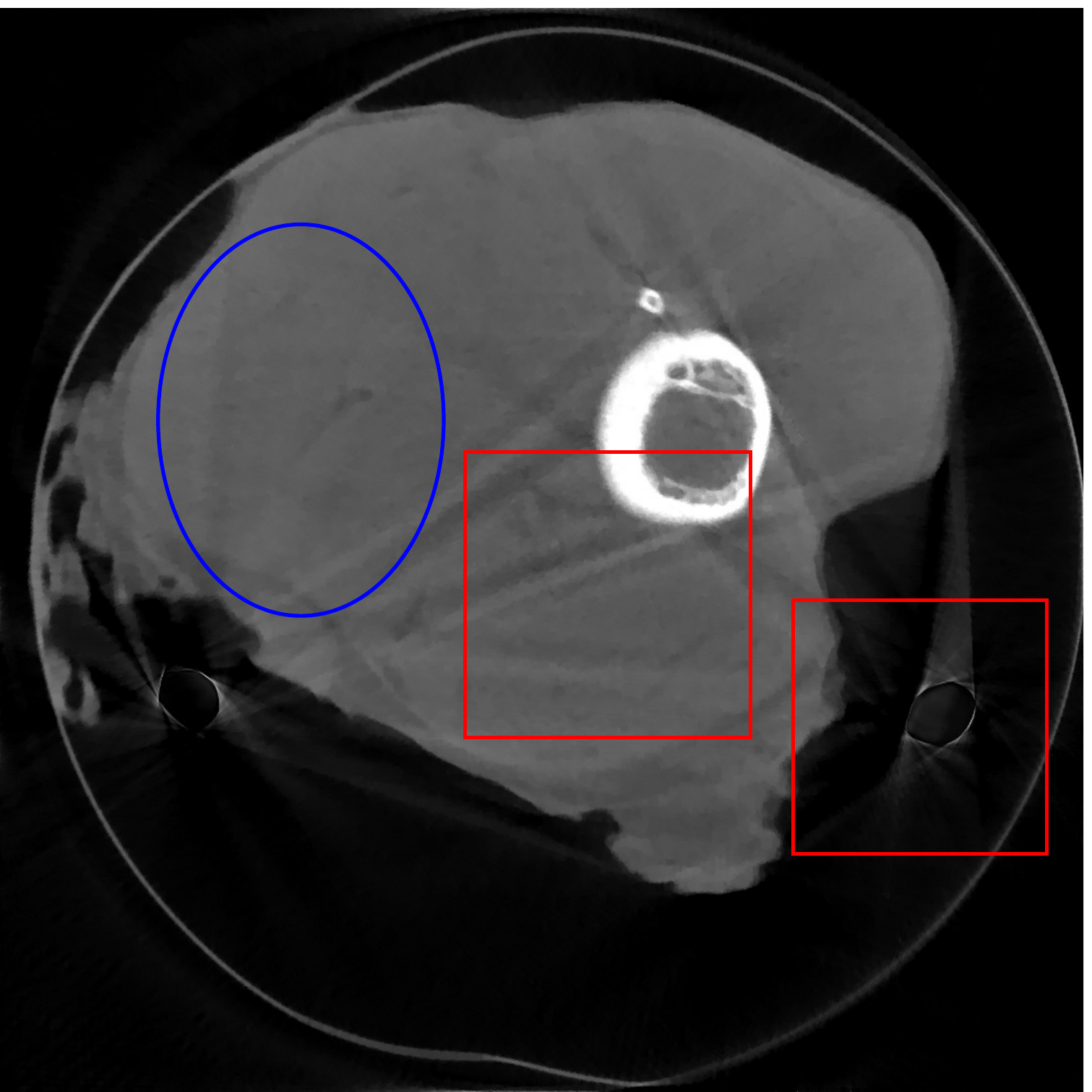}
  \end{minipage}
  }
  \subfigure[zoom-in of unweighted JSR]{
  \begin{minipage}{0.32\textwidth}\label{CW-JSR-zoom-in}
  \includegraphics[width=0.9\textwidth]{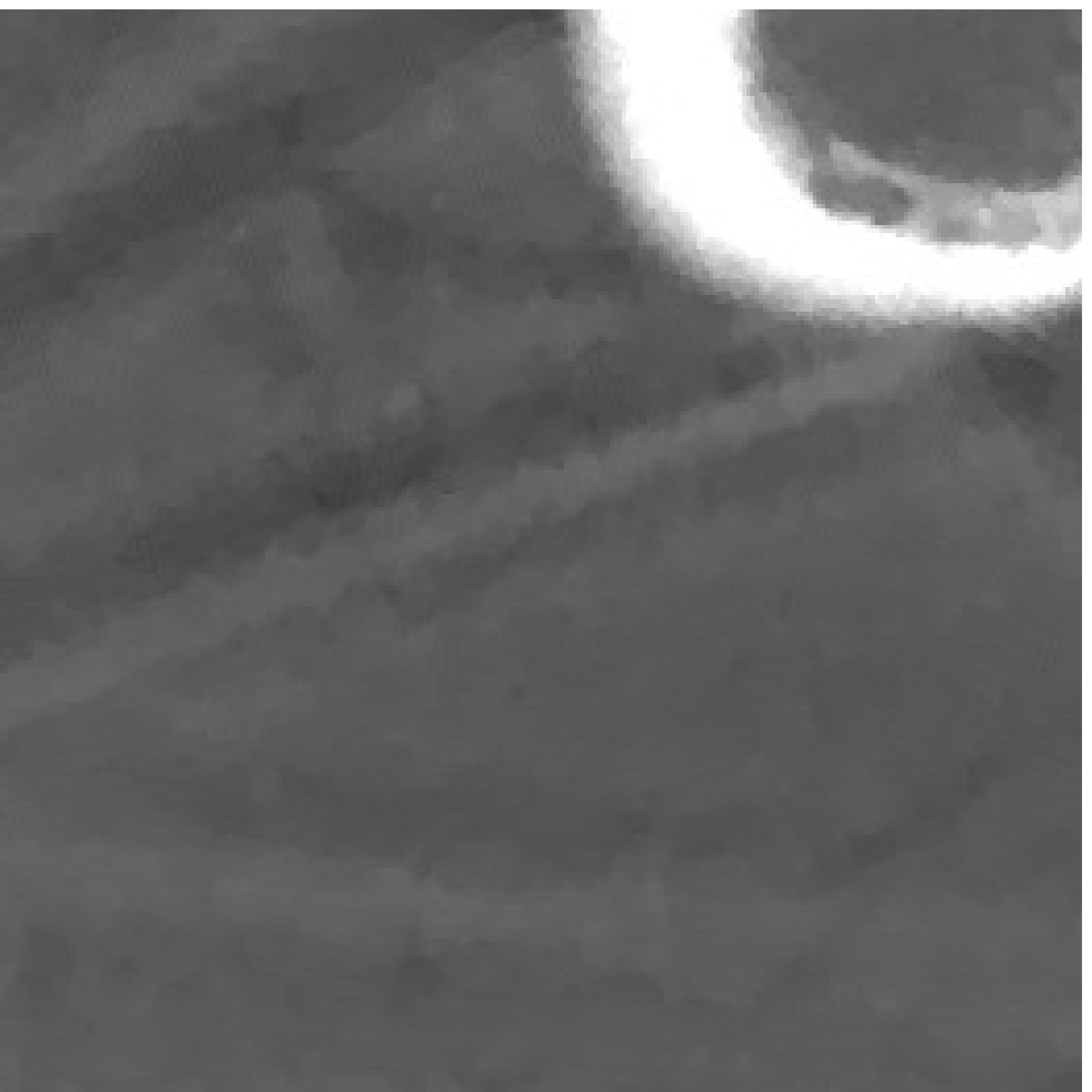}
  \includegraphics[width=0.9\textwidth]{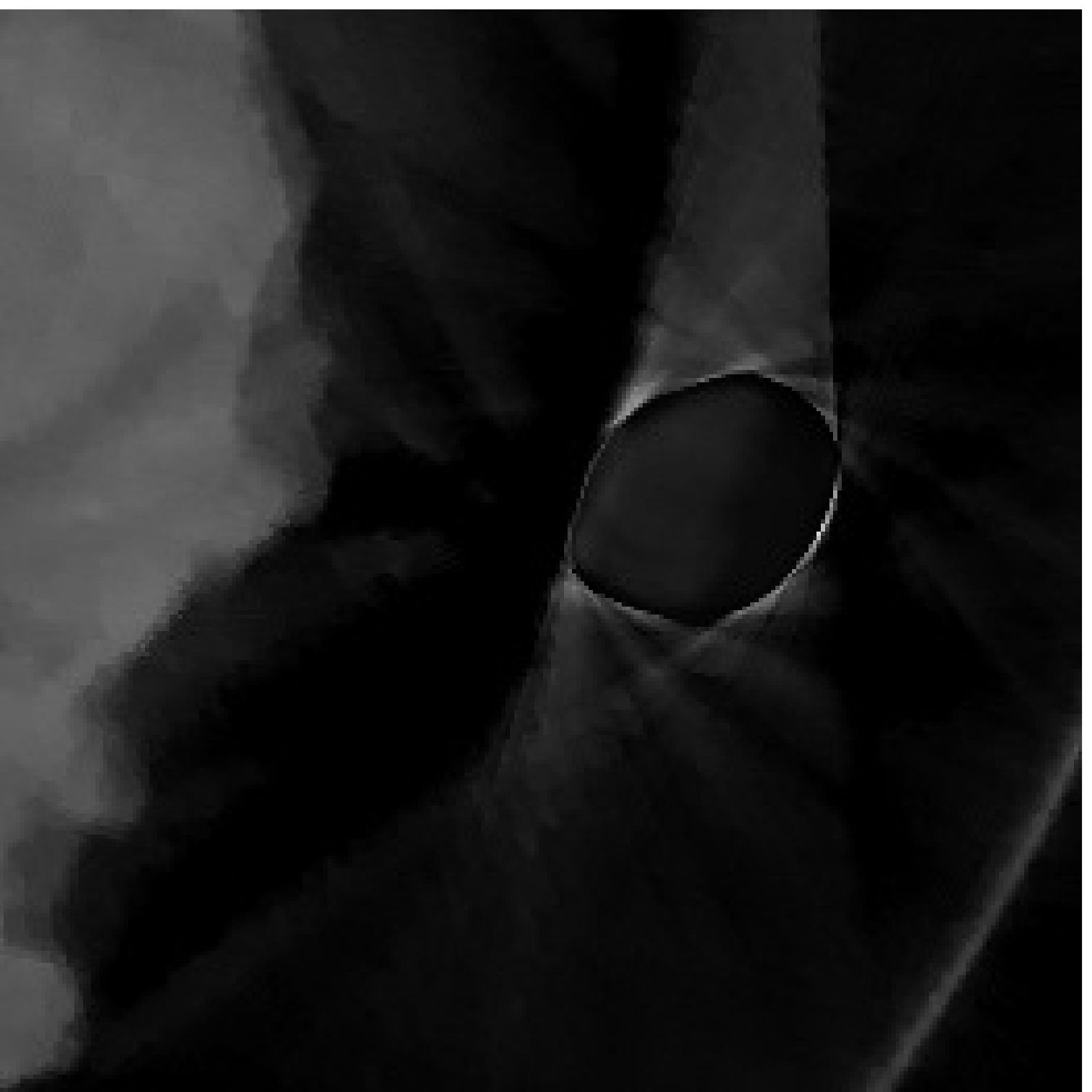}
  \end{minipage}
  }

  \subfigure[NMAR]{
  \begin{minipage}{0.64\textwidth}\label{CW-NMAR}
  \includegraphics[width=0.9\textwidth]{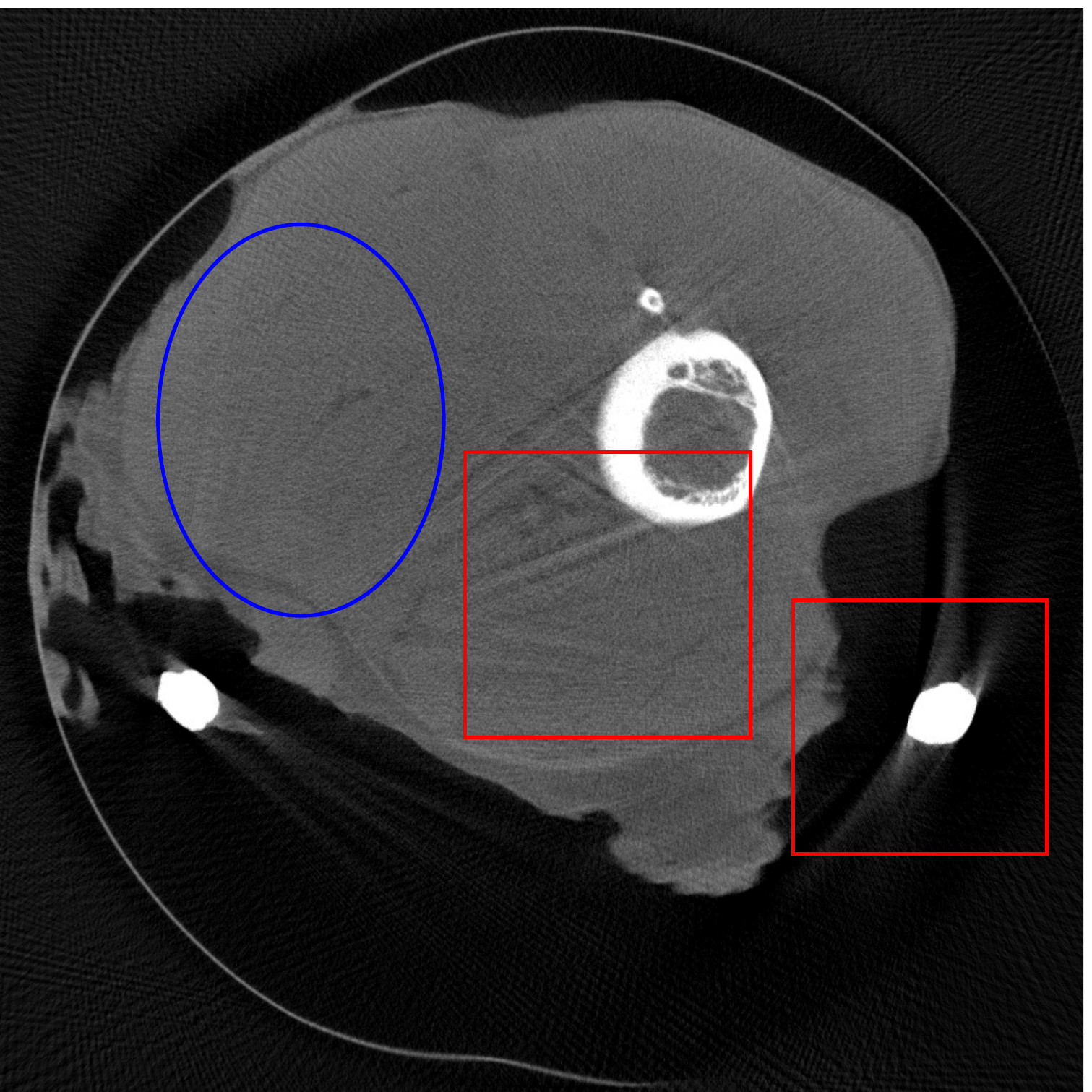}
  \end{minipage}
  }
  \subfigure[zoom-in of NMAR]{
  \begin{minipage}{0.32\textwidth}\label{CW-NMAR-zoom-in}
  \includegraphics[width=0.9\textwidth]{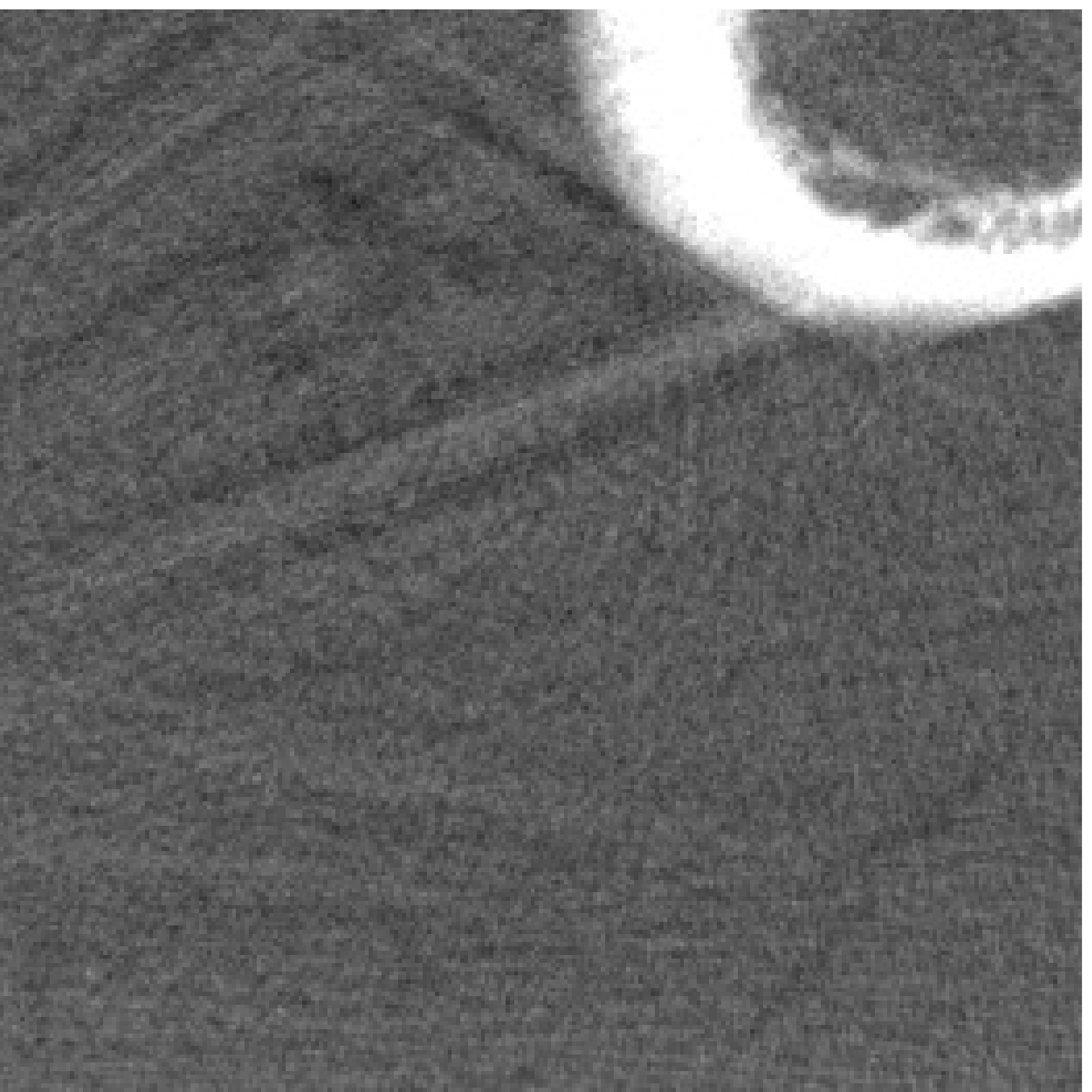}
  \includegraphics[width=0.9\textwidth]{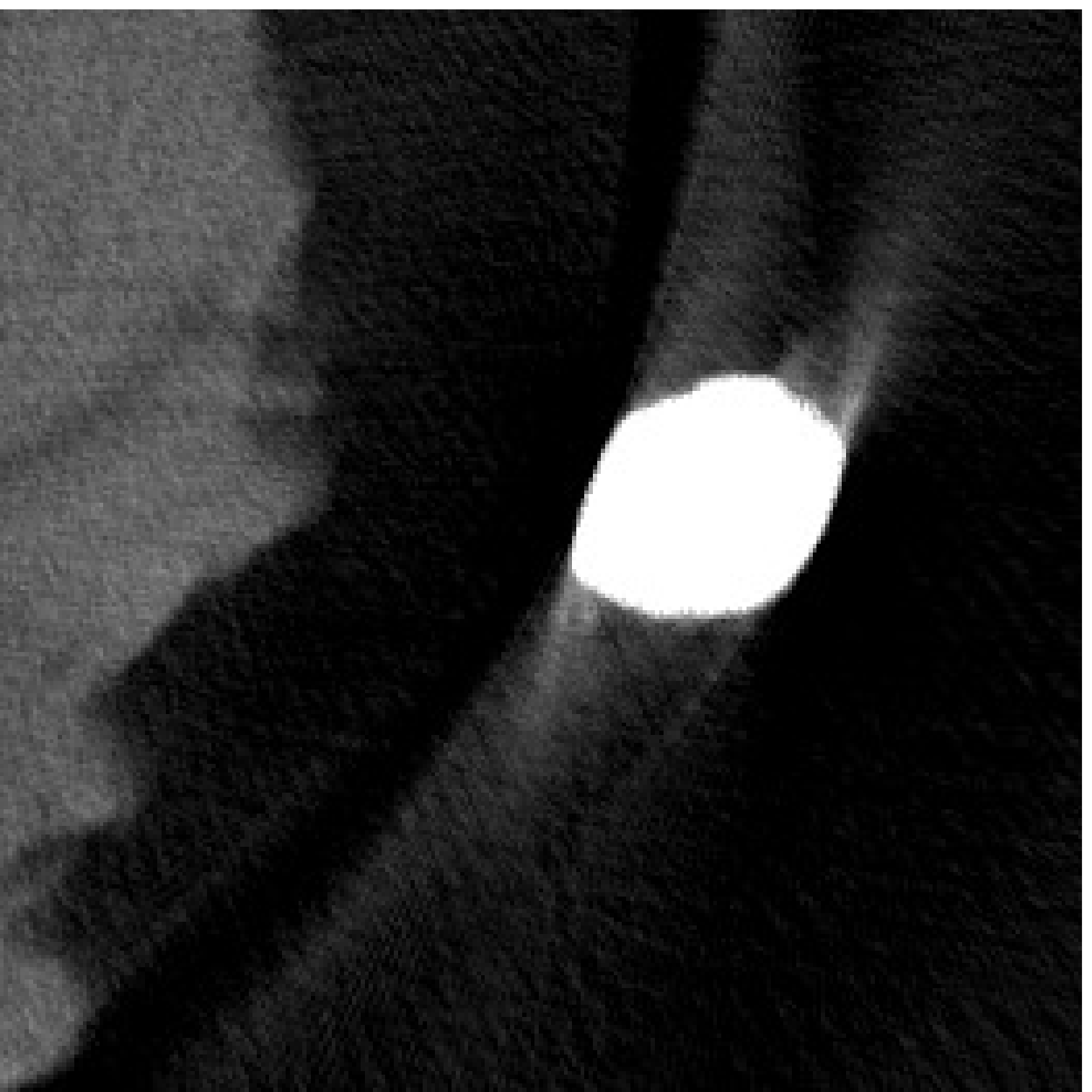}
  \end{minipage}
  }
  \caption{Comparisons of the reconstructed images.
  }
  \label{fig-CW-NMAR-JSR}
\end{figure}

\begin{figure}[p]
  \subfigure[TV-FADM]{
  \begin{minipage}{0.64\textwidth}\label{CW-TVFADM}
  \includegraphics[width=0.9\textwidth]{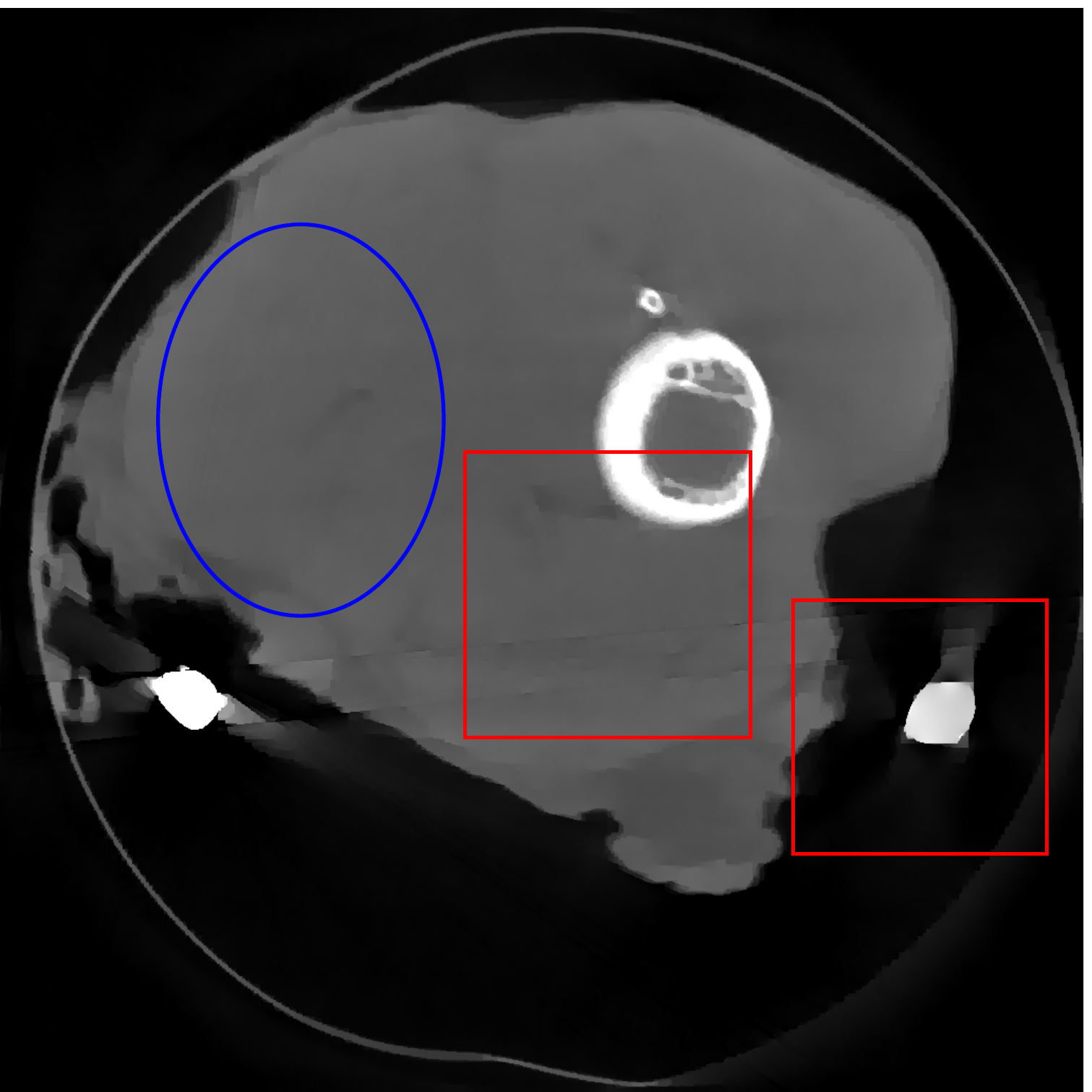}
  \end{minipage}
  }
  \subfigure[zoom-in of TV-FADM]{
  \begin{minipage}{0.32\textwidth}\label{CW-TVFADM-zoom-in}
  \includegraphics[width=0.9\textwidth]{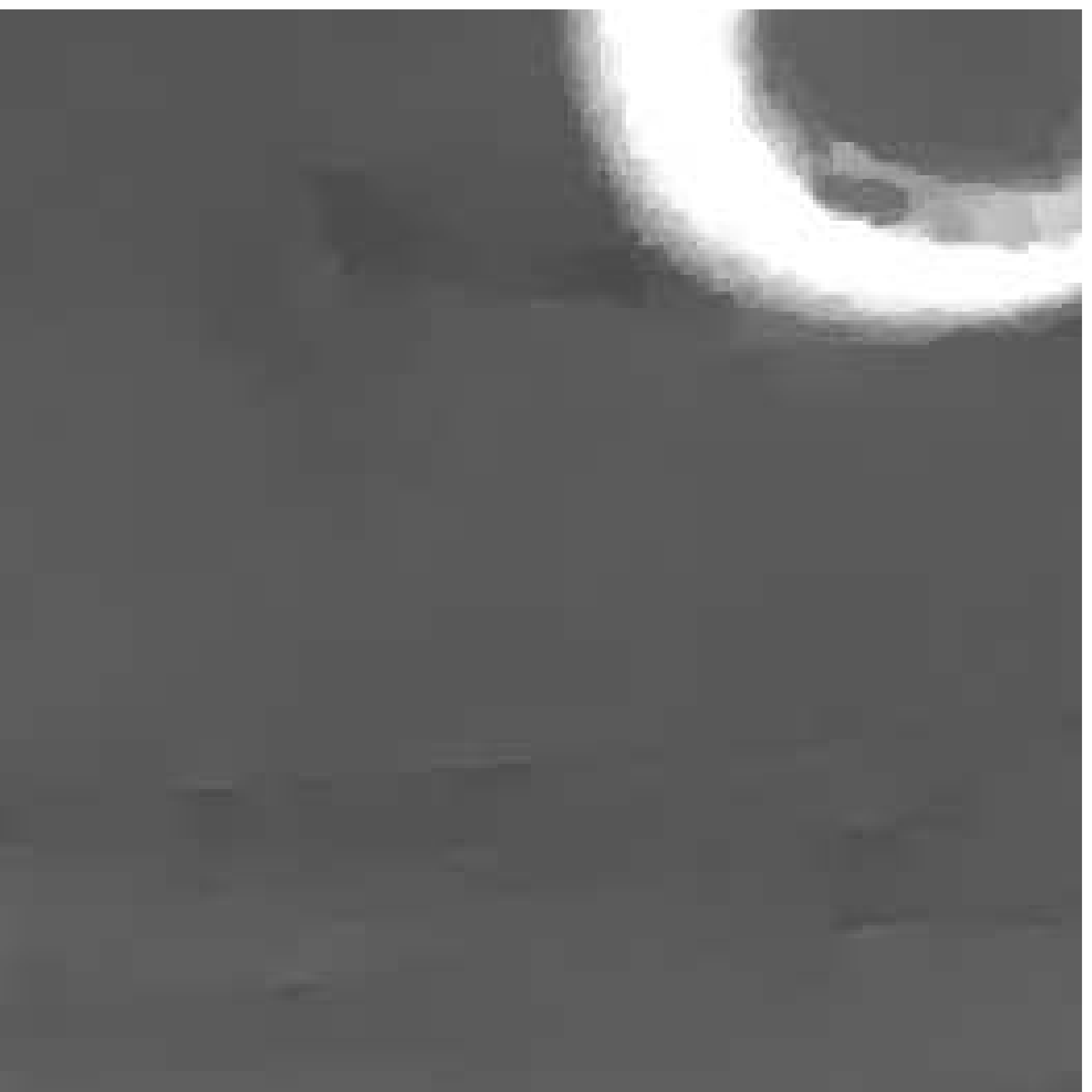}
  \includegraphics[width=0.9\textwidth]{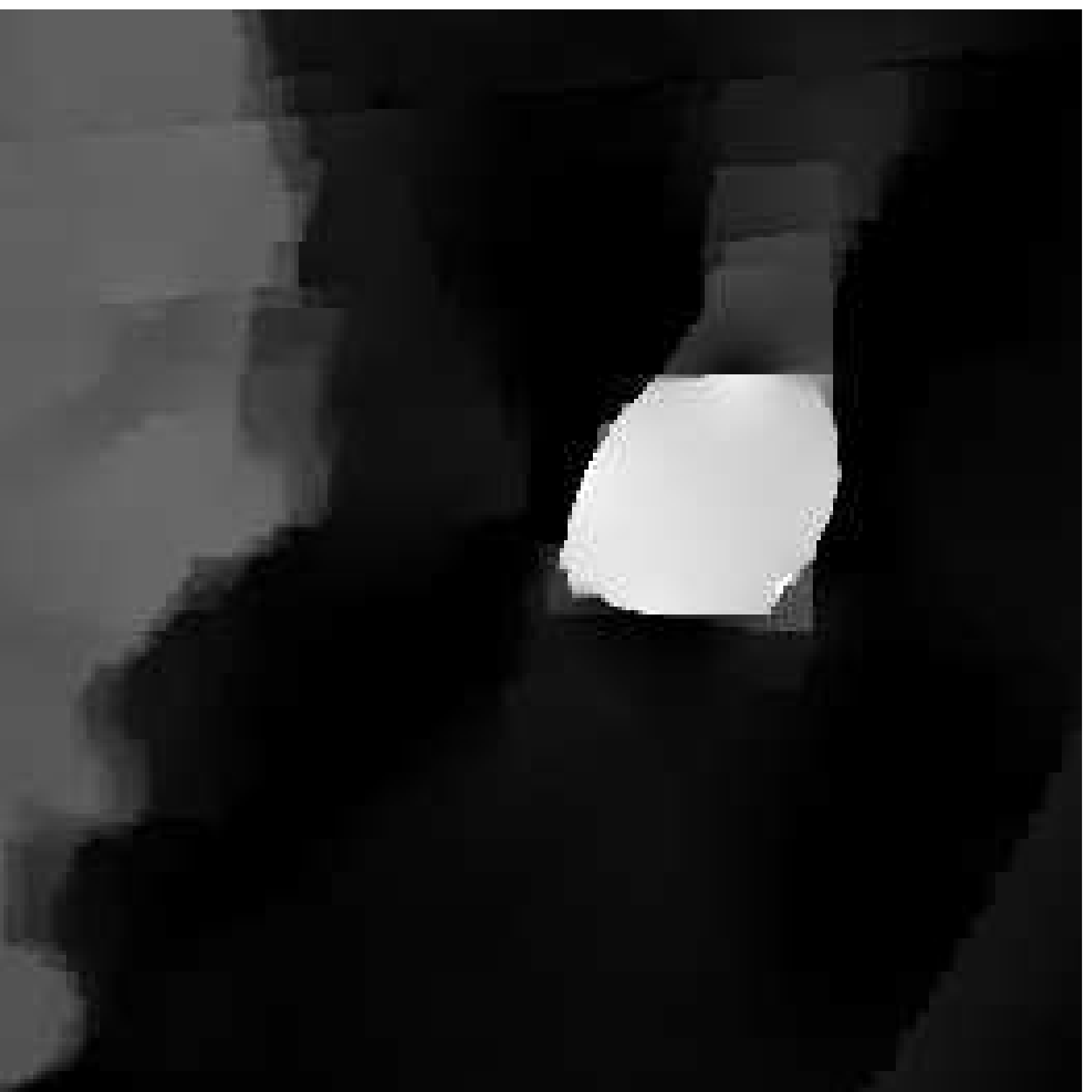}
  \end{minipage}
  }

  \subfigure[re-weighted JSR]{
  \begin{minipage}{0.64\textwidth}\label{CW-JSRDR}
  \includegraphics[width=0.9\textwidth]{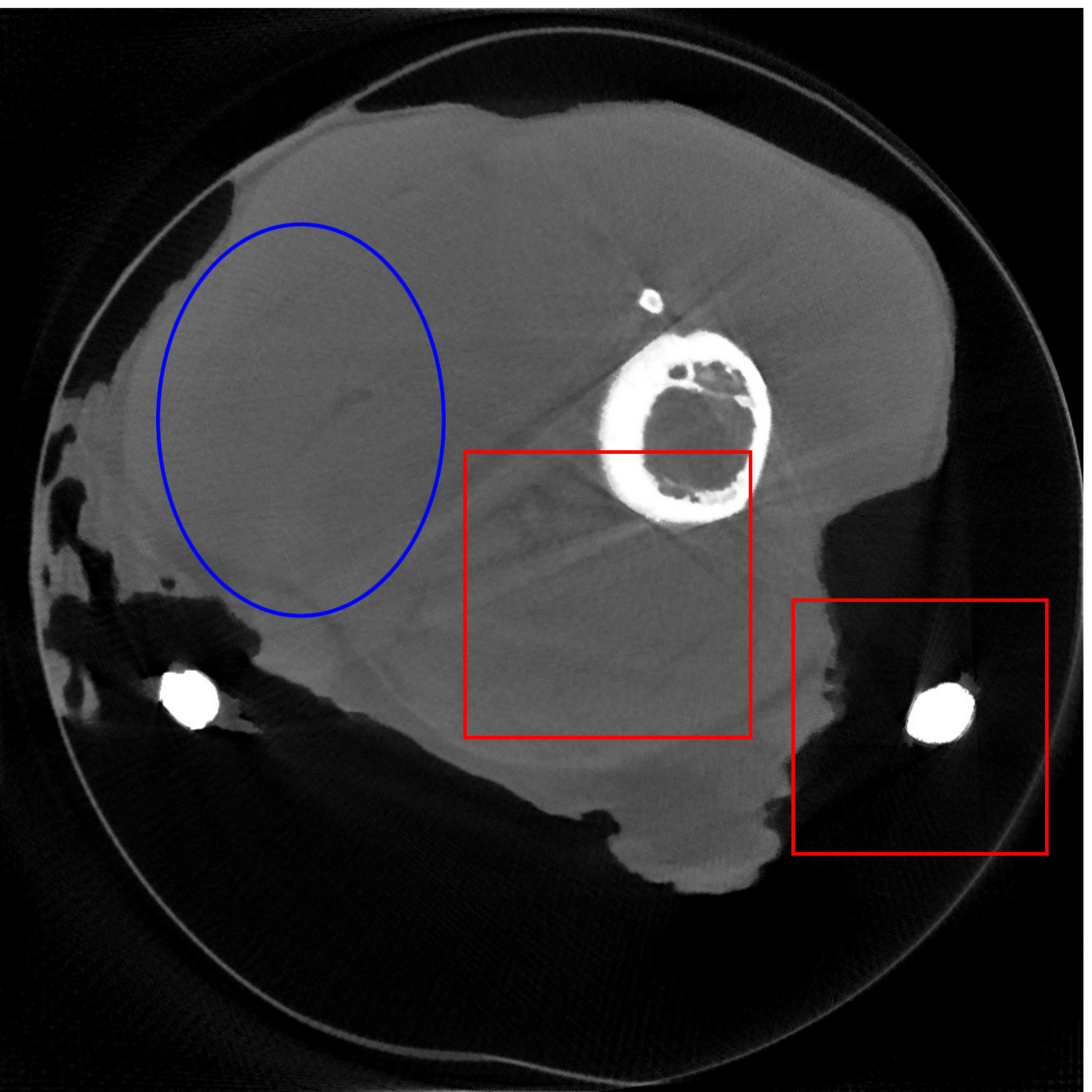}
  \end{minipage}
  }
  \subfigure[zoom-in of re-weighted JSR]{
  \begin{minipage}{0.32\textwidth}\label{CW-JSRDR-zoom-in}
  \includegraphics[width=0.9\textwidth]{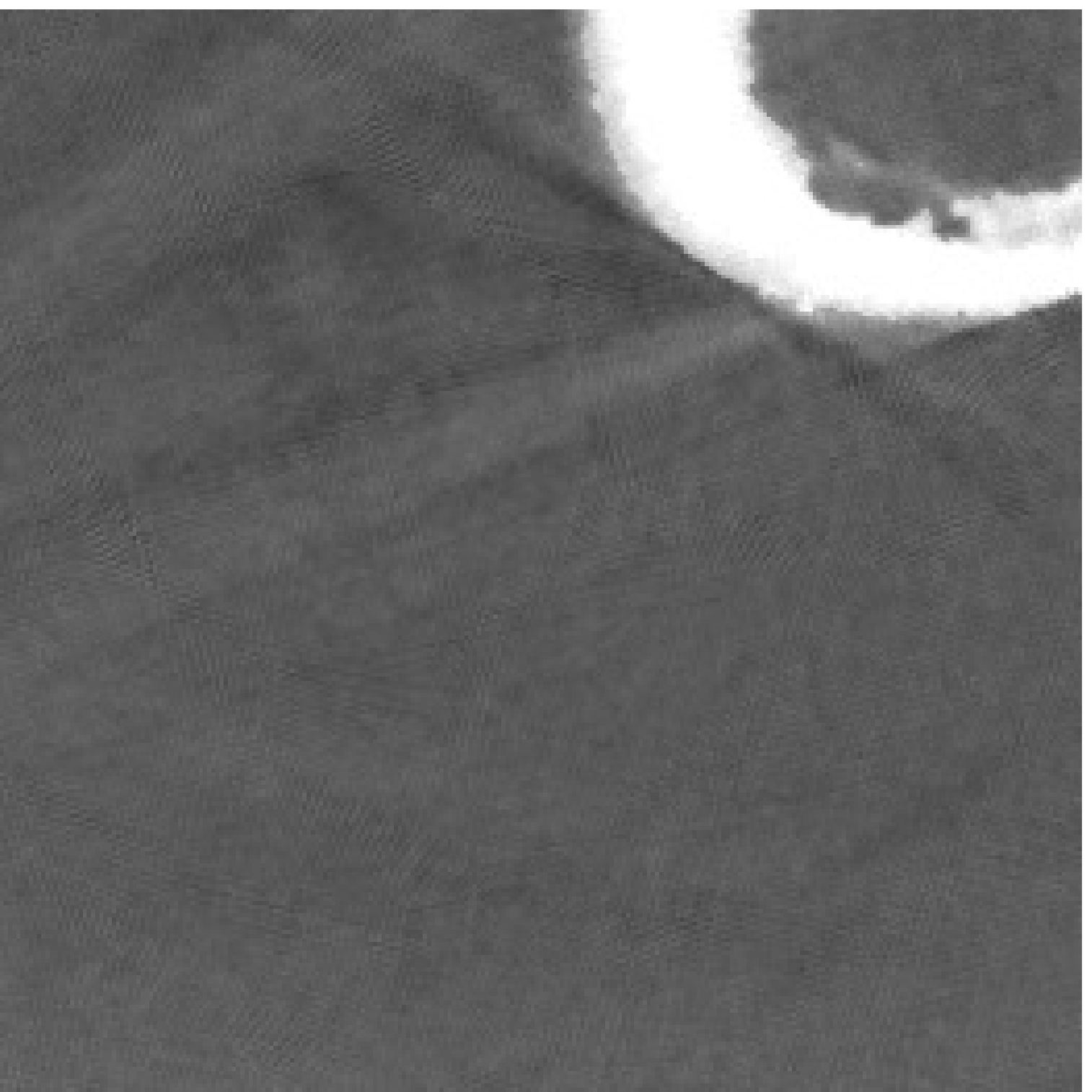}
  \includegraphics[width=0.9\textwidth]{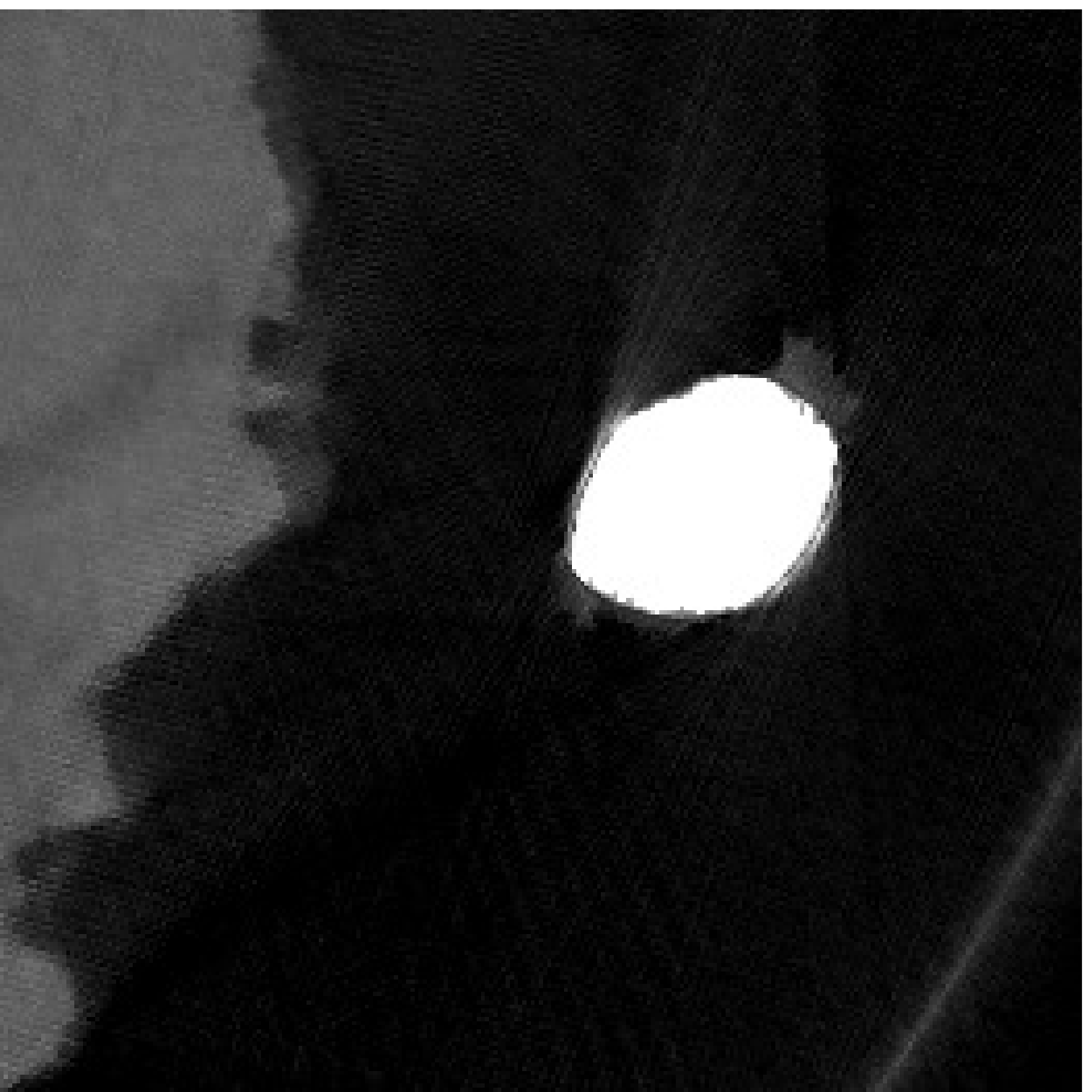}
  \end{minipage}
  }
  \caption{Comparisons of the reconstructed images.
  }
  \label{fig-CW-TV-reWJSR}
\end{figure}

Figure \ref{fig-CW-NMAR-JSR} shows a comparison between the reconstructed image from NMAR and the unweighted JSR model. Figure \ref{fig-CW-TV-reWJSR} shows a comparison between the reconstructed images from TV-FADM and the proposed re-weighted JSR model. Zoom-in views are provided in both Figure \ref{fig-CW-NMAR-JSR} and Figure \ref{fig-CW-TV-reWJSR} for a better visual assessment. As one can see that the reconstructed images from the unweighted JSR model and TV-FADM are less noisy than NMAR as indicated by the blue ellipse curve, whereas NMAR does a better job in preserving image features and suppressing metal artifacts. However, there are also new artifacts around the metal on the right as shown in Figure \ref{CW-NMAR-zoom-in}. The proposed re-weighted JSR model has best overall performance in terms of feature preservation, noise and metal artifact reduction.

\section{Conclusion}\label{sec-conclusion}

In this paper, we proposed a new model for metal artifact reduction in multi-chromatic X-ray CT imaging. The proposed model had a weighting and re-weighting mechanism naturally embedded in a framework of joint Radon domain inpainting and spatial domain regularization. Regularization by wavelet frame transforms was applied in both spatial and Radon domain to facilitate a high quality and stable image reconstruction. The proposed model was then rewritten into a more compact form so that the split Bregman algorithm could be directly applied to solve the model efficiently. Our numerical experiments using both image phantoms and real data showed that the proposed model was more effective than the classical FBP method \cite{katsevich2002theoretically}, the popular NMAR method \cite{meyer2010normalized}, and the recently proposed TV-FADM method \cite{zhang2016iterative}. Comparisons with the unweighted JSR model \cite{Dong2013} further illustrated the effectiveness of the weighting and re-weighting mechanism.

\section*{Acknowledgements}
We would like to thank the anonymous reviewers for their constructive suggestions and comments that helped tremendously in improving the presentation of this paper. We would also like to thank Professor Qibin Fan from the School of Mathematics and Statistics, Wuhan University for his support on this project.

\end{document}